\documentclass[useAMS,usenatbib,usegraphicx]{mn2e}
\usepackage{color}

\newcommand{\feoh}{\textrm{[Fe/H]}}
\newcommand{\cfe}{\textrm{[C/Fe]}}
\newcommand{\nfe}{\textrm{[N/Fe]}}
\newcommand{\ofe}{\textrm{[O/Fe]}}
\newcommand{\eufe}{\textrm{[Eu/Fe]}}
\newcommand{\bafe}{\textrm{[Ba/Fe]}}
\newcommand{\euba}{\textrm{[Eu/Ba]}}
\newcommand{\xfe}{\textrm{[X/Fe]}}
\newcommand{\abra}[2]{[{\rm {#1}}/{\rm {#2}}]}
\newcommand{\nbyc}{[{\rm N}/{\rm C}]}
\newcommand{\cemps}{CEMP-{\it s}}
\newcommand{\cempnos}{CEMP-no{\it s}}
\newcommand{\aelm}{$\alpha$-element}
\newcommand{\aelms}{$\alpha$-elements}
\newcommand{\is}{{\it s}}
\newcommand{\ir}{{\it r}}
\newcommand{\sit}{{\it s}-}
\newcommand{\rit}{{\it r}-}
\newcommand{\rps}{{\it r}+{\it s}}
\newcommand{\nucm}[2]{\ensuremath{{}^{#1}{\rm #2}}}
\newcommand{\msun}{\ensuremath{M_{\odot}}}
\newcommand{\loge}{\ensuremath{\log \epsilon}}
\newcommand{\logli}{\ensuremath{\log \epsilon ({\rm Li})}}
\newcommand{\teff}{\ensuremath{T_{\textrm {eff}}}}
\newcommand{\pow}[2]{\ensuremath{{#1} \times 10^{#2}}}
\newcommand{\ciso}{\ensuremath{\nucm{12}{C} / \nucm{13}{C}}}
\newcommand{\cpocket}{\nucm{13}{C} pocket}
\newcommand{\hyp}{\ensuremath{\, \mathchar`- \,}}

\title[SAGA Database II]{The Stellar Abundances for Galactic Archaeology (SAGA) Database II
-- Implications for Mixing and Nucleosynthesis in Extremely Metal-Poor Stars and Chemical
Enrichment of the Galaxy}
\author[T. Suda et al.]{Takuma Suda$^{1,2}$\thanks{E-mail:suda@astro.keele.ac.uk},
		Shimako Yamada$^{2}$,
        Yutaka Katsuta$^{2}$,
        Yutaka Komiya$^{3}$,
\newauthor % starts a new line in the author environment
		Chikako Ishizuka$^{1,2}$,
		Wako Aoki$^{3}$, and
		Masayuki Y. Fujimoto$^{2}$\\
$^{1}$Astrophysics Group, EPSAM, Keele University, Keele, Staffordshire ST5 5BG, UK\\
$^{2}$Department of Cosmoscience, Hokkaido University, Kita 10 Nishi 8, Kita-ku, Sapporo 060-0810, Japan\\
$^{3}$National Astronomical Observatory of Japan, Osawa 2-21-1, Mitaka, Tokyo 181-8588, Japan}
\begin{document}

\date{Accepted 2010 October 28. Received 2010 September 29; in original form 2010 February 24}

\pagerange{\pageref{firstpage}--\pageref{lastpage}} \pubyear{2002}

\maketitle

\label{firstpage}

\begin{abstract}
We discuss the characteristics of known extremely metal-poor (EMP) stars in the Galaxy using the Stellar Abundances for Galactic Archaeology (SAGA) database (Suda et al. 2008, PASJ, 60, 1159).
The analyses of carbon-enhanced stars in our sample suggest that the nucleosynthesis in AGB stars can contribute to the carbon enrichment in a different way depending on whether the metallicity is above or below $\feoh \sim -2.5$, which is consistent with the current models of stellar evolution at low metallicity.
We find the transition of the initial mass function at $\feoh \sim -2$ in the viewpoint of the distribution of carbon abundance and the frequency of carbon-enhanced stars.
For observed EMP stars, we confirmed that some, not all, of observed stars might have undergone at least two types of extra mixing to change their surface abundances.
One is to deplete the lithium abundance during the early phase of red giant branch.
Another is to decrease the C/N ratio by one order of magnitude during the red giant branch phase.
Observed small scatters of abundances for \aelms\ and iron-group elements suggest that the chemical enrichment of our Galaxy takes place in a well-mixed interstellar medium.
We find that the abundance trends of \aelms\ are highly correlated with each other, while the abundances of iron-group elements are subject to different slopes relative to the iron abundance.
This implies that the supernova yields of \aelms\ are almost independent of mass and metallicity, while those of iron-group elements have a metallicity dependence or mass dependence with the variable initial mass function.
The occurrence of the hot bottom burning in the mass range of $5 \la M / \msun \la 6$ is consistent with the initial mass function of the Galaxy peaked at $\sim 10 \hyp 12 \msun$ to be compatible with the statistics of carbon enhanced stars with and without \sit process element enhancement, and of nitrogen enhanced stars.
For \sit process elements, we find not only the positive correlation between carbon and \sit process element abundances, but the increasing slopes of the abundance ratio between them with increasing mass number of \sit process elements.
The dominant site of the \sit process is still an open question because none of the known mechanisms of the \sit process is able to account for this observed correlations.
In spite of the evidence of AGB evolution in observed abundances of EMP stars, we cannot find any evidence of binary mass transfer through the effect of dilution in the convective envelope.
We found the dependence of sulphur and vanadium abundances on the effective temperatures in addition to the previously reported trends for silicon, scandium, titanium, chromium, and cobalt.
\end{abstract}


\section{Introduction}

One of the fundamental problems in astrophysics is the primary site of element production.
Observing extremely metal-poor (hereafter EMP) stars is an assured method of tracing the nuclear history of the universe in connection with the stellar nucleosynthesis and chemical evolution.
Thanks to the many efforts of large-scale survey of metal-poor stars in the Galactic halo \citep{Bond1980,Beers1992,Christlieb2001,Christlieb2008}, we have several hundreds of stars having metallicity of $\feoh < -2$ so far.
According to the recent SDSS/SEGUE project \citep{Beers2010}, the number of identified stars in this metallicity range has increased to more than $> 25,000$.
Moreover, the follow-up observations with higher resolution spectra revealed the detailed chemical abundances of hundreds of halo stars.

The increasing amount of data confirms that the stellar abundances of EMP stars are not uniform, but rather largely dispersed.
This is not surprising if we assume the change of chemical abundances in the interstellar medium from which the stars were born.
The pollution of stellar surface or entire star is caused by supernovae from the previous generation(s) and/or the after-birth modification of surface abundances through the accretion of interstellar gas and by binary mass transfer.
Theoretically, it can be interpreted by the chemical evolution of the Galaxy;
the \aelm\ abundances of EMP stars are claimed to be explained by the first-generation core-collapse supernovae \citep{Shigeyama1998};
the individual supernovae are argued to be reflected in the scatter of neutron capture elements \citep{Tsujimoto2000,Ishimaru2004}.
Some of the abundance patterns can also be interpreted in terms of the nucleosynthesis of EMP AGB stars and the binary mass transfer.
\citet{Suda2004} and \citet{Komiya2007} pointed out the importance of the role of binaries in understanding the characteristics of low-mass survivors from the early epoch of the Galactic halo through the impact of distinctive evolution of EMP stars;
the stars of mass below $\sim 3 \msun$ give rise to hydrogen ingestion into the helium-flash convective zone and subsequent dredge-up of nuclear products in early TP-AGB phase for $\feoh \la -2.5$, while only third dread-up (TDU) can work for more massive EMP stars and for more metal-rich stars \citep{Fujimoto1990,Hollowell1990,Fujimoto2000,Iwamoto2004}.
This is called the He-Flash Driven Deep Mixing (hereafter He-FDDM), and later found by other groups \citep[Dual Shell Flash or Helium-Flash-Driven Mixing by][respectively]{Campbell2008,Lau2009}.

The observations of EMP stars have revealed many peculiar characteristics in elemental abundances.
These abundance patterns are the imprints of metal enrichment by supernovae and bring precious information on the yields of metal-poor supernovae in the early Universe \citep[e.g.][]{Umeda2005,Kobayashi2006,Tominaga2007}.
The pristine interstellar medium is also argued to be affected by mass loss from rotating massive stars especially for CNO elements \citep{Hirschi2007,Meynet2010}.
Furthermore, the remnant stars with different abundances give insight into the star formation history and the initial mass function in their host cloud.

One of the examples is the larger frequency of carbon-rich stars compared with stars with higher metallicity \citep{Norris1997a,Rossi1999}.
These stars are called ``CEMP stars'' (Carbon-Enhanced Metal-Poor stars) and are now divided into several subclasses such as ``CEMP-{\it s}'' (CEMP stars enriched with \sit process elements), ``CEMP-no'' (CEMP stars with normal abundances of neutron-capture elements), ``CEMP-{\it r}'' (CEMP stars enriched with \rit process elements), etc. \citep[see, e.g.][]{Beers2005a}.
The definition of these subclasses is determined mainly by the ratio of barium to europium abundances.
For \cemps\ stars, the observed chemical composition is reasonably explained by the nucleosynthesis in AGB stars \citep[see, e.g.][]{Bisterzo2009}.
These stars are thought to be born in binary systems at their birth and affected by mass transfer from intermediate-mass primaries.
In particular, this is the case when the primary companion is an EMP star of $M \la 3 \msun$ which experiences the He-FDDM event.
On the other hand, the CEMP stars, not enriched with \sit process elements, which we define as \cempnos, can result from the mass transfer in the binary systems with the primaries of $M > 3 \msun$ that experience only TDU without hydrogen ingestion nor without efficient formation of the so called \nucm{13}{C} pocket, as we proposed in \citet[][see also Suda \& Fujimoto 2010]{Komiya2007}.  
Based on these theoretical interpretation, \citet{Komiya2007} deduce the high-mass nature of IMF for the EMP stellar population that leaves behind the low-mass survivors, currently observed in the Galactic halo as EMP stars, and the predominance of low-mass members born in the binary systems among the latter from the statistics of observed carbon-enhanced EMP stars.   
This is also shown to be consistent with the observed number density of EMP stars in the Galactic halo \citep{Komiya2009a}. 
Furthermore \citet{Komiya2009a} and \citet{Komiya2009b} show that the metallicity distribution function observed from the Galactic halo can be reproduced by fully considering the secondary components of binary stars with the derived IMF peaked around $\sim 10 \msun$.

It is true, however, that there is argument that these groups are heterogeneous and contain the stars of other origins \citep{Ryan2005};
e.g., CEMP-no stars dominate over \cemps\ stars for $\feoh < -3$, as pointed out by \citet{Aoki2007b}.
The possible scenarios for these stars are discussed, though not conclusive yet, such as N-producing hypernovae \citep{Norris2001} and internal mixing in RGB stages to explain the abundance of CS22949-037, supernovae with small iron ejecta as suggested for CS29498-043 by \citet{Aoki2002d}.
As for the \cemps\ group, there are variations in the ratio between Ba and Eu abundances, some of which show the abundance ratio slightly smaller than realized by standard \sit process nucleosynthesis.
Furthermore, it is known that three most iron-poor stars all show great enhancement of carbon, one of which is a \cempnos\ star, while have only the upper limits of Ba abundances are available for other two stars.
For proper understandings of these characteristics, it may not suffice to work out only the evolution of stars and the nucleosynthetic sites that can reproduce the abundance patterns.
It is also necessary to reveal the formation processes and history of the relevant stars with given mass and metallicity.

In order to discuss the origins of elements and the star formation history in the Galaxy, we have developed a database for stellar abundances of Galactic stars holistically, mainly focused on halo stars \footnote{The Stellar Abundances for Galactic Archaeology (SAGA) database, available at http://saga.sci.hokudai.ac.jp} \citep[][hereafter Paper I]{Suda2008}.
We have assembled the data of the reported stellar abundances for any element species in the literature including the isotopic ratios, atmospheric parameters, photometric parameters, equivalent widths, binary status and period if known, observing log, and the basic information of the literature and individual objects.
Most of the compiled data are available on the web by using the data retrieval system that enables us to see the relationships graphically between two specified quantities.
The database is designed to compile all the stars having $\feoh \leq -2.5$.
In our compilations, all the reference stars with $\feoh > -2.5$ in the same literature are compiled except for the stars in the Galactic globular clusters.
The reason for the choice of this metallicity range is the theoretical correspondence that evolved stars with $\feoh \la -2.5$ are expected to undergo the hydrogen mixing into the helium-flash driven convective zone and to be enriched with nuclear products by the subsequent dredge-up as stated above.

In this paper, we extend the previous work on the global characteristics of EMP stars by exploiting the large sample stars collected by the SAGA database.
We intensively discuss the characteristics of EMP stars by visualizing the relationship between any combinations of two quantities stored in the database.
In order to take advantage of this large database, we also utilize statistical tools to obtain robust conclusions in the characteristics of EMP stars.
The present paper focuses on the evolutionary effect of low- and intermediate-mass stars on the possible change of elemental abundances.
The paper explores the chemical enrichment history of the Galactic halo as well.
We also discuss unknown correlations or properties among element abundances that have not been reported.

The paper is organized as follows.
In the next section, we refer to the updated version of the SAGA database since the publication of Paper I.
We also describe the sample of the SAGA database, the method of approach with the statistical model, and the overall characteristics of the sample in the database.
In \S \ref{sec:cemp}, we discuss the characteristics of CEMP stars.
The diversity and peculiarity of abundances observed for EMP stars are investigated with respect to $\alpha$, iron peak and neutron capture elements in \S \ref{sec:div}.
The chemical enrichment of the Galaxy and constraints on the supernovae yields that have promoted it are discussed in \S \ref{sec:enrich}.
We investigate the possibility of extra mixing and \sit process in helium-flash convective zones in EMP stars in \S \ref{sec:mixing} and \S \ref{sec:spr}, respectively.
\S \ref{sec:bin} is devoted to elucidating evidence of binary mass transfer in EMP stars.
In \S \ref{sec:unexp}, we report unexpected correlations and trends of abundances discernible among our EMP star sample.
Conclusions and summary follows in \S \ref{sec:sum}.

\section{Description of Sample Data}

In this section, we first present a brief summary of specifications of the SAGA database and the tools for analyses.
Then, we describe the characteristics of our sample stars in particular in relation to carbon-enhanced stars.
We reveal an abrupt change in the distribution of carbon abundances relative to iron around the metallicity of $\feoh \simeq -2$ in addition to the variation in the frequency of carbon-enhanced stars around the similar metallicity, as previously pointed out.
On the basis of these findings, we further argue the transition of the initial mass function of the Galactic halo stars to a low-mass one, typical to Populations I and II stars, from a high-mass one, derived for extremely metal-poor stars by \citet{Komiya2007}.

\subsection{Sample Selection from the SAGA Database}\label{sec:sample}

We extend the dataset compiled in the SAGA database by covering the papers published by the end of 2009, as compared with December 2007 in Paper I.
The data compilation traces back to 1994 depending on the availability of data tables in electronic format that is technically important for our compilation process.
We believe that we have completed the compilation of papers containing stars with $\feoh \leq -2.5$ at least for literature published since 2000.
We have compiled 158 papers containing 1386 unique stars through the periodical update of the database.
The total number of the independently reported element abundances is 23775 for which the data from multiple papers for the same element in the same object are counted separately.
The data are analyzed by the data retrieval system that we developed as an online tool to search and plot the stellar parameters and abundances.
A more detailed description of the data retrieval system is given in Paper I.

We have adopted different criteria for default data selection in plotting figures from those in Paper I, when multiple papers report the same quantities for the same objects.
In Paper I, we gave a priority, in order, on the latest publication, lower ionization states, and smaller values of error bars.
This ranking sometimes produces unreliable results because some of the recent observations concern statistical discussion using a large number of samples with low and intermediate resolution spectra.
In order to avoid these low-resolution based data in the default selection for plotting, we considered the spectral resolution in the observational setup.
Therefore, we estimated the quality of individual data based on information of the resolution and the other measures used in Paper I.
Using this data scoring, the data selection is automatically determined by imposing the preferences:
1) the highest resolution of observations, 2) the latest published year, and 3) the smallest values of errors (of course the data with upper limits are not preferred).
We also choose one datum for a specific element in the case of available element abundances derived by different ionization states, or by atomic and molecular lines.
We prefer a lower ionization state (neutral atom) for multiple data with different ionization levels.
For abundance data with atomic and molecular lines, we adopt the following ones: CH for carbon, CN for nitrogen, [O I] for oxygen.
These are treated in the same way as in Paper I.
In this paper, we rely on this default selection of data to produce figures, unless stated otherwise.

Here we note the importance of the line selection in determining the element abundances.
It is desirable to adopt the most reliable lines for individual data in plotting figures.
However, it is difficult to quantify the reliability of lines for each element because it can be different depending on atmospheric parameters and observational setups.
Therefore, the present method of data selection is still to be improved.
We try not to consider the abundances derived by non-LTE and 3D correction in an effort to use the homogenized data.

We use statistical tools to analyze the characteristics of the sample.
In the following figures, we use the Pearson coefficient to see the degree of global correlation between two element abundances or other attributes.
We also use a reduced major axis (RMA) regression using the statistics software R to see the relation between two element abundances in some of the figures\footnote{The R is available at http://www.r-project.org}.
This is used to quantify a slope and an intercept by linear fit, expressed by ``a'' and ``b'' in this paper, respectively.
It is to be noted that these values and their 95 \% confidence interval do not take into account the systematic errors probably mainly caused by the different setup of observations.

We also performed interval estimations for the dispersion of data in order to remove apparent scatter due to statistical effect.
We derived the interval for standard deviation under 95 \% confidence level at a certain metallicity range that is set at 0.5 dex in the current analyses.
The estimated intervals provide the intrinsic scatter of population (mainly halo stars) from the known sample mean and variance.
As seen below, the scatters of element abundances are obtained as a function of metallicity for data with hundreds of reported abundances.
Unless noted in the text, it is assumed that all the objects belong to the same population.
This assumption is in general not true, but will be reasonable as long as we discuss the chemical evolution of the Galactic halo such as abundance trends of \aelms\ and iron group elements for stars with $\feoh \la -1.5$.
The identification of the population to which the stars belong is outside the scope of the present paper, but should be considered in detail in the future.
Indeed, in some cases, we derived the mean and intervals of standard deviation for data without carbon-enhanced star groups because these groups can be considered to have different origins for some elements.

In using these statistical analyses, we excluded the data for which only upper limits are available.

\subsection{Characteristics of sample stars}\label{sec:imf}

\begin{figure}
  \includegraphics[width=84mm]{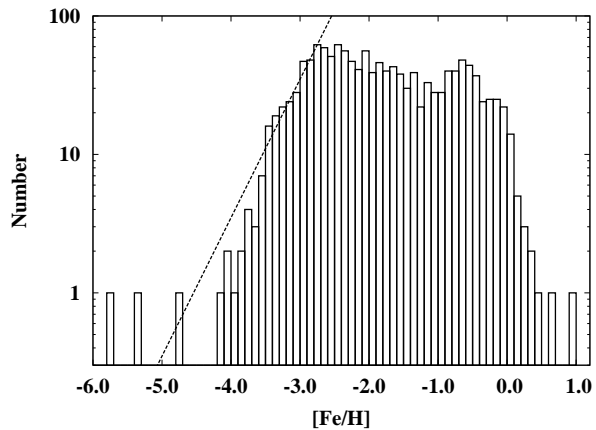}
  \caption{Metallicity distribution function of all the sample stars, collected by the SAGA database.
  Broken line denotes a linear relationship $\log n \sim \feoh$ expected from a closed box model of chemical evolution with a fixed initial mass function.
  }
  \label{fig:mdf}
\end{figure}

Fig.~\ref{fig:mdf} shows the metallicity distribution of all our 1386 sample stars with 0.1 dex binning that is the updated version of Fig.~5 in Paper~I.
The obtained MDF peaks around $\feoh \simeq -2.8$, which is an artefact of the preference of extremely metal-deficient stars in the surveys.
However, as stated in Paper~I, the sampling of target stars for abundance determinations can be regarded as unbiased below this metallicity since the metallicity of $\feoh \la -2.5$ can be determined only through high dispersion spectroscopy.
As previously pointed out \citep[e.g., see][]{Beers2005a}, the MDF of EMP survivors nearly follows a linear relationship with the iron abundance down to the metallicity $\feoh \simeq -3.5$, which is expected from a simple closed box model of chemical enrichment with constant initial mass function. 
For the metallicity below this range, the number of EMP stars tends to fall below the linear relationship.  
For still smaller metallicity than $\feoh \simeq -4.5$, on the other hand, there are three low-mass survivors in excess of the linear relationship, which are sometimes called as hyper or ultra metal-poor (HMP/UMP) stars.

% Figure C-abundance distribution
\begin{figure}
  \begin{center}
    \includegraphics[width=80mm]{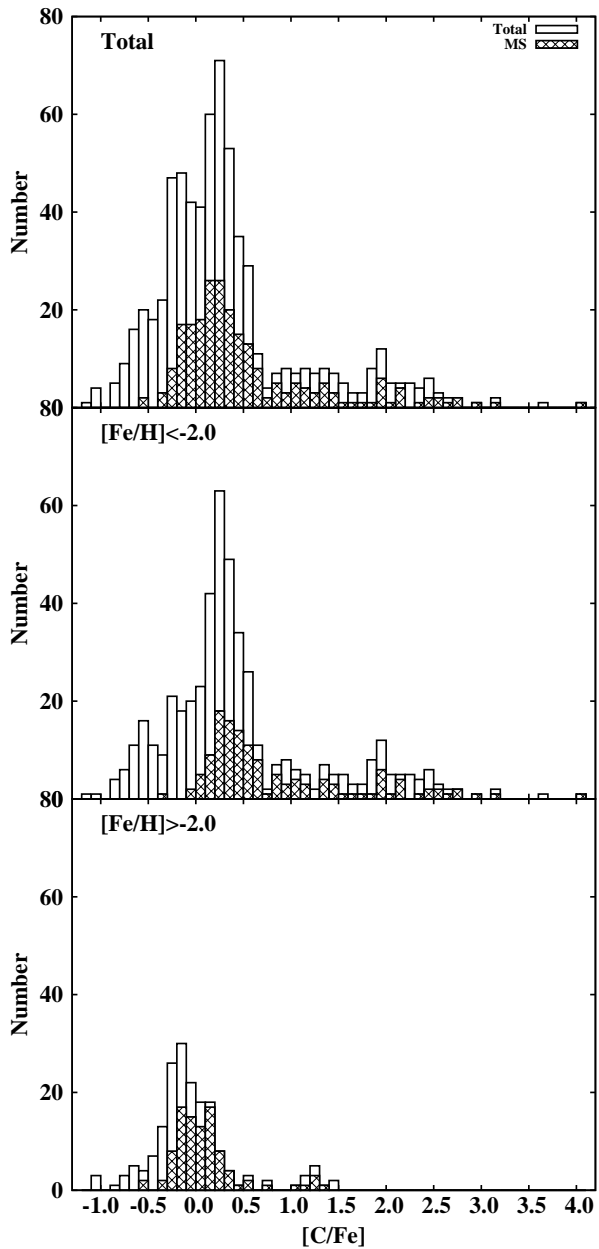}
  \end{center}
  \caption{Distribution of carbon enrichment, \cfe, relative to iron, plotted by the data retrieval system for 678 sample stars with the carbon abundances registered in the SAGA database (top panel).
	      Hatched boxes denote the distribution for dwarf stars classified as ``MS'' in the database.
	     Middle and bottom panels show the same as top panel but for the metallicity of $\feoh \le -2$ and $\feoh > -2$, respectively.
  }\label{fig:chist}
\end{figure}

\begin{table*}
  \caption{Classification of the sample in the SAGA database}\label{tab:class}
    \begin{minipage}{140mm}
    \begin{tabular}{*{4}{l}}
      \hline
      Class & Criteria & Origin of carbon & n-capture elements \\
      \hline
      MP   & $\feoh > -2.5$, $\cfe < 0.7$  & supernovae  & \rit process dominant \\
      EMP  & $\feoh \le -2.5$, $\cfe < 0.7$  & supernovae  & \rit process dominant \\
      C-rich  & $\feoh > -2.5$, $\cfe \geq 0.7$, $\bafe \geq 0.5$, $\euba < -0.2$ & TDU & \sit process dominant \\
      \cemps  & $\feoh \le -2.5$, $\cfe \geq 0.7$, $\bafe \geq 0.5$, $\euba < -0.2$ & He-FDDM w/wo TDU & \sit process dominant \\
      \cempnos & $\feoh \le -2.5$, $\cfe \geq 0.7$, $\bafe < 0.5$ & TDU or unknown & \rit process dominant \\
      NEMP & $\abra{C}{N} \la -1$, $\abra{C+N}{H} \ga -2$, $\cfe \ga 0.5$ & He-FDDM and/or TDU & depending on stars \\
      \hline
   \end{tabular}
   \end{minipage}
\end{table*}

We classify the objects according to the same criteria as adopted in Paper I except for carbon-enhancement, as summarized in Tab.~\ref{tab:class};
``EMP'' for $\feoh \le -2.5$ and ``MP'' for $\feoh > -2.5$: 
``CEMP'' and ``C-rich'' stars for the carbon enhancement of $\cfe \geq 0.7$ with $\feoh \leq -2.5$ and $\feoh > -2.5$, respectively: 
``RGB'' if $\teff \leq 6000$ K and $\log g (\rm{cm} / \rm{s}^{2}) \leq 3.5$, and otherwise, ``MS'', the latter of which includes a small number of blue HB stars (3 Crich stars and 1 MP star).
Note that the classification is based on the average value of all available data for the same object.
Therefore, the plotted figures can be inconsistent with these classes because we adopt the ``most reliable'' data using the data scoring as stated above.

As for the carbon enhancement of ``CEMP'' and ``Crich'', we define $\cfe \ge 0.7$, while some authors prefer $\cfe > 1.0$ \citep[e.g.,][]{Beers2005a}.
Fig.~\ref{fig:chist} shows the distribution of carbon abundance, $\cfe$, relative to iron for 678 stars with the carbon abundances out of the 1386 total sample. 
The distribution of carbon abundances consists of two distinct components with $\cfe = 0.7$ as the boundary:
for $\cfe < 0.7$, the stars assemble to form a broad top, or bimodal peaks, around $\cfe \simeq -0.2 \hyp 0.3$, rapidly decreasing in number for higher carbon abundances, while for $\cfe \ge 0.7$, the stars exhibit a wide and shallow distribution, stretching out up to $\cfe \simeq 3.0$ with two HMP stars as extremes beyond it.  
These two components are thought to result from the different origins of carbon enhancement, and we may well adopt a boundary at $\cfe = 0.7$ as the criterion of carbon enhancement, i.e., the carbon enhancement for $\cfe \ge 0.7$ and the normal abundance for $\cfe < 0.7$.  
In the following, we deal with the stars of $\cfe \ge 0.7$, instead of $\cfe \ge 0.5$ in Paper~I, when discussing the properties of stars with the carbon enhancement in order to avoid the contamination of the stars with the normal carbon abundances.

\begin{figure}
    \includegraphics[width=84mm]{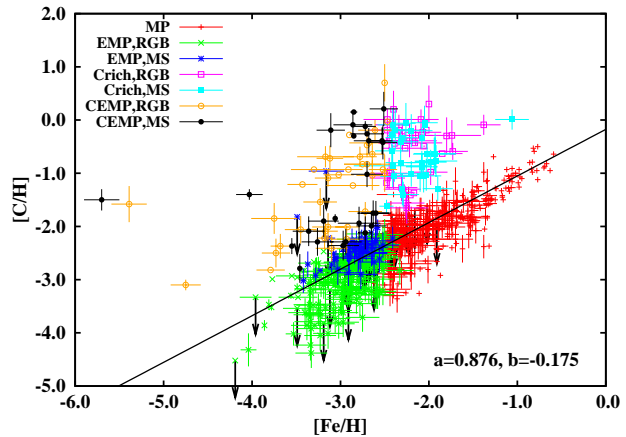}
  \caption{Relation between [Fe/H] and [C/H] taken from our sample in the SAGA database and plotted by the data retrieval system.
  Arrows indicate upper limit derived for the abundance.
  All stars are classified as ``C-rich'' stars for $\cfe \geq 0.7$, ``EMP'' stars for $\feoh \leq -2.5$, and ``RGB'' stars for $\teff \leq 6000$ K and $\log g \leq 3.5$.
  Stars are classified as ``CEMP'' if they are both ``C-rich'' and ``EMP''.
  If a star is not classified as ``RGB'', it is tagged as ``MS''.
  Stars classified as neither ``C-rich'' nor ``EMP'' are classified as ``MP''.
  Solid line gives the RMA regression for data excluding carbon-enhanced stars (``C-rich'' and ``CEMP'' groups) and mixed stars, the latter of which are defined by $\cfe < -0.1$ for $\feoh \leq -2$ and $\cfe < -0.4$ for $\feoh > -2$.
  The slope $a$ and intercept $b$ of the regression line are shown in the bottom right corner.
  See discussion in \S~\ref{sec:enrich} and \S~\ref{sec:cn}.
  }
  \label{fig:cbyh}
\end{figure}

Two lower panels of Fig.~\ref{fig:chist} presents a comparison of the distributions of carbon enhancement in the different ranges of metallicity of $\feoh \le -2$ and $\feoh > -2$, respectively.
For carbon-enhanced stars, both CEMP and C-rich stars exhibit similar flat distributions, though the largest enhancement depends on metallicity.
This gives a support the common origin of their carbon enhancement, i.e., the wind accretion of envelope matter ejected from erstwhile AGB primaries in binary systems, since the amount of accreted matter is related to the distribution of binary separations, as discussed in \citet[][also see Komiya et al.~2007]{Suda2004}.  
For the carbon-normal stars, on the other hand, the distributions in the two panels are different.
They are decomposed into two separate peaks across the metallicity of $\feoh \simeq -2$.
For the smaller metallicity of $\feoh \le -2.0$ (middle panel), a narrow peak forms at $\cfe \simeq 0.2 \pm 0.1$, while for the larger metallicity of $\feoh > -2$ (bottom panel), the peak shifts by $0.3 \hyp 0.4$ dex toward a smaller carbon enhancement of $\cfe \simeq -0.2 \hyp -0.1$.  
In addition, we see that for both EMP and MP giants, the distributions are asymmetric and develop a shoulder in lower side of carbon enhancement, extending to as small as $\cfe \simeq -1$.
In contrast, dwarfs show the distributions rapidly decreasing in the left side of the peak among the carbon-normal stars, indifferently to the metallicity;  
it is true that for dwarfs, we should take into account the detection limit of carbon lines around $\abra{C}{H} \simeq -3$ (see below Fig.~\ref{fig:cbyh}), but this lack of dwarfs with small carbon enhancement is real at least for metal-rich stars of $\feoh \ga -2$. 
These differences in the carbon enhancement may reflect the variations in the primary sources, such as supernova yields, and/or the process of nucleosynthesis and mixing in the stellar interiors. 
We will return to these points later. 

\subsection{Frequency of Carbon-Enhanced Stars}

Fig.~\ref{fig:cbyh} represents the classification of our sample taken from the database.
It includes 678 stars out of 1386 total sample.
Other stars not appeared in the figure belong to ``EMP'' or ``MP'' groups depending on the metallicity because of the lack of derived carbon abundance.
Therefore, it should be warned that $\sim 600$ stars classified as ``carbon-normal'' are not judged from their carbon abundance, although there is a possibility that the spectral lines of carbon were not detected by the observations.

For the carbon-normal stars (the EMP and MP groups), the carbon abundance increases nearly in a linear relationship to the iron abundance, with a slight difference in the peak of the distribution of $\cfe$ for $\feoh <-2$ and $> -2$, as stated above.  
The scatter in the abundance is contributed by small enhancement of $\cfe \la -1$, which is observed more or less continuously in all the metallicity range except for $\feoh \ga -1.5$ where the number of the sample stars greatly decreases.
For dwarfs, few stars are observed below $\cfe \simeq -0.5$ even for $\feoh \ga -2.5$, although the carbon abundances are well above the detection limit.

The carbon-enhanced stars (the CEMP and C-rich groups) gather mostly in the metallicity range between $-3 \la \feoh \la -2$ where the largest carbon abundances reach and level off at $\abra{C}{H} \simeq 0$.
In the lowest metallicity range, the decrease in the number of CEMP stars is due to the decrease in the number of total stars.  
Among 18 sample stars of the metallicity blow $\feoh \simeq -3.3$, the maximum enhancement of carbon abundance reduces, and in particular, no stars are observed with the carbon enhancement exceeding $\abra{C}{H} \simeq -1$.
Because of their smaller metallicities, on the other hand, the stars with smaller carbon abundances can be registered as CEMP stars and exhibit a rather flat distribution of carbon enhancements, which is again indicative of their origin through the wind accretion.

\begin{figure}
  \begin{center}
    \includegraphics[width=84mm]{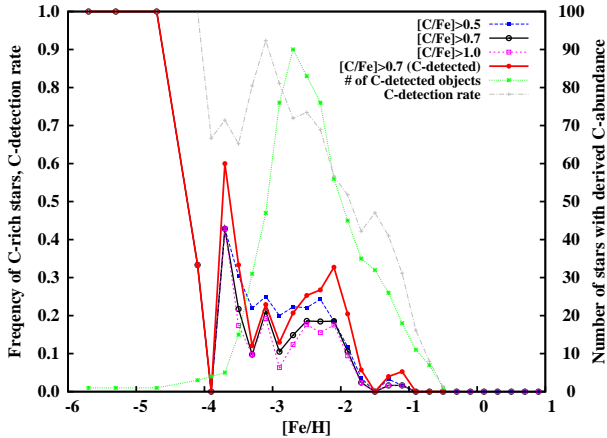}
  \end{center}
  \caption{The frequency of carbon-enhanced stars in total of 1386 stars as a function of metallicity for the criterion, $\cfe \geq 0.5$ (solid line with triangle), $0.7$ (short-dashed line with circles) and $1.0$ (dotted line with open squares), respectively.
           The fraction in the 678 carbon-measured stars is also plotted for the criterion of $\cfe \geq 0.7$ (dash-dotted line with filled squares).  
           The number of stars with measured-carbon abundances is also given as a function of metallicity (dotted line with asterisks).
		   Dash-dotted line represents the number of carbon detected sample divided by the number of sample in each metallicity bin.
  }\label{fig:cfrac}
\end{figure}

Our sample stars show the frequency of carbon-enhanced stars among the EMP stars by far larger than among the Population I and II stars, as reported previously \citep[][]{Rossi1999,Beers1999}.
Fig.~\ref{fig:cfrac} shows the proportion of carbon-enhanced stars among our sample as a function of metallicity with 0.2 dex bins.
In order to see the dependence on the criterion for carbon enhancement, we show the results for three different criteria, $\cfe \geq 0.5, 0.7$, and $1.0$;
the average frequency of CEMP stars is 22.7, 16.7, and 14.3 \% among the total EMP samples.
About a half of the sample stars lack the carbon abundance and are classified as ``EMP'' or ``MP'' groups depending on the metallicity, which little affects the results since most of them are among the stars of high metallicity of $\feoh > -2$, as shown by dotted line in the figure.
The average frueqency of CEMP stars is 21.0 \% for the criterion of $\cfe \geq 0.7$ if limited to the sample with the detected carbon abundance.  
For still smaller metallicity, the frequency of CEMP stars increases, and in particular, all the three stars of the lowest iron abundances $(\feoh < -4.5)$ exhibit a large carbon enhancement.
The number of objects with the carbon abundance is, however, less than 5 per bin for $\feoh < -3.5$ and may not be statistically reliable for the lowest metallicity range.
On the other hand, we have more than 15 samples per each bin with the carbon abundance for $-3.5 \le \feoh \le 0.1$.

For the metallicity of $\feoh > -2$, the frequency of C-rich stars decreases steeply and becomes negligibly small for $\feoh \ga -1.8$.
The latter is consistent with the previous results for the metal-rich equivalents, CH stars and Ba stars, which are found to account for only small fractions $(\sim 1 \%)$ among their respective populations \citep[][]{Tomkin1989,Luck1991}.
According to \citet{Komiya2007}, the difference in the fraction between CEMP stars and CH stars is brought about by a change of the initial mass function (IMF) from a high-mass to low-mass one as well as by the difference in the parameter range of primary mass and binary separation that allows the low-mass member to acquire carbon enhancement.  
In addition, we may well take into account the efficiency of the third dredge-up, which decreases for EMP stars because of smaller metallicity \citep{Suda2010}.  
In this figure, however, the frequency of C-rich stars with $-2.5 \le \feoh < -2$ remains nearly in the same level as that of CEMP stars.
It is likely, therefore, that this decrease in the frequency of C-rich stars around $\feoh \simeq -2$ stems from the change in the initial mass function rather than to the difference in the metallicity.
For the EMP population, a high-mass IMF with typical mass $\sim 10 \msun$ is derived by \citet{Komiya2007}, while low-mass IMF with typical mass $0.22 \msun$ is derived for the stellar halo of average metallicity, $\feoh \simeq -1.7 \hyp -1.4$ \citep{Chabrier2003}.
We evaluate the effect of the IMF change on the carbon-enhancement in the similar method to that used by \citet{Komiya2007}, to find that the frequency of C-rich stars decreases from $14\%$ to $0.7\%$ when the IMF changes from a log-normal form with the medium mass $M_{\rm md} = 10 \msun$ to one with $M_{\rm md} = 0.22 \msun$ both with the dispersion of $\Delta_M = 0.33$;
the criteria of carbon enhancement is taken to be $\cfe = 0.7$ and other parameters are the same as in Fig.~8 of \citet{Komiya2007}.
This can be taken as an evidence that the drastic change in the frequency of carbon-enhanced stars observed around $\feoh \simeq -2$ is well explicable in terms of the change of the IMF.
In this connection, it is worth noting again that the distribution of carbon abundances differs at the either side of this metallicity for the carbon-normal stars with $\cfe < 0.7$. 
This result also suggests that the averaged supernova yields may vary as a function of the IMF (see Yamada et al. 2010 in preparation).

% Figure C-rich frequency

\section{Carbon-enhanced stars and neutron-capture elements}\label{sec:cemp}

The origins of carbon-enhanced groups are intriguing open questions among many mysteries of EMP stars.
In this section, we revisit the known features of CEMP and C-rich stars. 
We first discuss the classification into two subclasses, \cemps\ and \cempnos, according to whether the \sit\ or \rit process contributes to the abundances of neutron-capture elements.
Then we turn to the different mechanisms of carbon enhancement between stars with the metallicity below and above $\feoh \approx -2.5$.
This is predicted from the theory \citep[see, e.g.][and references therein]{Suda2010} in which the critical metallicity is argued to cause the variations of \sit process nucleosynthesis between CEMP and C-rich stars, which results in the different frequencies of carbon-enhanced stars without the enhancement of \sit process elements for a large sample of stars. 

\subsection{Subdivision of Carbon-Enhanced Groups}\label{sec:subclass}

\begin{figure}
  \begin{center}
    \includegraphics[width=84mm]{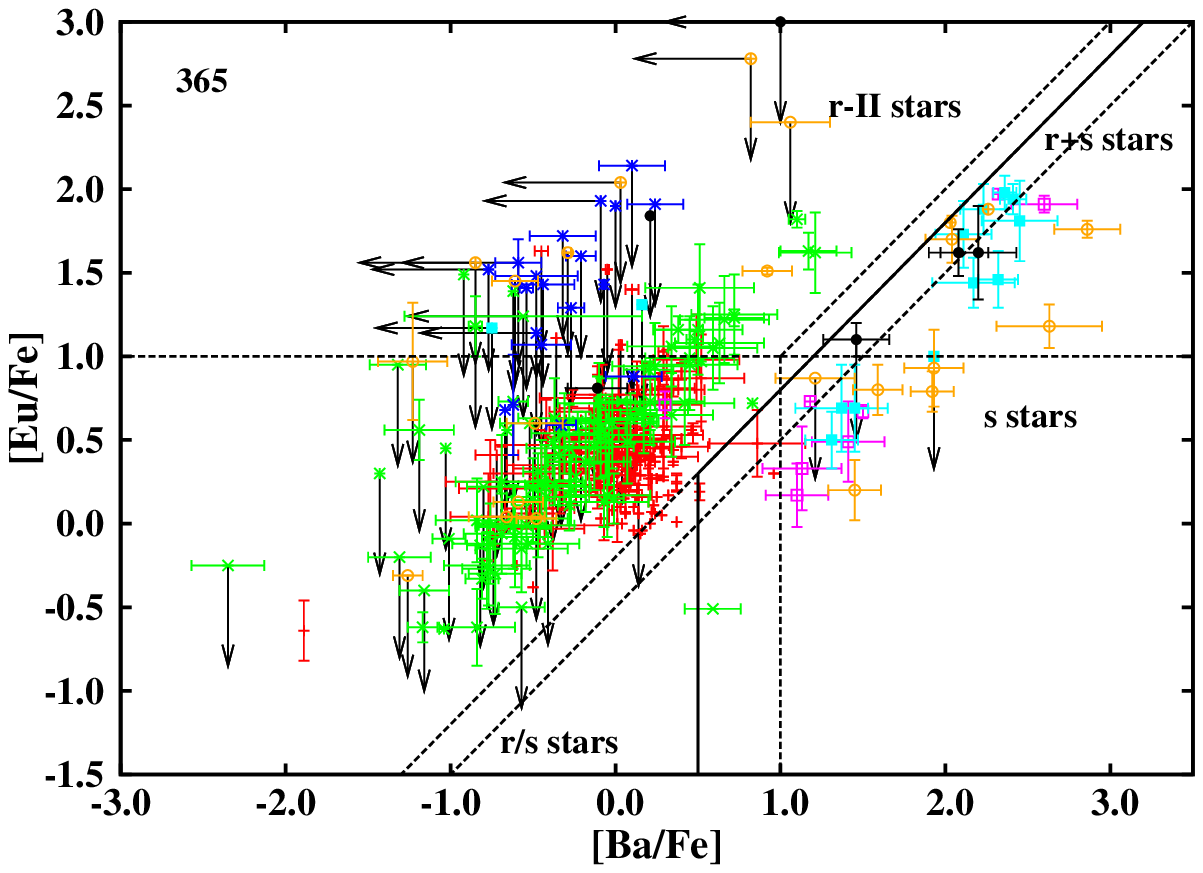}
  \end{center}
  \caption{Distribution of sample stars on the $\bafe$ and $\eufe$ diagram.
           The production site of heavy elements is separated by a solid line: $\bafe \geq 0.5$ and $\euba < -0.2$.
		   The heavy element abundances plotted to the right of the solid line come from the \sit process.
		   Dashed lines divide ``\rps'' and ``\sit'' stars defined by \citet{Jonsell2006}, ``\rit II'' stars defined by \citet{Christlieb2004a}, and ``CEMP-{\sl r/s}'' stars according to the definition by \citet{Beers2005a}.
			The definition of the symbols is the same as in Fig.~\ref{fig:cbyh}.
  }\label{fig:euba}
\end{figure}

We define the subclass \cemps\ and its metal-rich counterpart, C-rich-\is, in the C-rich group as having the enhancement of \sit-process elements by the criteria of $\bafe \geq 0.5$ and [Eu/Ba] $< -0.2$.
These criteria are consistent with the previous works considering the contribution to the abundance pattern by the \rit\ or \sit process nucleosynthesis \citep{Sivarani2004,Beers2005a,Jonsell2006,Masseron2010}.
Fig.~\ref{fig:euba} shows the distribution of stars on the diagram of $\bafe$ and $\eufe$. 
We see that the sources of heavy elements are clearly separated by the solids line representing the above criteria.
In the following discussion, we use these criteria for the enhancement of \sit-process elements, and otherwise, attribute the source of neutron capture elements to the \rit process nucleosynthesis, calling them as ``\sit-[\rit-](process) dominant'', respectively.
We have 365 stars in total in Fig.~\ref{fig:euba}, among which 274 stars are determined as \rit\ or \sit process dominant for neutron capture elements.
The stars for which the origin of neutron-capture elements can not be determined have only the upper limits for Eu and/or Ba.
For the MP and EMP RGB groups, almost all the stars are \rit dominant except for a few stars such as a barium stars HD15096 ($\bafe = 0.96$).
For the EMP MS group, all the Eu abundance data are constrained by upper limit, but should be \rit dominant because no stars have $\bafe > 0.5$.
These \sit dominant, carbon-enhanced stars are thought to be born in binary systems and affected by mass transfer from their intermediate-mass primaries in their AGB phase.

% Figure [Eu/Fe] vs. [Ba/Fe]

In addition to these subclasses, we define \cempnos\ stars and the metal-rich equivalence, C-rich-no\is, in the C-rich group as carbon-enhanced stars other than confirmed as \sit dominant.
They are defined as CEMP stars with $\bafe \le 0.5$ or with $\euba \ge -0.2$ or both.
Thus, \cempnos\ includes both CEMP-no stars without the enhancement of neutron-capture elements and CEMP-\ir\ stars enriched with \rit process elements, as defined by  \citet{Beers2005a}.
All these subclasses are listed in Tab.~\ref{tab:cemp} and \ref{tab:crich}.

Among CEMP stars with the derived abundances both for Ba and Eu, 11 out of 21 giants and 3 out of 6 dwarfs are confirmed as \sit dominant and the others are \rit dominant.
A CEMP-\ir\ star, CS22892-052, which is observed with $\bafe = 0.92$ and $\eufe = 1.51$ \citep{Honda2004b}, is classified as CEMP RGB and also called ``\rit II'' stars.
A CEMP-no star, CS22949-037, which is observed with $\bafe = -0.66$ \citep{Cohen2008} and $\eufe = 0.04$ \citep{Depagne2000}, is also located in the area of \rit dominant origin for neutron capture elements.  
These two stars are the only \cempnos\ stars, confirmed as \rit dominant. 
In the \rit dominant region, there are several stars with the carbon enrichment of $\cfe \simeq 0.5$, some enriched with the \rit process elements but other not. 
One of them, CS30325-028, is located in the area of \rit process origin without any enhancement of neutron-capture elements, for which possible binarity is suggested \citep{Aoki2005}.

For C-rich group stars with the derived Ba and Eu abundances, 8 out of 9 giants and 10 out of 11 dwarfs are the \sit dominant for neutron capture elements, resulting from the third dredge-up and the \nucm{13}{C} pocket.
A giant HD135148 with $\cfe = 0.8$ and $\euba = 0.41$ \citep{Burris2000,Simmerer2004} is recognized as a CH star \citep{Carney2003}, though the carbon abundance is not determined in high precision \citep{Simmerer2004}.
It is also a known binary whose period is estimated at $1416 \pm 28$ days \citep{Carney2003}.
Because of small estimated mass of the companion, however, they insist that the binary system did not experience mass transfer of carbon-rich matter, and if this is true, the abundance pattern of neutron-capture elements for this star may be pristine with the \rit dominant origin.
The \rit dominant C-rich dwarf is CS 22878-027 whose carbon enhancement is $\cfe = 0.8$ \citep{Lai2008}.
This star has [Fe I/H] $= -2.48$ and [Fe II/H] $= -2.68$ and may be classified as CEMP MS.

% Figure CENP-s/C-rich-s and CEMP-nos/C-rich-nos.
\begin{figure}
  \begin{center}
    \includegraphics[width=84mm]{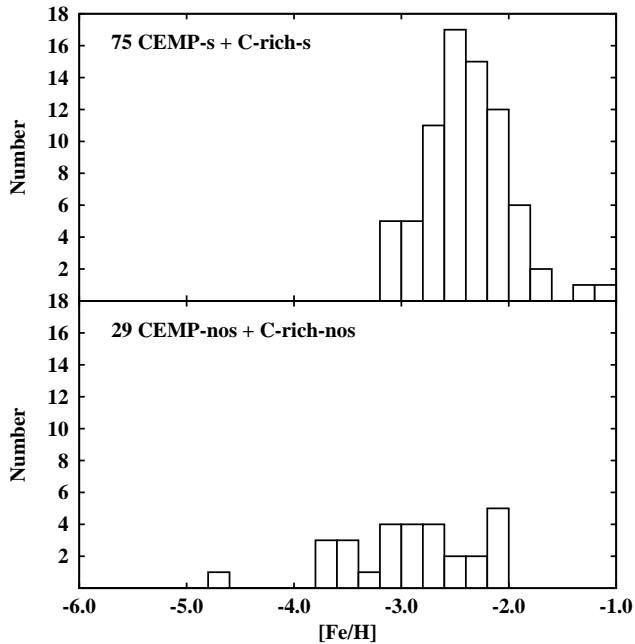} 
  \end{center}
    \caption{The metallicity distribution of carbon-enhanced stars, enriched with \sit-process elements (CEMP-$s$ and C-rich-$s$ groups, top panel) and without the enhancement of \sit-process elements (CEMP-no$s$ and C-rich-no$s$ groups, bottom panel).
    }\label{fig:cempsnos}
\end{figure}

Accordingly, the enhancement of \sit process nucleosynthesis differs between the CEMP and C-rich groups.
Among the above sample stars having the derived abundances of both Ba and Eu, the fraction of carbon-enhanced stars showing the enhancement of \sit process elements among the entire carbon-enhanced stars is $\hbox{\cemps} / \hbox{CEMP} \simeq 0.6$ for CEMP groups, while the corresponding fraction is $\hbox{C-rich-\is} / \hbox{C-rich} \ga 0.8$ for the C-rich group.
Since we have more sample stars with measured Ba abundance, we can estimate the fraction of \sit dominant stars by assigning $\bafe \geq 0.5$.
The fraction is $\hbox{\cemps} / \hbox{CEMP} = 27/47$ for CEMP group and $\hbox{C-rich-\is} / \hbox{C-rich} = 48/57$ for C-rich group, respectively, i.e., much larger fraction for C-rich stars, as seen from Tables~\ref{tab:cemp} and \ref{tab:crich}.
Fig.~\ref{fig:cempsnos} compares the metallicity distributions of carbon-enhanced stars with and without the enhancement of \sit-process elements.
The \cemps\ stars follow approximately the metallicity distribution of our total sample stars in Fig.~\ref{fig:mdf} up to the metallicity $\feoh \simeq -2$.
The peak of distribution shifts slightly to larger metallicity around $\feoh \simeq -2.5$ compared with the peak of the total MDF at $\feoh \simeq -2.8$.
For the metallicity larger than $\feoh \sim -2$, the number of stars decreases much more rapidly than the entire distribution does.
For $\feoh > -2$, no C-rich-no\is\ are found, while more than 10 C-rich-\is\ stars are observed.
This is consistent with the current observations that all the CH stars having $\feoh \ga -1$ are all observed to be enriched with \sit process elements \citep[see, e.g.][]{Vanture1992}.
On the contrary to the distribution of \sit dominant stars, \cempnos\ and C-rich-no\is\ stars have a rather flat distribution for $\feoh \la -2$.
In the lowest metallicity, it is pointed out that the \cempnos\ stars tend to overweigh the \cemps\ stars in number \citep{Aoki2002b}, and in particular, only \cempnos\ stars are confirmed below the metallicity of $\feoh \simeq -3.3$ including HE0557-4840 of $\feoh = -4.75$ \citep{Norris2007}.
These differences may be associated with the difference in the evolutionary characteristics between the stars with $\feoh \la -2.5$ and with $\feoh > -2.5$ \citep{Fujimoto2000,Suda2004,Suda2010}, which is discussed in the following subsection.

As for the neutron-capture elements, some \cemps\ and C-rich-\is\ stars show small abundance ratios of $\abra{Ba}{Eu} < 0.5$, seemingly too small to be realized by the current models of \sit process nucleosynthesis.
These stars, as classified into subclass ``CEMP-{\sl r/s}'' \citep{Beers2005a}, lie in the narrow range of $0.2 < \abra{Ba}{Eu} <0.5$, but distribute more or less continuously within \cemps\ and C-rich-\is\ stars.
Their property and origin require further investigations both theoretically and observationally \citep[e.g.,][]{Masseron2010}.

% Figure [Eu/H] vs. [Ba/H] with r+s and r-II stars highlighted.
\begin{figure}
  \begin{center}
    \includegraphics[width=84mm]{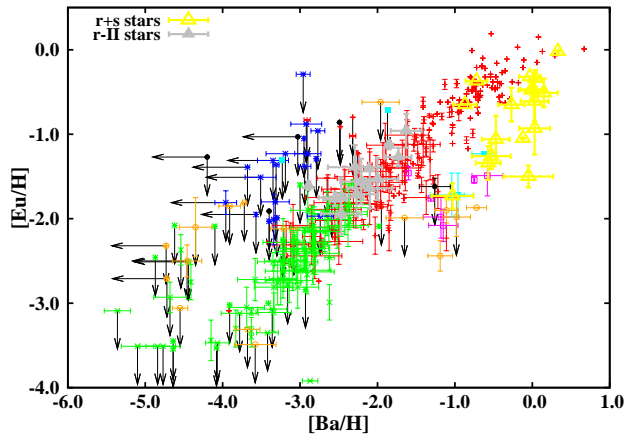}
  \end{center}
  \caption{The relation  between the $\abra{Ba}{H}$ and $\abra{Eu}{H}$.
			The definition of the symbols is the same as in Fig.~\ref{fig:cbyh}, but \rps\ and \rit II stars are shown by separate symbols.
  }\label{fig:euba2}
\end{figure}

In Fig.~\ref{fig:euba}, we also show the classification of neutron capture elements by dashed lines, according to the definition of \citet{Jonsell2006} by dashed line.
They define two subclasses, \is\ and \rps\ stars, both as satisfying the condition $\bafe > 1.0$ but as distinguished by the condition of $\eufe < 1.0$ and $\eufe > 1.0$, respectively, in addition to \rit II stars defined by $\eufe > 1.0$ and $\euba > 0$ \citep{Christlieb2004a}.
In our sample, all stars with $\bafe \sim 2$ belong to \rps\ stars.
This corresponds to the previous discussion that the enhancement in neutron capture elements is larger in \rps\ stars than in \rit II stars \citep{Jonsell2006,Aoki2006a}, which is clearly seen in Fig.~\ref{fig:euba2} of $\abra{Ba}{H} - \abra{Eu}{H}$ diagram.
In our sample, there are 17 \rps\ stars, 14 \rit II stars, and 14 \is\ stars.
Fig.~\ref{fig:euba} shows that most of the C-rich MS stars are \rps\ stars ($8 / 11$), whereas most of the C-rich RGB stars are \is\ stars ($7 /9$).
This may not be the case for CEMP stars from the sample, since CEMP RGB stars are equally distributed over the \rps\ and \is\ areas in the figure.
As for CEMP MS stars, we cannot conclude anything about the distribution due to the small number in the sample.
The different distributions between the C-rich MS and RGB stars may be interpreted as a consequence of dilution in surface convection for C-rich RGB stars.
This can be due to the observational bias, however, since it is difficult to detect mildly enhanced Eu abundances for dwarfs.
Indeed, it seems that carbon enhancement does not depend on the depth of surface convection of observed EMP stars, as discussed below.
Attempt to measure the Eu abundance of EMP dwarfs have been made mainly by \citet{Cohen2004}, but only upper limits are available even with their high resolution spectra.
It is necessary to determine the Eu abundance for more EMP dwarfs to discuss further the origins of their neutron capture elements.

\subsection{Two modes of \sit process nucleosynthesis}

It is shown that two different mechanisms work to enrich the surface with carbon during the thermal-pulsating AGB (TPAGB) phase for the extremely metal-poor stars \citep[][]{Fujimoto2000,Suda2010};
the stars of mass $M \la 3 \msun$ come across hydrogen mixing into the helium-flash convective zone to trigger He-FDDM during the early TPAGB phase, while the stars of mass $M > 3 \msun$ undergo only TDU after the thermal pulses grow sufficiently.
The stars in the mass range of $1.5 \msun \la M \la 3 \msun$ undergo TDU after He-FDDM raises the surface CNO abundance above $\abra{CNO}{H} \ga -2.5$ and continue to enrich the surface with carbon.
Correspondingly, two sites are proposed for the \sit process nucleosynthesis, i.e., convective and radiative nucleosynthesis.
The former is triggered by hydrogen engulfment into the helium convective zone during the helium shell flashes to produce \nucm{13}{C} as neutron source \citep{Suda2004,Komiya2007,Nishimura2009}.
The latter is the \cpocket\ at the top of helium zone during the inter-pulse phases \citep{Straniero1995}.
The \cpocket\ is assumed to be formed by the injection of hydrogen during the preceding third dredge-up probably owing to the overshooting from the bottom of the surface convection, although there is no reliable theory so far to estimate its efficiency or even to predict whether it occurs or not.
For metal-rich stars, TDU and radiative \nucm{13}{C} burning model are believed to work with the lower mass limit around $M \ga 1.5 \msun$, and regarded as the standard mechanism of \sit process nucleosynthesis \citep[e.g.][]{Busso1999}.

We have seen in \S~\ref{sec:subclass} that the CEMP and C-rich stars differ in the enhancement of \sit process elements.
Accordingly, their different behaviours give an insight into the efficiency of \cpocket\ and its dependence on mass and metallicity within the current framework of \sit process nucleosynthesis in AGB stars.
\citet{Suda2004} propose that \cemps\ stars are formed in the binary systems through the mass transfer from AGB companions of mass $M \la 3 \msun$ that have undergone the \sit process nucleosynthesis of convective \nucm{13}{C}-burning \citep[see also][]{Komiya2007,Nishimura2009}.
For the companions more massive than $M \simeq 1.5 \msun$, TDU follows after He-FDDM \citep{Suda2010} to enrich the surface with carbon and \sit process elements.
Then, there is a possibility to form \cpocket\ under some favourable case \citep{Suda2004,Nishimura2008}.
As for the \cempnos\ stars, \citet{Komiya2007} identify their origin in the binary systems with the companions of $M > 3 \msun$, which experience only TDU without the He-FDDM events.
This interpretation presumes that the \cpocket\ is inefficient for massive intermediate-mass stars and the \sit process nucleosynthesis will not work in substantial fraction of stars.
On the other hand, for the metallicity of $\feoh > -2.5$ where the He-FDDM events do not occur, the radiative \cpocket\ model is only the viable mechanism of \sit process nucleosynthesis and should work to explain the C-rich-\is\ stars.
As for the metal-rich stars, it is usually assumed that the formation of \cpocket\ should be efficient for relatively low-mass AGB stars in order to operate the \sit process under the low-mass peaked IMF \citep{Busso1999}.
Accordingly, for C-rich stars, we expect that the \cpocket\ is inefficient for massive intermediate-mass stars, similarly to the above presumption for CEMP stars, although the mass ranges are likely to be different.
The efficiency of \cpocket\ as a function of metallicity can be measured observationally by plotting \abra{Pb}{Ba} versus \feoh\ as in Fig.~7 of \citet{Suda2004}.
The overall relation and discussions in \citet{Suda2004} still holds in our sample of 28 stars compared with 13 stars in the previous work.
Again, it is observationally not true that the number of neutrons per seed nuclei is larger for CEMP stars than for C-rich stars if the efficiency of \cpocket\ is comparable.
Therefore, it is speculated that the efficiency of \cpocket\ varies with the metallicity and decreases with decreasing metallicity because of the above observational evidence in connection with the fact that the frequency of \cemps\ stars among the CEMP stars is smaller than that of correspondences among C-rich stars.

% Figure [Ba/Fe] vs. [C+N/Fe]
\begin{figure}
  \begin{center}
    \includegraphics[width=84mm]{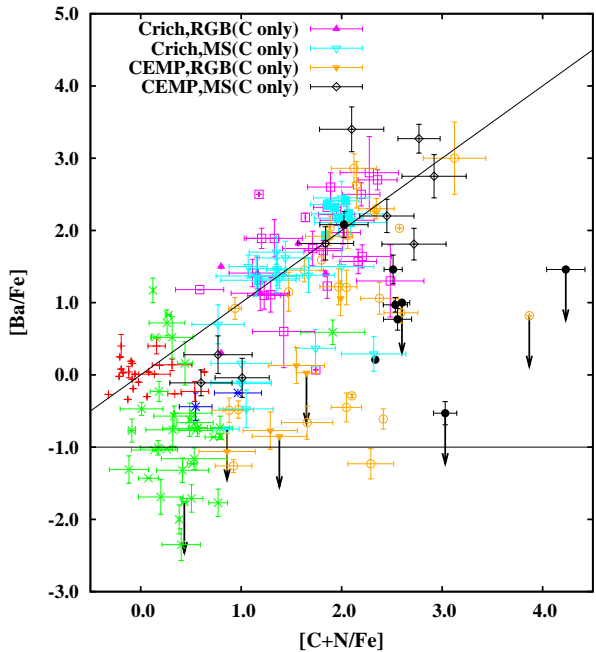}
  \end{center}
  \caption{Abundance trend of $\bafe$ as a function of $\abra{C+N}{Fe}$ for 126 stars available from the SAGA database.
			We also add 40 data of $\cfe$ for which nitrogen abundances are not available from C-rich and CEMP groups.
			These stars are denoted by smaller symbols shown in left top corner.
			In calculating $\abra{C+N}{Fe}$, we assumed solar abundances of \citet{Grevesse1996}.
           There are two branches for stars with large $\abra{C+N}{Fe}$ denoted by thin solid lines.
		   The upper one is assumed as ``CEMP-{\it s}'', and lower one as ``\cempnos''.
           The meanings of symbols not presented here are the same as in Fig.~\ref{fig:cbyh}.
  }\label{fig:bacn}
\end{figure}

% Figure [Ba/C] histogram
\begin{figure}
  \begin{center}
    \includegraphics[width=80mm]{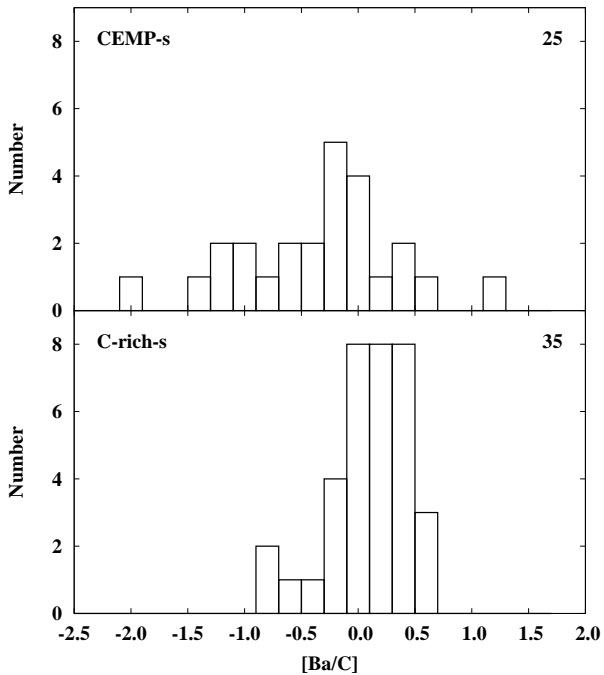}
  \end{center}
  \caption{Distribution of $\abra{Ba}{C}$ for \cemps (top panel) and Crich-\is\ (middle panel);
           only plotted are the stars with the abundance data obtained by spectroscopy of higher resolution than $R = 20000$.
  }\label{fig:bachist}
\end{figure}

With respect to the classification of \cemps\ and \cempnos\ stars, the correlation between carbon (plus nitrogen) and \sit process elements has been used as a test for classifying CEMP stars \citep[see, e.g.][]{Aoki2002b}.
It is pointed out that they apparently separate into two branches in the relation between $\bafe$ and $\abra{C+N}{Fe}$ (and $\cfe$), as seen from Fig.~8 in \citet{Aoki2002a} and Fig.~6 in \citet{Suda2004}.
In Fig.~\ref{fig:bacn}, we plot 34 (plus 15) CEMP stars and 32 (plus 25) C-rich stars with the carbon and nitrogen abundances (carbon abundance only) on the same diagram as theirs.
In this figure, we show the two branches, the liner correlation of $\abra{Ba}{C} = 0$ and a loci of constant $\bafe = -1$, as same as in \citet{Suda2004}.
The separation into two branches is still discernible, but becomes more ambiguous compared with the previous works, along with the increased number of CEMP stars with relatively small Ba enrichment in the sample.
In Fig.~\ref{fig:bachist} compares the distributions of $\abra{Ba}{C}$ for \cemps\ and C-rich-\is\ stars.
Across the metallicity $\feoh \simeq -2.5$, the distributions differ;
\cemps\ stars display a rather flat and broad distribution down to $\abra{Ba}{C} \simeq -2$, while C-rich-\is\ stars show a strong concentration to $\abra{Ba}{C} \simeq 0$.
This is indicative that the \sit process nucleosynthesis differs in \cemps\ stars from that in the metal-rich counterparts, as argued by \citet{Suda2004}.

For the most iron-poor stars, HE0107-5240 and HE1327-2326, it is not concluded that both stars belong to either \cemps\ or \cempnos\ because of the only available upper limits of barium abundance.
For HE1327-2326, two resonance lines of Sr II are clearly detected and the star may have a large value of \abra{Sr}{Ba} \citep{Aoki2006b}, possibly larger than expected from the standard \sit process nucleosynthesis.
On the other hand, the abundances of \sit process elements are not determined for HE0107-5240, but we cannot exclude the possibility that the very efficient \sit process produces the large amount of lead whose lines are still very difficult to detect with current instruments \citep{Suda2004}.
At present, both HE0107-5240 and HE1327-2326 are still to be classified as \cemps\ or \cempnos\ by future observations.

\section{Diversity of abundances}\label{sec:div}

In this section, we briefly discuss the abundance trend with respect to metallicity for \aelms\ and titanium, iron-group elements, and neutron-capture elements.
We obtained the consistent results with previous works for the abundance scatters for elements with $12 \la Z \la 90$ at $-4 \la \feoh \leq -1$, and for the operation of the ``weak-\ir'' process in some of EMP stars.
We also argue that stars with peculiar $\alpha$-element abundances are likely to be affected by binary mass transfer because peculiar abundance patterns can be explained by the relative enhancement of Mg and because these peculiar stars are CEMP stars.

\subsection{\aelms\ and titanium}\label{sec:alpha}

% Figure [Fe/H] vs. [a/Fe] -- put before the section 5 for layout
\begin{figure*}
\centering
    \includegraphics[width=0.7\textwidth]{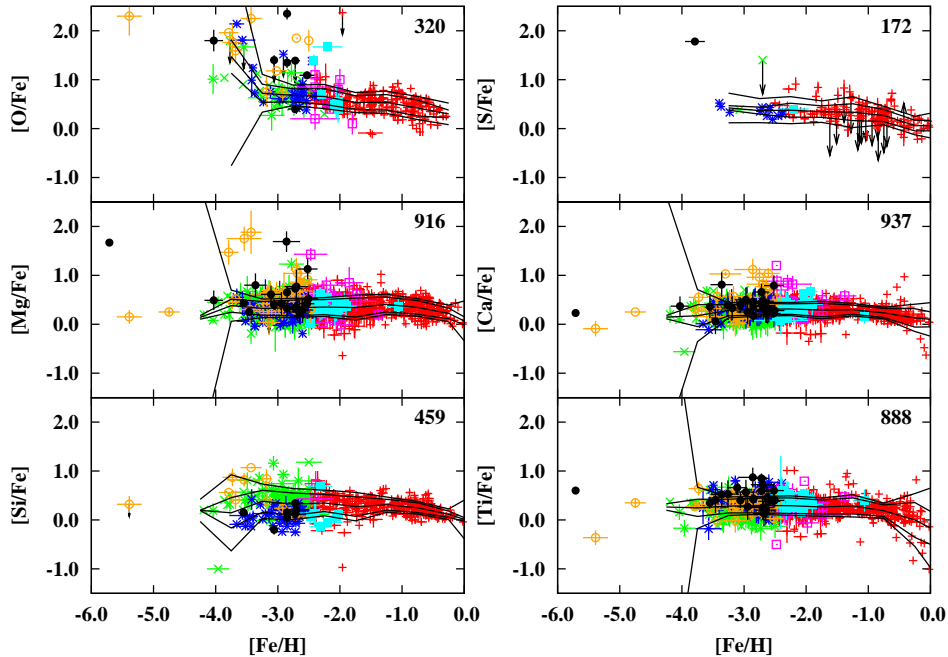}
  \caption{$\alpha$-element (O, Mg, Si, S, Ca, Ti) abundances as a function of metallicity.
  			The number of plotted data is labelled in each panel.
		   The definition of the symbols is the same as in Fig.~\ref{fig:cbyh}.
		   Carbon-rich stars (C-rich and CEMP groups) are excluded from the statistical treatment.
		   Each line denotes, from top to bottom, the upper and lower limit of confidence interval for the upper standard deviation, the sample mean, and the lower and upper limit of confidence interval for the lower standard deviation, respectively.
		   These values are estimated from the sample in the metallicity range of 0.5 dex.
  }\label{fig:afe}
\end{figure*}

% Figures [alpha/H] vs. [alpha/H]
\begin{figure*}
\centering
    \includegraphics[width=1.0\textwidth]{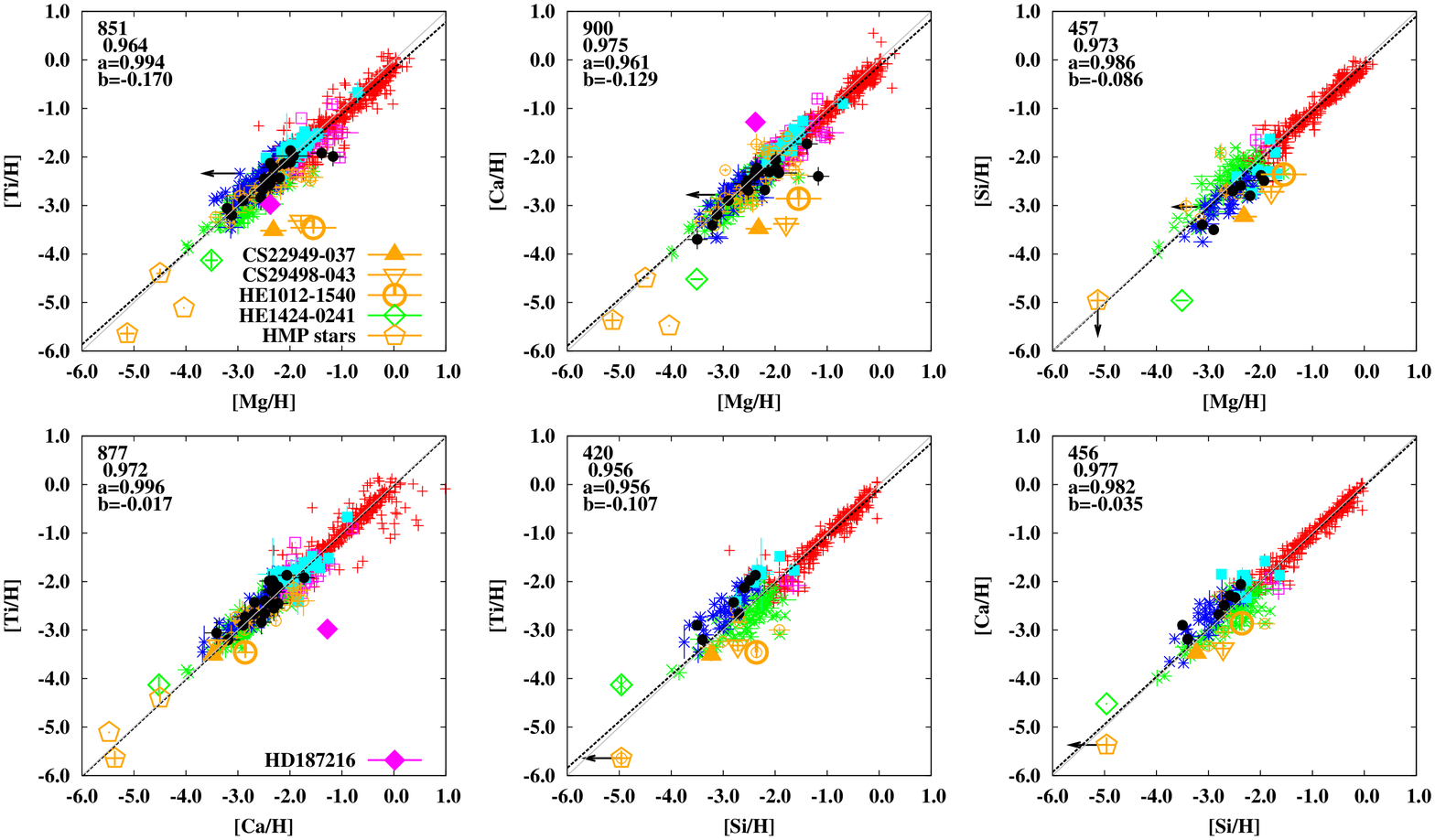}
  \caption{Abundance correlations between \aelms.
		   The number of sample stars, the Pearson coefficient, and the slope $a$ and intercept $b$ of the RMA regression are shown in top left corner.
		   Black-dashed line is the regression line for data excluding the Crich and CEMP groups.
		   Grey-solid line shows the linear relationship between two elements.
		   Stars with peculiar abundances and hyper/ultra metal-poor (HMP/UMP) stars are labelled separately with larger symbols.
		   For stars with peculiar \aelm\ abundances, CS22949-037, CS29498-043 and HE1012-1540 are originally in the CEMP RGB group.
		   HE1424-0241 is in the EMP RGB group, while HD187216 is in the C-rich RGB group.
		   For other symbols, the definitions are the same as in Fig.~\ref{fig:cbyh}.
  }\label{fig:alpha}
\end{figure*}

The magnesium and heavier \aelms\ are produced in massive stars and can be used as useful probes for chemical evolution especially at low-metallicity, and hence, the abundances of these elements are expected to be influenced by peculiar supernovae (SNe) at the lowest metallicity.
Abundances of \aelms\ and titanium are derived by many authors in the literature.
Fig.~\ref{fig:afe} shows the abundances of \aelms\ from oxygen through calcium and titanium relative to the iron abundance as a function of iron abundance for our sample stars.
These elements show almost constant trends with decreasing metallicity for $\feoh < -1$ except for oxygen, which is in agreement with the latest result \citep{Lai2008}, but at variance with the previous reports that they increase with decreasing metallicity \citep{Stephens2002,Gratton2003,Cayrel2004,Jonsell2005}.
These trends are statistically supported by the regression analyses as discussed in \S~\ref{sec:enrich}.

In this figure, we also plot the standard deviations of abundances in \xfe.
They indicate that the scatters of element abundances are rather small and remain typically within 0.2 dex, comparable to the observational errors,
except for oxygen and silicon, for which somewhat larger scatters are found at $\feoh \la -2.5$.
An important suggestion from the figure is that there should be weak mass dependence of SN yields for stars at the lowest metallicity where a single or a few SNe are expected to contribute to the enrichment of \aelms.
In addition, for metallicity of $\feoh \ga -3$ where the abundances are expected to be averaged over the initial mass function, we cannot see any significant variations caused by the metallicity dependence of SN yields.
In particular, there is no signature of the change of the IMF around $\feoh \simeq -2$, discussed in \S~\ref{sec:imf}, which implies the similar mass dependence of SNe yields for these elements.

For oxygen abundance, the increasing scatter with decreasing metallicity is statistically meaningful at least as low as $\feoh \sim -3.5$.
It is to be noted, however, that oxygen abundance is subject to a large uncertainty especially due to the usage of different lines such as O I, [O I], and OH.
The determination of oxygen abundance is still in controversy and requires further improvement of abundance analyses as well as larger sample stars \citep[see, e.g.][]{GarciaPerez2006b}.
For the silicon abundance, the relatively large scatter at $\feoh < -2$ comes from the discrepancy of derived abundance between dwarfs and giants, which is discussed in \S~\ref{sec:unexp}.
The apparent trend of \abra{Si}{Fe} increasing with decreasing metallicity is due to the separation of abundances between dwarfs and giants as clearly seen in Fig.~\ref{fig:afe}.
The larger number of giants than dwarfs among EMP groups by a factor of two causes the upward trend of the average abundance.
This argument also applies to the titanium abundance, although the abundance scatter is not so large compared with silicon.

Fig.~\ref{fig:alpha} shows the correlations between the abundances of \aelms\ and titanium.
As seen from the figure, most of the EMP stars correlate linearly with each other.
The RMA regression for data excluding carbon-enhanced stars shows that all the relations in the figure have the slopes close to unity.
The intercepts for these regressions are close to zero except for the relation between \abra{Ca,Ti}{H} and \abra{Mg}{H} that gives $\la -0.1$.
It may be remarkable that the largest deviation from zero is for the relation between the lightest element (Mg) and the heaviest one (Ti) among the elements in Fig.~\ref{fig:alpha}.
These overall trends mean that the chemical yields were almost homogenized by the time the next generation stars were born.
The slopes of nearly unity in the abundance ratios also suggest that the chemical yields of \aelms\ by SNe are independent of the initial abundances of progenitors.
It is in sharp contrast to the case of iron group elements, as shown in the next subsection.

A few peculiar stars with peculiar $\alpha$-element abundances are shown separately in Fig.~\ref{fig:alpha}.
The well-known peculiar stars are CS22949-037 and CS29498-043.
CS22949-037 has a very large enhancement of Mg, Ca \citep{McWilliam1995b}, Si \citep{Norris2001}, and S \citep{Depagne2002}.
CS29498-043 and HE1012-1540 also show large excesses of Mg and Si \citep{Aoki2002c,Cohen2008}.
In contrast to these $\alpha$-enhanced stars, HE1424-0241 shows significant depletions of Si and Ca \citep{Cohen2007}.
Interestingly, all these stars have iron abundances as low as $\feoh < -3.5$, as pointed out by \citet{Aoki2007a}.
In addition, three $\alpha$-enhanced stars belong to the \cempnos\ group showing a large enhancement of nitrogen.
It is only HE1424-0241 that is a carbon-normal star with upper limit available for carbon abundance \citep[$\cfe < 0.62$, ][]{Cohen2007}.
In this figure, there is an extraordinary star, HD187216 in the C-rich RGB group that shows peculiar abundance pattern for \aelms \citep{Kipper1994}, but it is not discussed here since the data are among the oldest sample in the database.

It is intriguing that the peculiar stars stated above are distributed around the linear correlations in Fig.~\ref{fig:alpha}.
The slight deviation from the linear correlation between Ti and Si comes from the discrepancy between dwarfs and giants for EMP groups that is clearly discernible in the figure.
Three $\alpha$-element enhanced stars are located greatly apart from the main paths in the relations of Mg with other elements, while falling on or close to the latter in the relation between Ti and Ca.
It is remarkable that HE1424-0241 is in good agreement with the overall trend of \aelms\ except for the correlation between Mg and other elements.
In the viewpoint of the relation between \aelm\ abundances, HE1424-0241 can be regarded as a Mg-enhanced star rather than an $\alpha$-depleted star.
This figure also includes the three most iron-poor, HMP/UMP stars \citep[HE0557-4840, HE0107-5240, and HE1327-2326,][]{Norris2007,Christlieb2002,Frebel2005}.
These stars are almost on the main path in this diagram, although HE1327-2326 apparently deviates from the main correlations in the Ti-Mg and Ca-Mg relation.
It should be noted, however, that HE 1327-2326 differs greatly in the enhancement of Mg relative to carbon from the $\alpha$-enhanced stars, i.e., $\abra{Mg}{C+N} \simeq -2.5$ vs. $-0.19 \hyp -0.24$.

If this is the case, what causes the different behaviour in the chemical enrichment of \aelms\ and iron-group elements?
Considering the non-existence of the significant enhancement/depletion of iron-group element abundances in EMP stars, decoupling of chemical yields from SNe certainly happened before the formation of stars with $\feoh \la -3.5$.

Accordingly, the peculiarity among \aelm\ and titanium abundances are essentially ascribed to the enhancement of Mg abundance.   
The abundance ratios of Mg to carbon are too large to be realized by means of the currently known nucleosynthesis in AGB stars \citep[see eq.~(A19) in][]{Nishimura2009}.
There is a possibility that these stars were born from the ejecta of peculiar SNe as discussed in \citet{Aoki2002d}, although \citet{Cohen2008} pointed out the difficulty in reproducing the abundance pattern of HE1424-0241 with nucleosynthesis by type II SNe.
One of these \aelm\ enhanced stars belongs to NEMP stars (CS22949-037, see \S~\ref{sec:nemp} below) with large enhancement of oxygen, as large as, or even exceeding, the enhancement of carbon and nitrogen enhancement.  
The large enhancement of CNO elements is also the case for other $\alpha$-enhanced stars.
Another possibility for the origin of these stars is the dredge-up of products of carbon burning during the super AGB evolution with ONeMg core.
In any case, the existence of stars with peculiar $\alpha$-element abundances only at low metallicity poses a mystery in the sense that the majority of the sample shows no deviation from the average value irrespective of metallicity.

% Figure [Fe/H] vs. [Sc,V,Cr,Mn,Co,Ni,Cu,Zn/Fe]
\begin{figure*}
  \begin{center}
    \includegraphics[width=0.7\textwidth]{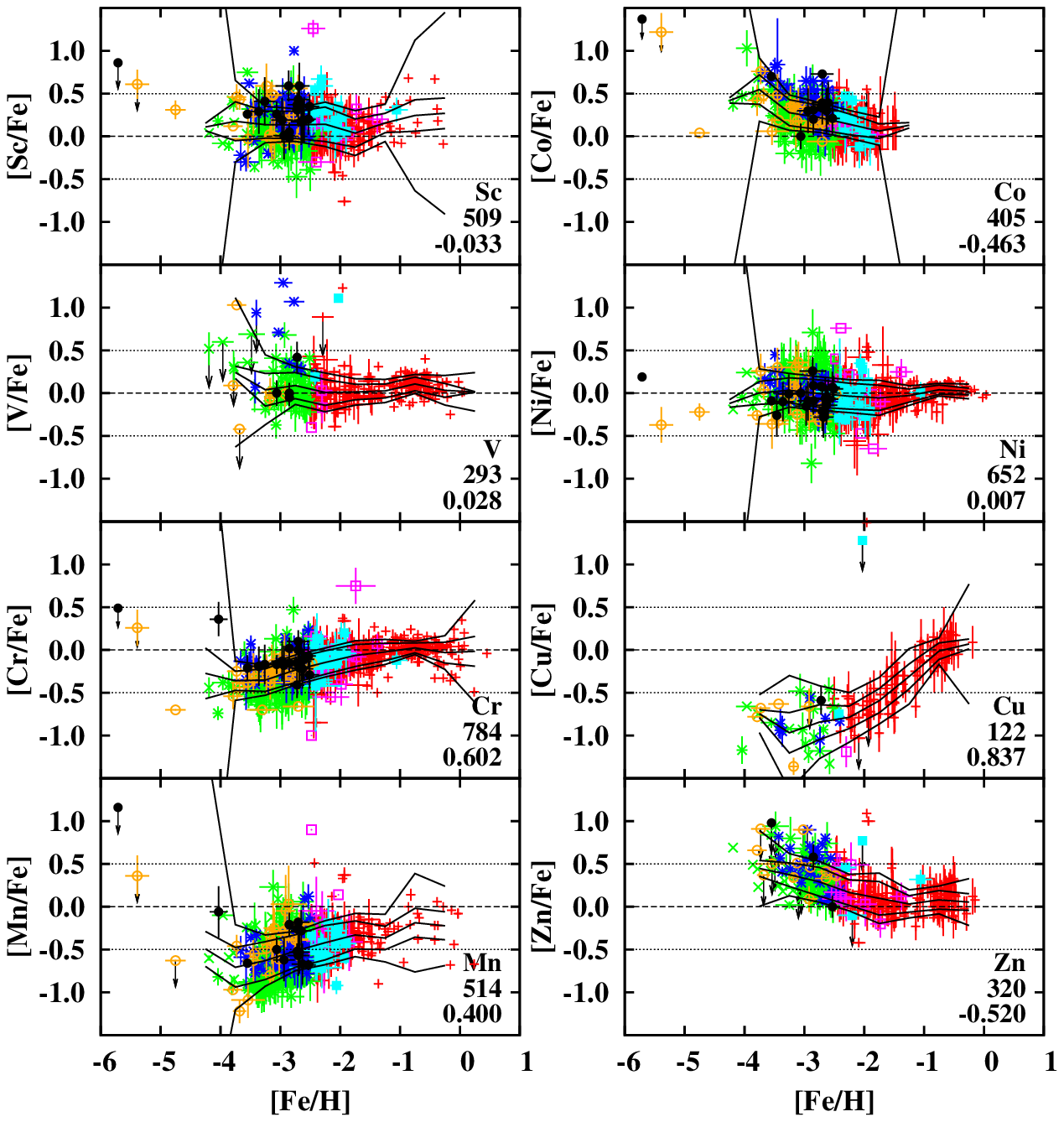}
  \end{center}
  \caption{Iron group element (Sc, V, Cr, Mn, Co, Ni, Cu, Zn) abundances as a function of metallicity.
		   The number of sample stars and the Pearson coefficient are shown in bottom right corner in each panel.
		   The definition of the symbols is the same as in Fig.~\ref{fig:cbyh}.
		   We exclude the data for which the abundances are weakly constrained by upper limits.
		   The scaled solar value and [X/Fe] $= \pm 0.5$ are shown by dashed and dotted lines for visibility of the difference between elements.
		   The five solid lines are a sample mean and standard deviations of population as described in \S~\ref{sec:sample} and in Fig.~\ref{fig:afe}.
  }\label{fig:iron}
\end{figure*}

% Figure [n/Fe] vs. [Fe/H]
\begin{figure*}
  \begin{center}
    \includegraphics[width=\textwidth]{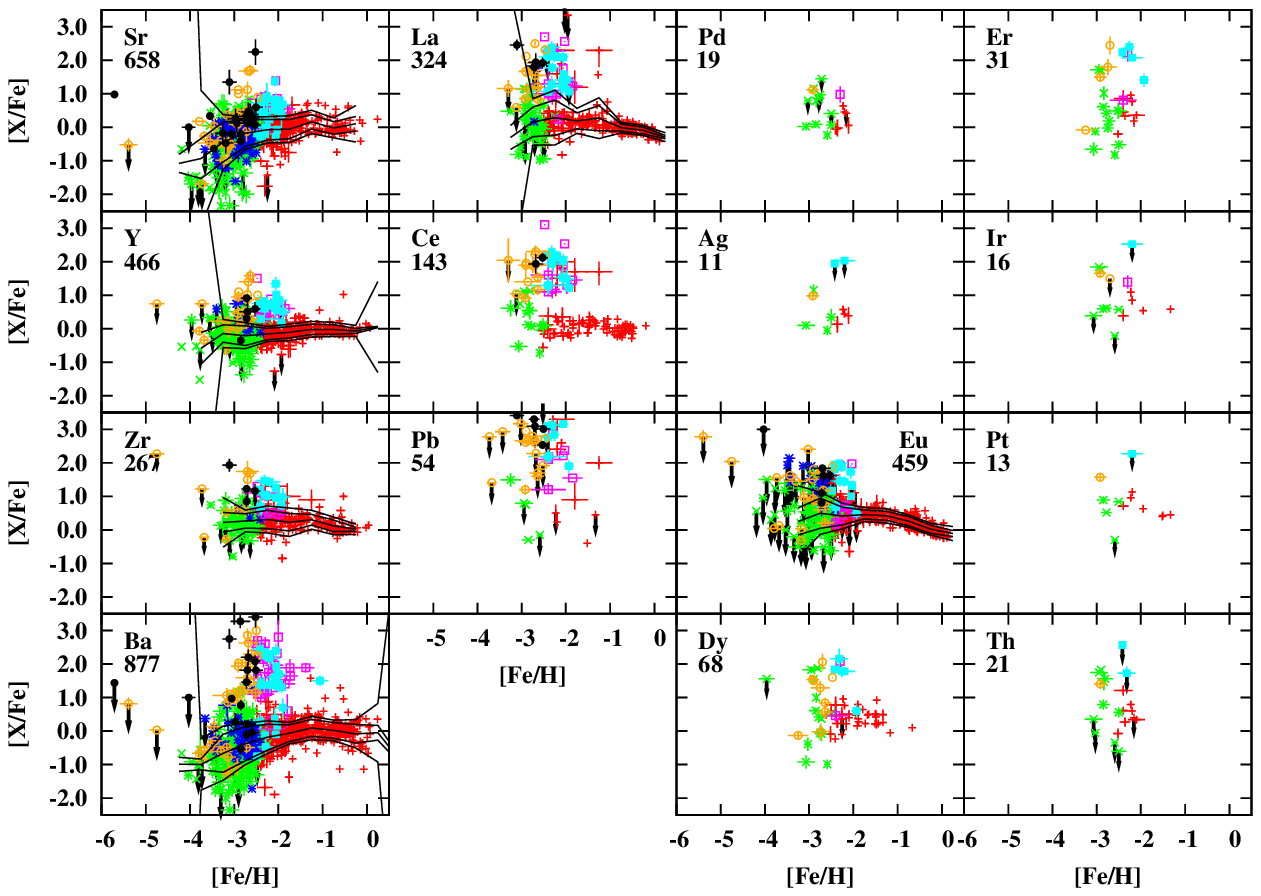}
  \end{center}
  \caption{Enhancement of neutron-capture elements relative to iron as a function of metallicity.
           Element species and the number of data are shown in each panel.
		   Seven elements in the left hand side are the \sit process elements, and the eight elements in the right hand side is the \rit process elements, which are assigned according to the contributions of these two processes of nucleosynthesis to the solar abundances.
		   The meaning of solid lines for selected elements is the same as in Fig.~\ref{fig:afe}.
		   For more detail, see the description in the caption of Fig.~\ref{fig:afe} or in \S~\ref{sec:sample}.
		   The definition of the symbols is the same as in Fig.~\ref{fig:cbyh}.
  }\label{fig:nfe}
\end{figure*}

\subsection{Iron-group elements}

Iron group elements ($21 \leq {\rm Z} \leq 28$) are considered good tracers of chemical evolution as well as the diagnosis for the yields of SNe.
These elements are thought to be produced by massive stars during explosive burning by supernovae, although some previous observations suggest the possibility of the contribution by \sit process for copper and zinc \citep{Sneden1991,Mishenina2002}.
Since the lines of these elements are relatively easily detected in EMP stars, hundreds of abundances are available in the database.
In the following, we present the general trend for iron group elements.

Fig.~\ref{fig:iron} shows the enhancement of iron group elements relative to iron as a function of metallicity.
In the database, hundreds of abundances are available, and the global trends are consistent with all previous works for the increased number of samples in the SAGA database.
According to the estimate of the confidence interval of the population variance, it is confirmed that the scatter of the element abundances is as small as the observational errors.
In particular for the elements with large number of samples, it does not vary for metallicity below $\feoh \sim -2$ down to $\feoh \simeq -3.5$.
As in the case of \aelms, we cannot see the possible scatters caused by the strong mass dependence of SN yields in this metallicity range.  
Differently from \aelms, however, the iron group elements display different tendencies of the mean abundances varying with the metallicity.  
These behaviours may give an insight into the metallicity and mass dependences of SNe yields of elements.
We return to this discussion in detail in \S~\ref{sec:enrich}.
For $\feoh \ga -1$, it is not clear if the scatter exists or not for some elements like vanadium and chromium, due to the small number of samples.

As previously reported, the abundance ratios of Cr and Mn decrease with decreasing metallicity, while Co abundance shows an opposite trend \citep{McWilliam1995b}.
As discussed in \citet{Lai2008} and also later in \S \ref{sec:unexp}, derived Cr and Co abundance depends on the effective temperature especially at metallicity below $-2.5$.
In Fig.~\ref{fig:iron}, the abundances data of Cr are unbiased in the effective temperature over the whole metallicity range including abundances at $\feoh < -2.5$ where the dependence of effective temperature appears.
However, it is possible that the decreasing trend of Cr abundance is caused by a number of the abundance data with potentially smaller values for the EMP and CEMP RGB groups plotted in the figure.

Copper abundances show an increasing trend with metallicity as first pointed out by \citet{Sneden1991} and confirmed by \citet{Mishenina2002}.
For the Zn abundance, we can see an increasing trend with decreasing metallicity at $\feoh < -2$ in this figure, as confirmed for $\feoh \la -3$ by \citet{Cayrel2004} and \citet{Nissen2007} irrespective of the NLTE correction.
On the other hand, for higher metallicity, it is suggested that the value of [Zn/Fe] is constant at $\feoh \ga -3$ \citep{Mishenina2002}, or at $\feoh \ga -2.0$ \citep{Nissen2004,Nissen2007}, and slightly enhanced at $\feoh \sim -1$ \citep{Saito2009}.
These trends are thought to be explained by the combination of the contributions from the $\alpha$-rich freeze-out process for zinc at low metallicity \citep{Cayrel2004}, from hypernovae at $\feoh < -3$ \citep{Nomoto2006}, and from type Ia SNe for zinc and copper abundance trends \citep{Mishenina2002}.

% Figure [Sr,Y,Zr/H] vs. [Ba,Eu/H]
\begin{figure*}
  \begin{center}
    \includegraphics[width=1.0\textwidth]{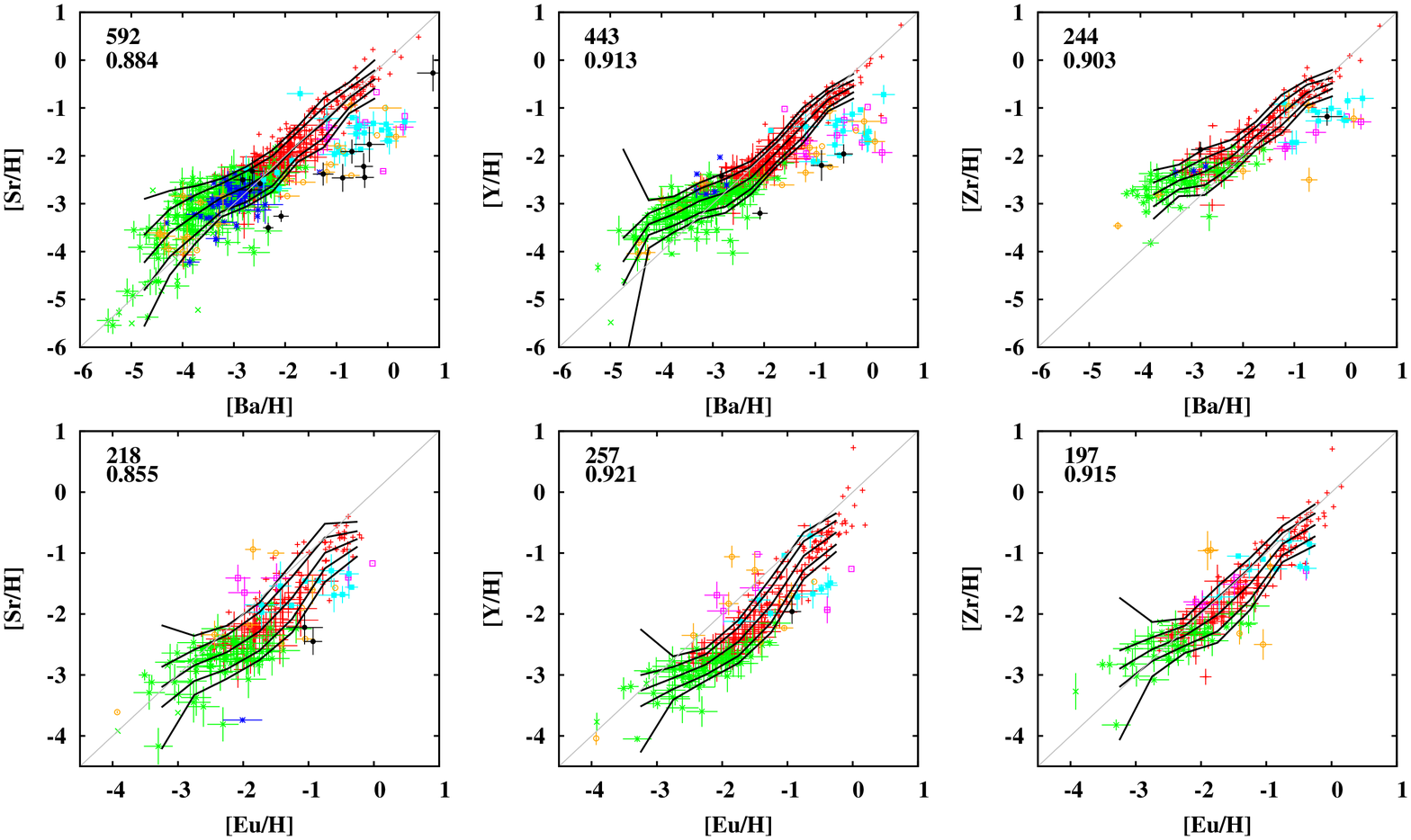}
  \end{center}
  \caption{Abundance relations between neutron-capture elements for the sample excluding the data for which only upper limits are available.
		   The definition of the symbols is the same as in Fig.~\ref{fig:cbyh}.
		   Black solid lines show the sample mean of \abra{X}{H} (X is Sr, Y, or Zr) in the range of 0.5 dex in \abra{Ba}{H} or \abra{Eu}{H} and the 95 \% confidence interval of standard deviation for stars excluding the carbon-enhanced groups.
		   Grey lines have the slope of unity and the intercept of zero.
  }\label{fig:weakr}
\end{figure*}

\subsection{Neutron-capture elements}\label{sec:ncap}

It is well known that the abundances of neutron capture elements have a large scatter around $\feoh \sim -3$, first proposed by \citet{Gilroy1988} for halo stars with $\feoh < -2$.
This is shown in Fig.~\ref{fig:nfe}.
In our sample, almost all the \sit\ and \rit process elements show similar scatter at $-3.5 \la \feoh \la -2$.
For the stars in the carbon-normal groups, we see that the derived abundances display the scatters as large as 3 dex or even more, depending on the sample sizes and/or on the detection limits.   
Barium, with the largest sample size and with the smallest detection limit as low as $\abra{Ba}{H} \sim -5.5$, exhibits the greatest scatters exceeding the extent of $\sim 3.5$ dex.  
Another \sit process element strontium, with the second largest sample also shows the scatters beyond 3 dex.  
Such behaviour is common to the \rit process element like europium, but with slightly smaller scatter of $\sim 2.5$ dex because of relatively larger detection limit of $\abra{Eu}{H} \simeq -4$.  
For other \rit process elements like dysprosium and erbium, the scatters are observed as large as, or even larger than europium, despite much smaller number of sample stars.  
Even thorium shows the scatters beyond 2 dex.
This simply suggests that the abundances of neutron-capture elements for these stars are dominated by the \rit process rather than by the \sit process.

In this figure, we plot the mean values and the interval of estimated standard deviation of the population excluding the carbon-enhanced groups for Sr, Y, Zr, Ba, La, and Eu.
We can see that the enrichment \xfe\ decreases with decreasing metallicity for $\feoh < -2$ for these elements, though the tendency is affected by the large detection limit for Y, Zr, and Eu.
In addition to these decreasing trends, we see the large scatter greater than the typical observational errors.
For strontium and barium abundances, the standard deviation becomes $0.60 \hyp 0.82$ and $0.57 \hyp 0.80$, respectively, in the metallicity range of $-3.5 \leq \feoh < -3$, respectively, where more than 60 stars are included in this range.
The range of the standard deviation is $0.45 \hyp 0.63$ for europium in the metallicity range of $-3 \leq \feoh < -2.5$.
These are thought as the consequence of intrinsic scatter and/or the mass dependence of yields in the former nucleosynthetic sites where they were produced.

Fig.~\ref{fig:weakr} shows the correlation of abundances between heavy (the atomic number of $56 \leq Z \leq 70$) and light ($38 \leq Z \leq 46$) neutron-capture elements.
The argument that the abundance scatter of strontium decreases with increasing barium abundance, as pointed out by \citet{Truran2002} and discussed by \citet{Honda2004b} and \citet{Aoki2005}, still holds in our database containing 521 stars without carbon enhancement.
This may be taken as evidence for the so called ``weak \rit [\sit]'' process that produce more light neutron-capture elements than heavy neutron-capture elements, as discussed in the above papers.
For other combinations of neutron-capture elements, the above trend is not confirmed because of the lack of data for $\abra{Ba,Eu}{H} \la -4$.

On the other hand, the mean abundances of \abra{Sr,Y,Zr}{H} against \abra{Ba,Eu}{H} in 0.5 dex bin clearly deviates from the relationship with the slope of unity to which they are nearly subject to $\abra{Ba,Eu}{H} \ga -2$.
This is confirmed by the confidence intervals of the mean abundances for all the panels in Fig.~\ref{fig:weakr}.
Obviously, we can see the trend of the slope gradually increasing with increasing abundances for $\abra{Ba,Eu}{H} \ga -3$, especially for the relations between light neutron-capture elements and europium.
This may suggest that the production efficiency of light or heavy \rit process elements varies with the chemical enrichment history of the Galactic halo.

Another interesting feature is the changes of the slopes for low \abra{Sr}{H} and \abra{Y}{H} at $\abra{Ba}{H} \la -4$.
This trend may hold for all the elements presented in the figure, although the number of the sample with such low abundances is too small to be conclusive.
The interpretation of this trend can be either that there is another transition of the production efficiency of \rit process elements or that the very small abundances of neutron-capture elements are the result of the accretion of interstellar medium which have the abundances of $\abra{Ba}{H} \ga -3$.
For the relations between light neutron-capture elements and europium, some stars show relatively small abundances of light neutron-capture elements having $\abra{Sr,Y,Zr}{H} < -3$ with respect to \abra{Eu}{H}.
In any case, we need more data, and more study with the mass and metal dependence of SNe yields, and also, with the modification by interstellar accretion, to discuss about the very beginning or the stages of the smallest abundances, of the chemical enrichment processes of neutron-capture elements in the Galactic halo.

\section{Chemical Enrichment of our Galaxy}\label{sec:enrich}

The chemical enrichment process of our Galaxy is one of the major concerns about the evolution of our Galaxy.
In the next subsection, we discuss the chemical imprints that are left in the surface chemical composition of stars.
We analyze the relationship of abundance variations of elements with those of iron, in order to reveal the characteristics of enrichment history for each element.
In \S~\ref{sec:yield}, we derive chemical yields relative to iron ones for type II SNe of Pop.~III and Pop.~II from the abundance trends available in the database.

% Figure [X(iron)/H] vs. [Fe/H]
\begin{figure*}
  \begin{center}
    \includegraphics[width=\textwidth]{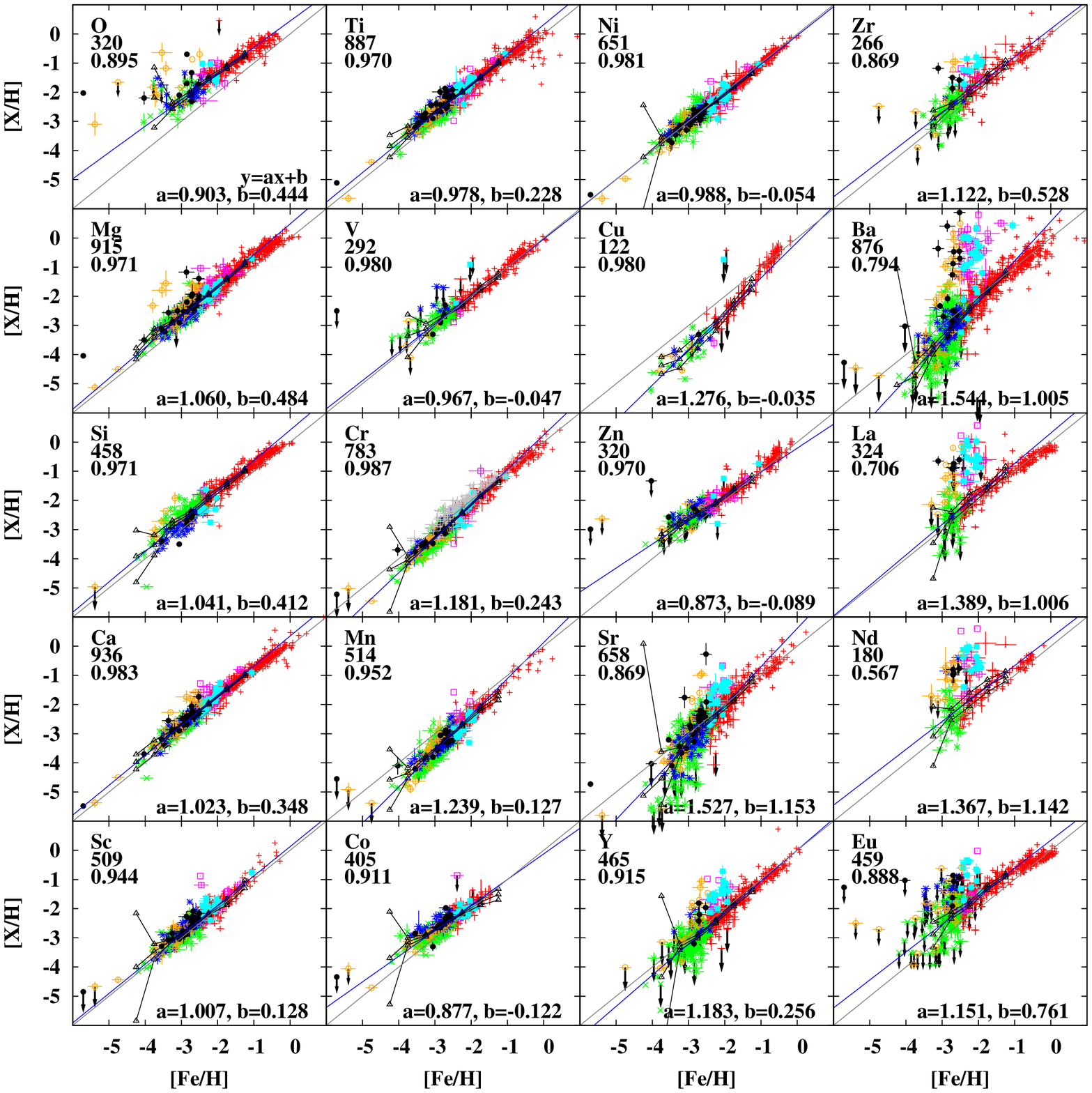}
  \end{center}
  \caption{Abundances in the form of [X/H] as a function of \feoh.
  			The meanings of numbers and symbols are the same as in Fig.~\ref{fig:iron} except for grey points in Cr abundances for which these abundances are determined by Cr~II lines only.
			A power law fit with slope $= 1$ is indicated by the solid lines.
			The slope and intercept of the RMA regression are given in bottom right corner of each panel.
			These values and their uncertainties are given in Table~\ref{tab:rma}.
			Three solid lines show the average value with a metallicity interval of 0.5 dex and its 95 \% confidence interval.
			In these analyses, we exclude data with upper limits for any elements and data belonging to the carbon-rich groups for elements except for vanadium to zinc.
  }\label{fig:enrich}
\end{figure*}

\subsection{Chemical enrichment during $-4 \la \feoh \la -1$}

Fig.~\ref{fig:enrich} shows the chemical enrichment in our Galaxy for \aelms, iron-group elements, and neutron-capture elements for which we have more than 100 data for one element in the database.
The enrichment is thought to be the effect of the combination of type II and Ia SNe \citep{Matteucci1986}, i.e., typical type II SNe contribute to the enrichment of \aelms, while type Ia SNe produce much iron to reduce the value of \abra{$\alpha$}{Fe} for $\feoh \ga -1$.
The labels in the top left corner represent the element species, the number of data, and the Pearson coefficient.
We plot in each panel the average values of abundances in 0.5 dex bin width and its 95 \% confidence interval by solid lines.
In addition, we plot by solid lines the linear relationship, $\abra{X}{H} = a \feoh + b$, between the element and iron, obtained by the RMA regressions.
The slopes and intercepts are given in the bottom right corners of each panel and also in Table~\ref{tab:rma}, the latter of which provides the element species, the slope and its standard error, and the intercept and its standard error from the first to the fifth columns.
In the calculation, we deal with the stars with the metallicity of $\feoh \leq -1$ in order to remove the possible effects of Type Ia supernovae.
For \aelms\ and neutron-capture elements, we removed the sample in the carbon-enhanced groups to derive these values since these elements are possibly affected by carbon-enrichment processes other than supernovae.
For carbon abundance, the abundance trend and the regression analysis are given in Fig.~\ref{fig:cbyh}.

The enrichments of carbon and oxygen relative to iron both decreases with increasing metallicity appreciably with the slopes significantly below unity, i.e., $a \simeq 0.876 \pm 0.017$ and $0.903 \pm 0.023$ if we remove the possible contamination by CN cycles for carbon abundance (see Fig.~\ref{fig:cbyh} and \S~\ref{sec:cn}).

In contrast, the overall trends for \aelms\ like Mg, Si, Ca, and Ti are almost the same, exhibiting the abundances in nearly exact proportion to the iron abundance with the slopes $a$ close to unity;  
the slopes decrease from $\alpha =1.060$ for Mg to 1.023 and 0.970, respectively, for Ca and Ti.
for example, this is true for Si and Ca ($\alpha= 1.04$ and $1.02$, respectively), which may pose the severe constraint on the SN models and their mass and metallicity dependences and also on the evolutionary variations of IMF, since the nucleosynthetic sites are thought to be different.
The slight deviations from unity may be reconciled by the systematic errors that is not taken into account in these estimates.
This may pose severe constraints on the SN models and their mass and metallicity dependences and also on the evolutionary variations of IMF.

\begin{table}
	\begin{center}
    \begin{minipage}{70mm}
    \caption{Slopes and intercepts with their one $\sigma$ error obtained by the RMA regression for chemical enrichment.}
    \label{tab:rma}
    \begin{tabular}{lcccc}
      \hline
       Element & Slope $a$ & $\sigma_{a}$ & Intercept $b$ & $\sigma_{b}$ \\
         C  & 0.876 &  0.017 & $-0.175$ & 0.041 \\
         O  & 0.903 &  0.023 &  0.444 & 0.048 \\
         Mg & 1.060 &  0.012 &  0.484 & 0.029 \\
         Si & 1.041 &  0.020 &  0.412 & 0.047 \\
         Ca & 1.023 &  0.009 &  0.348 & 0.021 \\
         Sc & 1.007 &  0.016 &  0.128 & 0.043 \\
         Ti & 1.015 &  0.011 &  0.318 & 0.027 \\
         V  & 0.967 &  0.023 & $-0.047$ & 0.053 \\
         Cr & 1.181 &  0.010 &  0.243 & 0.026 \\
         Mn & 1.239 &  0.020 &  0.127 & 0.054 \\
         Co & 0.877 &  0.018 & $-0.122$ & 0.050 \\
         Ni & 0.988 &  0.011 & $-0.054$ & 0.030 \\
         Cu & 1.276 &  0.044 & $-0.035$ & 0.105 \\
         Zn & 0.873 &  0.020 & $-0.089$ & 0.049 \\
         Sr & 1.527 &  0.035 &  1.153 & 0.090 \\
         Y  & 1.183 &  0.026 &  0.255 & 0.062 \\
         Zr & 1.122 &  0.044 &  0.528 & 0.100 \\
         Ba & 1.544 &  0.031 &  1.005 & 0.073 \\
         La & 1.389 &  0.073 &  1.006 & 0.154 \\
         Nd & 1.367 &  0.091 &  1.142 & 0.202 \\
         Eu & 1.151 &  0.043 &  0.761 & 0.096 \\
      \hline
   \end{tabular}
   \end{minipage}
   \end{center}
\end{table}

From the current understanding of chemical evolution in our Galaxy, the values of intercept for \aelms\ represent the typical contribution by type II supernovae.
For O, Mg, Si, Ca, and Ti, the values of the intercepts means that the value of $\xfe$ becomes $\simeq 0.3 \hyp 0.4$ for $-4 \la \feoh \la -1$ for elements except for oxygen.
For oxygen abundance, the result suggests that $\ofe \approx 0.5$ at $\feoh = -1$ and $\approx 0.8$ at $\feoh = -4$.

The iron group elements have various slopes with respect to metallicity at variance with \aelms.
They are clearly separated into three groups by the slopes; Co and Zn with the slopes of $< 0.9$, Cr, Mn, and Cu having slope of $> 1.1$, and Ni having slope of $\approx 1$ by 95 \% confidence level.
As stated above, the Cr abundance may be affected by the possible underestimate of its value at low metallicity as well as a discrepancy between abundances determined by Cr~I and Cr~II \citep[see e.g., ][]{Sobeck2007}, the latter of which is shown in the figure by grey points.
The values of intercepts are approximately zero except for elements having the slopes deviated from unity.
Therefore, total enrichment by iron group elements proceeds linearly with metallicity like \aelms.

For any neutron-capture elements, the slopes clearly change with the metallicity in Fig.~\ref{fig:enrich}.
For elements like Sr and Ba for which more than 100 abundance data are available at $\feoh < -3$, the abundances drop rapidly at $\feoh \la -3$ in Fig.~\ref{fig:enrich}.
Therefore, using linear regression does not make sense for these elements in the range of small metallicity, although linear relationship holds good for large metallicity of $\feoh \ga -2$. 
For other neutron-capture elements, linear fit looks better, because of the lack of stars with small abundances owing to detection limit for$\feoh \la -3$.
.

The slopes and diversity in Fig.~\ref{fig:enrich} may give us the general picture of the evolution of our Galaxy.
The slopes remain almost constant between $-4 \la \feoh \la -1$ except for neutron-capture elements and the constant slope is not good approximation probably for copper and zinc, as can also be seen in Fig.~\ref{fig:iron}.
There are several factors that may affect the changing or constant slopes, and small or large scatters in element abundances; ISM mixing in host halos after SN explosions, the metal and initial mass dependences of chemical yields of SNe, and the IMF.

The element-to-element variations of the slope in Fig.~\ref{fig:enrich} suggests that the average yields from SNe depend on metallicity at $\feoh \ga -4$ for the elements with non-unity slopes as discussed below in this section.
For elements having the slope of unity, on the other hand, it is not expected to have severe metallicity nor mass dependences for chemical yields.
It is likely therefore that the chemical yields of \aelms\ have very little dependence on metallicity to the extent that the change of slope is not detectable in the trends.
The chemical yields of \aelms\ have little dependence on the initial mass of stars.
The deviations of slopes from unity can be brought about not only by the metallicity but also by the mass dependence of SN yields with the change of the IMF or both.
For iron group elements with the sloes different from unity, therefore, the SN yields vary with the metallicity or with the change of IMF and the mass dependence.  
At present, it is not possible to distinguish the metallicity dependences of yields from the effect of changing the IMF if either or both of them happened.

For neutron-capture elements, the observed trends require strong mass dependence for the nucleosynthetic site of \rit process elements.
The obvious changes of slopes with respect to metallicity as well as the large scatters may imply the dependence on metallicity and different types of supernovae such as hypernovae and electron-capture supernovae for which the production of Zn and light neutron-capture elements is suggested by the models of massive stars and chemical evolution \citep[see, e.g.][]{Qian2008,Wanajo2009}.

In conclusion, the slopes of very close to unity imply that the SN yields are independent on the metallicity or the initial mass of progenitor stars for \aelms.
For most of iron group elements and neutron capture-elements, the deviations of slopes from unity can be explained in terms of the metallicity dependence of SN yields, or result from the combination of the mass dependence of SN yields with the change in IMF.

In \S~\ref{sec:cemp}, we pointed out the possibility that the IMF may vary at $\feoh \sim -2$.
Clear evidence of the changing the IMF cannot be seen in Fig.~\ref{fig:enrich}, probably because the changes of trends and scatters caused by the IMF are not so significant.
However, a detailed analyses of abundance trends for oxygen, cobalt, and zinc reveals the possible transition at $\feoh \sim -2$ (Yamada, Suda, Fujimoto et al. in preparation).
It is to be pointed out that these elements including carbon show the common slopes of $a < 1$ by the RMA regression, though it is unknown if this is a coincidence or not.

It is to be noted, however, that there is another possibility that the formation history of the Galactic halo affects the chemical abundances of ISM by changing the total mass of halo through the merging events.
This can be related to the lifetime of mini-halo to construct the Galactic halo.
According to the results of \citet{Komiya2009b}, the typical lifetime of mini-halo is the order of $\sim 10$ Myr at $z = 30$ and $\sim 100$ Myr at $z = 10$, the former of which is comparable to the lifetime of massive stars. 
It also seems likely that the formation history of the Galactic halo reproduce the large scatter of only \rit process elements if the production site of these elements are localized to $8 \hyp 10 \msun$ as suggested by the previous works as stated above.
More detailed discussion for this topic will be given in the forthcoming paper (Yamada, Suda, Fujimoto et al., in preparation).

\subsection{Chemical yield ratios derived from observations}\label{sec:yield}
The chemical enrichment with different slopes can be interpreted as evidence of completely different chemical yields at $\feoh < -4$ from those at higher metallicity, i.e., different yields from probably Pop.~III SNe.
This is because simple chemical evolution begins with $X_{i} = 0$ ($i$ is the element species of the iron group) at the big bang chemical composition and ends with $X_{i} = X_{i,\sun}$.
A non-unity slope means different sources for initial enrichment for the relevant element.
If we assume that all Pop.~III SNe results in $\feoh = -4$ in the second generation, we can estimate the ratio of averaged chemical yield from Pop.~III SNe using the coefficient of linear fit.
The yield ratio averaged over the IMF for species $i$, $\langle Y_{i} \rangle$ to that for iron yield $\langle Y_{\rm Fe} \rangle$ is given by $X_{i,\sun} / X_{{\rm Fe},\sun} 10^{(a - 1) \feoh + b}$ where $X_{i,\sun}$ is the solar abundance of element species $i$.
The results are given in Tab.~\ref{tab:yield}.

\begin{table*}
\begin{center}
    \caption{Estimated ratio of averaged chemical yields by Pop.~III supernovae}\label{tab:yield}
    \begin{minipage}{\textwidth}
    \begin{tabular}{{l}*{7}{c}}
      \hline
       Element & C & O & Mg & Si & Ca & Sc & Ti \\
	   $\langle Y \rangle / \langle Y_{\textrm Fe} \rangle$
	   & \pow{5.50}{0} & \pow{5.59}{1} & \pow{9.91}{-1} & \pow{1.07}{0}
	   & \pow{9.76}{-2} & \pow{4.19}{-5} & \pow{4.51}{-3} \\
	   \hline
       Element & V & Cr & Mn & Co & Ni & Cu & Zn \\
	   $\langle Y \rangle / \langle Y_{\textrm Fe} \rangle$
	   & \pow{3.93}{-4} & \pow{5.02}{-3} & \pow{1.69}{-3} & \pow{6.74}{-3}
 	   & \pow{6.19}{-2} & \pow{5.22}{-5} & \pow{4.49}{-3} \\
      \hline
   \end{tabular}
   \end{minipage}
\end{center}
\end{table*}

\begin{figure}
  \begin{center}
    \includegraphics[width=84mm]{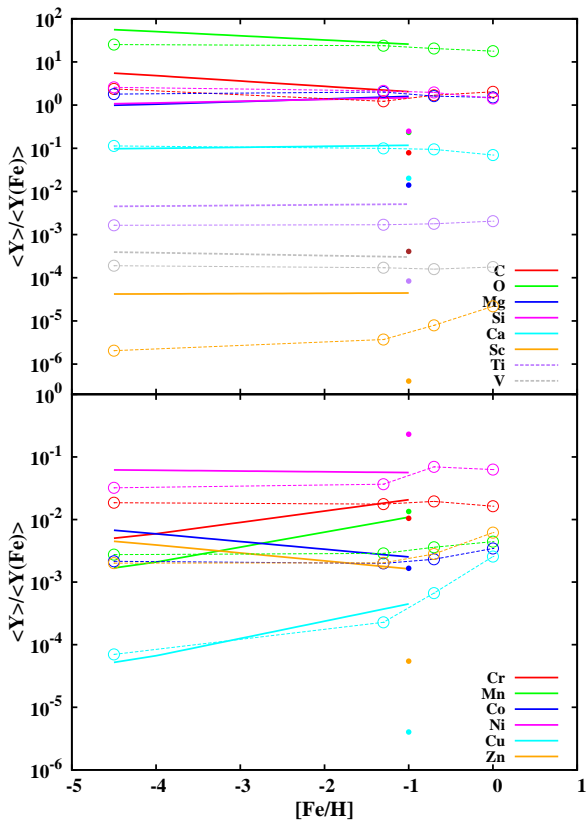}
  \end{center}
  \caption{Estimated metallicity dependence of the chemical yield ratio from SNe derived from a power law fit in Fig.~\ref{fig:enrich} (denoted by lines).
  			Open circles represent IMF averaged chemical yields considering SNe and hypernovae taken from \citet{Kobayashi2006}.
			Note that the chemical yields from stars at $Z = 0$ are plotted at $\feoh = -4.5$.
			Filled circles represent yield ratios by type Ia SNe at $\feoh = -1$.
  }\label{fig:yield}
\end{figure}

This simple estimate can be applied to stars with $\feoh \ga -4$ based on the assumption discussed in the previous subsection.
The metallicity dependence of SN yields of element species $i$ having slope $a$ and intercept $b$ can be written by
\begin{equation}
\frac{\langle Y_{i} \rangle}{\langle Y_{\rm Fe} \rangle} = \frac{d X_{i,{\rm ISM}}}{d X_{\rm Fe,ISM}}
= a \frac{X_{i,\sun}}{X_{{\rm Fe},\sun}} 10^{(a - 1)\feoh + b}
\end{equation}
where $\langle Y_{i} \rangle$ is the IMF averaged yield of element species $i$ and $X_{i,{\rm ISM}}$ is the mass fraction of the interstellar medium in the Galactic halo.
Here we adopt the solar chemical composition by \citet{Anders1989}.
The ratio of chemical yields using this equation for each element is  shown in Fig.~\ref{fig:yield}.
The results are compared with SN models provided by \citet{Kobayashi2006}.
The ratios for IMF averaged yields are given by open circles in the figure.
Here we plot the results for $Z = 0$ at $\feoh = -4.5$.
For SN models at $\feoh = -1.0$, we plot the yields from type Ia SNe by filled circles.
For most elements, the values of yield ratios for type II models and observations are consistent with each other, while Type Ia models are inconsistent with observations.
The metallicity dependence for type II SNe shows an opposite trend for iron-group elements, except for copper.

The different chemical yields deduced from observations between Pop.~II and Pop.~III can be explained by either or the combinations of the following assumptions:
(1) The IMF at Pop.~III is different from the one at Pop.~II,
(2) the timescale of changing the environment of interstellar medium is comparable to or shorter than the lifetimes of massive stars that end with supernovae,
(3) the interstellar medium is not well mixed only at the formation epoch of Pop.~III stars,
or (4) the evolution of massive stars or the mechanism of supernova explosions is different from those in finite metallicity, which results in a different chemical yields.
For the first three assumptions, it requires an appropriate dependence of the chemical yields on initial mass.

% Figure <Yi><Y(Fe)> vs. [Fe/H]

\section{Possible extra mixing in EMP stars}\label{sec:mixing}

In this section we examine the possible extra-mixing in EMP stars from the following three aspects.
First, we deal with the lithium depletion observed in most of giants in relation to the second dredge-up, with the conversion of carbon to nitrogen by the CN cycles in the RGB stars and with nitrogen enhancement among the carbon-enhanced stars.
They are related to extra-mixing that may occur during the ascent of red-giant branch as well as to the hot-bottom burning and binary mass transfer from the erstwhile AGB primary stars.
Finally, we compare the peculiar abundances of EMP stars with the abundance anomalies observed for the stars of globular clusters to discuss the distinction between their extra-mixing mechanisms.

\subsection{Lithium depletion in surface convection}\label{sec:li}

\begin{figure}
  \begin{center}
    \includegraphics[width=84mm]{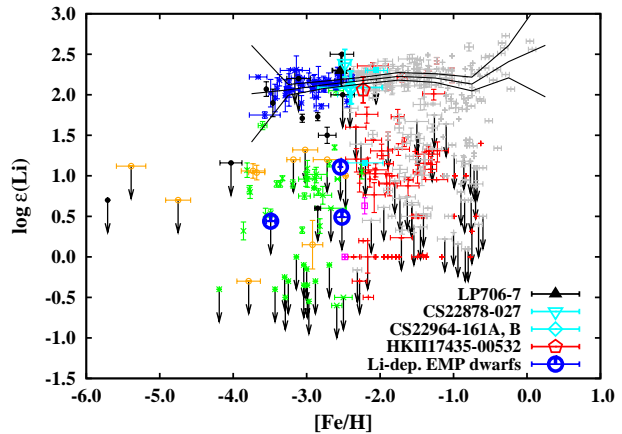}
  \end{center}
  \caption{Lithium abundance as a function of metallicity for 415 stars.
		   The data with arrows show upper limits.
		   The definition of the symbols is the same as in Fig.~\ref{fig:cbyh} except for the MP group.
		   We divide the MP group into RGB (shown by red symbols) and MS (shown by grey symbols) stars using the same criterion for the RGB in this study.
		   Stars having a Li-plateau value in the CEMP-MS group (LP706-7), C-rich-MS group (CS22878-027 and CS22964-161), and the MP group (HKII 17435-00532) are shown by separate symbols.
		   Another symbol for Li-depleted EMP dwarfs refers to G158-100, G82-23, or HD340279.
		   See text for more detail about these stars.
  }\label{fig:li}
\end{figure}

The \nucm{7}{Li} abundance in EMP stars is a diagnostic for the effect of binary mass transfer as well as the big-bang nucleosynthesis.
It can also be used as a check for lithium production in AGB stars.
Fig.~\ref{fig:li} shows the lithium abundance of sample stars in the database as a function of metallicity.
The so-called ``Spite plateau'' of $\logli \simeq$ 2.1 is clearly seen in the figure for MP dwarfs and EMP MS population with only a few exceptions.
The dispersion of the plateau is rather narrow.
In this figure, we draw the sample means and the standard deviations obtained with 0.5 dex binning by taking the Spite plateau above $\logli >1.7$.
The spread is as small as the observational errors of abundances.
In addition, we recognize a tendency of s the mean value decreasing slightly for smaller metallicity from $\logli = 2.23 \pm 0.14$ at $\feoh = -1.75$ to $\logli =2.12 \pm 0.11$ at $\feoh = -2.75$.
The tendency persists for sill smaller metallicity with the mean value decreasing to $\logli =2.06 \pm 0.13$ at $\feoh = -3.25$ and further $\logli =2.01 \pm 0.20$ at $\feoh <-3.5$, although the sample number grows too small for statistical significance.

For carbon-enhanced dwarfs, most of the CEMP MS (9 of 10) and C-rich MS (4 of 7) stars are below the plateau, including the stars with upper limits.
This suggests that these stars are likely to be affected by binary mass transfer whose ejecta are devoid of lithium.
Among the turn-off stars or subgiants, CS22958-042 and HE1327-2326 show the Li abundance much smaller than the Spite plateau values \citep{Sivarani2006,Frebel2008}.
These two stars are common in that they have large enhancements of C, N, and Na and no enhancement of \sit process elements like Ba.
This can be the case for binary mass transfer without the production of Li in AGB companions, although although radial velocity variations are yet to be observed.
Another lithium-depleted dwarf, G 77-61, is also carbon-enhanced \citep{Plez2005}, which is a low-mass and wholly convective star belonging to a binary system with a period of $\sim 245$ days with an unseen primary.
CS29528-041 and CS31080-095 show mild depletion of lithium as small as $\log \epsilon ({\rm Li}) \approx 1.7$ \citep{Sivarani2006}.
These stars can also be affected by binary mass transfer as inferred from the enhancement of other elements like C and Ba.
HE0024-2523 shows $\logli = 1.5$ with the enhancement of carbon and \sit process elements \citep{Lucatello2003} and is known to be a spectroscopic binary.

On the other hand, there are some carbon-enhanced stars without depletion of the lithium abundance having a metallicity of $\feoh \simeq -2.5$.
One of these exceptional stars, LP706-7 classified as CEMP MS, has $\logli = 2.3$ with $\feoh = -2.53$ \citep{Aoki2008}.
For the C-rich MS group, three stars have nearly or as large as the Spite plateau value; CS22878-027 \citep[$\logli = 2.39$ and $\feoh = -2.48$, ][]{Lai2008}, CS22964-161A and B \citep[$\logli = 2.09$ for both and $\feoh= -2.39$ and $-2.41$, respectively, ][]{Thompson2008}.
As discussed in \citet{Sivarani2006}, the source of Li in these CEMP-MS stars may be from the unseen companion AGB stars.
The variations and frequencies of Li abundances among the carbon-enhanced MS stars may reflect the efficiency of Li-production in the envelope of AGB stars.
In our sample, there are 7 CEMP MS stars with $\logli > 1.7$ among 17 stars that have the Li abundances, including upper limits, measured out of 34 CEMP MS stars.
Simple inference may then give the frequency of Li production during AGB at $41 \hyp 20$ \%, although it demands an explanation for the coincidence of the largest Li production during AGB phase with the abundance of the Spite plateau values.

For CEMP and Crich giants, the lithium abundances are depleted with the scatter of more than two dex.
There are, however, a few stars classified as giants in the database that do not show any depletion from the Spite plateau value.
HKII 17435-00532 is a moderately carbon-enhanced giant ($\cfe = 0.68$ and $\log g = 2.15$) having $\logli = 2.06$ with $\feoh = -2.23$ studied by \citet{Roederer2008}.
This star also shows a moderate enhancement of \sit process elements.
The origin of this object is yet to be established, although they argue extra mixing in the red giant branch to enrich the surface lithium abundance in addition to the mass from an erstwhile AGB companion.
Another lithium-rich star, G255-32, is a spectroscopic binary \citep{Latham2002} with $\logli = 2.05$ and $\feoh = -2.60$ \citep{Charbonnel2005}, but without any information for element abundances.
This star is classified as EMP RGB in the database, but may be considered as a turn-off star according to its stellar parameters \citep[$\teff =5925$ and $\log g = 3.50$, ][]{Charbonnel2005}.
Two metal-poor stars, HD160617 and HD345957 show $\logli > 2.1$, while other elements are normal \citep{Jonsell2005,Fulbright2000}.

In addition, there are two or more EMP giants with moderate depletion of lithium.
One is G238-30 \citep[$\logli = 1.62$,][]{Boesgaard2005} that is a less evolved giant located at $\log g = 3.43$, $\teff = 5383$ K.
Another star, CS22966-057 \citep[$\logli = 1.35$,][]{Spite2005}, is a more evolved giant ($\log g = 2.2$, $\teff = 5300$ K).

% Figure [Fe/H] vs. log-e(Li)

\begin{figure}
  \begin{center}
    \includegraphics[width=84mm]{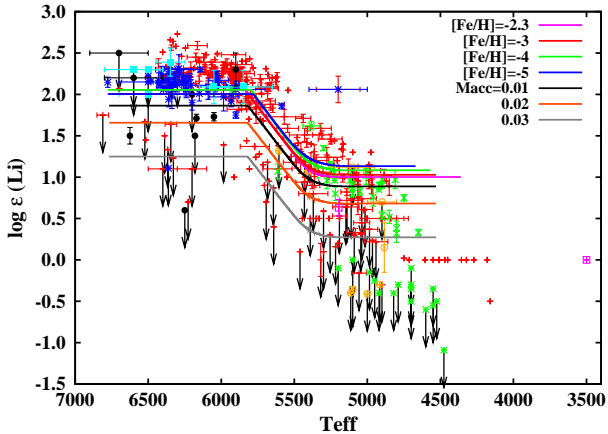}
  \end{center}
  \caption{Lithium abundance as a function of effective temperature.
           Lines represent the expected lithium abundance using stellar models by assuming that the lithium is completely burnt at $T \geq \pow{2.5}{6}$ K.
		   The bottom three lines represent the models assuming the accretion of gas devoid of lithium onto the top of the envelope.
		   These models are assumed to be $0.8 \msun$ with $\feoh = -3$.
		   The accreted masses are shown in the panel in units of $\msun$.
		   Sample stars and the definition of the symbols are the same as in Fig.~\ref{fig:li}.
		   The definition of the points is the same as in Fig.~\ref{fig:cbyh}.
		   Three cool dwarfs (G77-61, G158-100, and G82-28) are excluded from the figure to separate from giants.
  }\label{fig:liteff}
\end{figure}

The Li depletion in the RGB may be explained by the deepening of the convective envelope, but not necessarily so are its extent and scatter.
Fig.~\ref{fig:liteff} shows the dependence of the lithium abundance on the evolutionary status, where we plot the Li abundances as a function of the effective temperature.
In this figure, we superposed the expected variation of the surface lithium abundance due to the dilution by surface convection under the simple assumption that lithium is completely burned in the shell where the temperature is above $\pow{2.5}{6}$ K.
Stellar models are taken from $0.8 \msun$ and $\feoh = -3, -4$, and $-5$ of \citet{Suda2010}.
We obtain the depleted lithium abundance as $\log \epsilon ({\rm Li}) = 2.10 + \log [ (M_{\rm crit} /  \Delta M_{\rm conv}) ]$ where $M_{\rm crit}$ is the mass shell that the temperature reaches $\pow{2.5}{6}$ K and $\Delta M_{\rm conv}$ is the maximum mass in the surface convection after it attains deeper than $M_{\rm crit}$.
For simplicity, we start with an initial lithium abundance as the Spite plateau value of $\logli = 2.10$ \citep{Bonifacio2007}.
Each evolutionary line begins from the turn-off point and ends at the tip of the red giant branch.
The maximum depletion of the lithium abundance occurs soon after the core contraction during the subgiant branch when the surface convection attains at the maximum depth.
This leads to lithium isochrones for subgiants and giants as discussed in \citet{Deliyannis1990}.
The result agrees well with the models without molecular dilution in Fig.8 of \citet{Pilachowski1993}, although the metallicity range is different.
Our procedure is acceptable because the temperature at the base of surface convection is always well below the critical temperature for lithium burning in these low-metallicities.
The temperature at the bottom of surface convection remains below $\sim \pow{5}{5}$ K during the red giant branch phase.
This means that lithium cannot be burnt in the surface convection and is diluted into the entire convective region on the red giant branch in the current 1D models of low-mass stars.

As for the stars of the EMP RGB group for which the lithium abundance is determined, most of them are nearly on the lithium isochrones.
On the other hand, there are many stars only with upper limits well below the lithium isochrones.
Such a large depletion in the surface requires further mixing in the stellar interior below the convective envelope.
The possibility of extra mixing in red giants has long been suggested for the field giants \citep[e.g., see][]{Gratton2000} and for the globular cluster giants but to different extent (see below).
The mixing process has been discussed in relation to the rotation and/or to the turbulence mixing \citep[see. e.g.,][and the references there]{Fujimoto1999,Suda2006}.
As for the lithium depletion in globular cluster NGC6397 \citep{Korn2006,Lind2009}, it is also attributed to the existence of thermohaline mixing in red giants, proposed originally to explain the abundances of field giants \citep{Charbonnel2007} and to explain the inconsistency of \nucm{3}{He} abundance between observations and Big Bang nucleosynthesis \citep{Eggleton2006}.

Another possibility to explain lithium depletion is the binary mass transfer that pollutes the outer shell of the main sequence stars with matter totally, or almost, devoid of lithium.
In order to examine this assumption, we show three accretion models in Fig.~\ref{fig:liteff}.
In these models, we put an accreted matter of $0.01$ to $0.03 \msun$ onto the stellar surface so that the total mass becomes $0.8 \msun$.
These models can partly explain the lithium depletion in red giants by covering the possible range of lithium abundance.
However, this scenario cannot explain all of these stars since the frequency of such deep lithium-depletion is smaller in dwarfs than in giants.
In our sample stars, we have 86 out of 284 dwarfs (30 \%) and 54 out of 121 giants (45 \%) whose surfaces are deeply depleted with $\logli  < 1.7$ for dwarfs and $\logli  < 0.7$ for giants.
For EMP stars, the difference is even larger because the frequency of lithium-depleted stars is 11 \% (7 of 61 stars) and 50 \% (26 of 52 stars) for dwarfs and giants, respectively.
Consequently, there should be an extra-mixing mechanism to delete Li during the evolution from dwarfs to giants.
As for the proportion of stars affected by the binary mass transfer, the above frequencies of dwarfs with the deep Li-depletion give a measure although we should take into account the ABG stars with Li production.

% Figure T_eff vs. log-e(Li)

\subsection{Signature of CN cycles in EMP stars}\label{sec:cn}
It is reported that some halo stars show the signature of CN-cycles processing in their surface abundances.
These stars are thought to dredge-up nuclear products to the surface of observed stars or to undergo binary mass transfer whose companion polluted the envelope by hot bottom burning during the AGB phase.
In particular, \citet{Spite2005} demonstrate that EMP giants without carbon-enhancement are divided into ``mixed'' and ``unmixed'' groups that stars are enriched with nitrogen or not (see their Fig.~6), arguing that the CN cycles operate in observed EMP giants.

% Figure [N/C] vs. [C/Fe]
\begin{figure}
  \begin{center}
    \includegraphics[width=84mm]{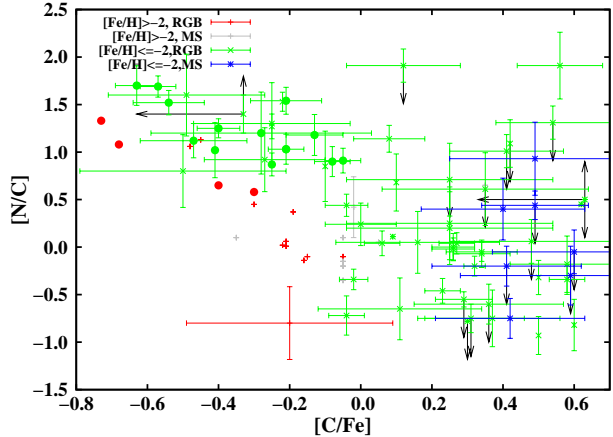}
  \end{center}
  \caption{Abundance relation between \nbyc and \cfe\ for 94 carbon-normal ($\cfe < 0.7$) stars taken from the SAGA database.
		   The definition of the symbols is shown in the top-left corner.
		   Note that we set the border between the EMP and MP groups at $\feoh = -2$.
		   Large filled circles denote stars with $\ciso \leq 10$ in the EMP RGB and MP groups.
  }\label{fig:nccfe}
\end{figure}

Fig.~\ref{fig:nccfe} shows the abundance ratios of carbon to nitrogen as a function of the carbon abundances for our sample of carbon-normal stars.
For our sample of 94 carbon-normal stars, the both ``mixed'' and ``unmixed'' groups are distributed continuously rather than separately as compared with Fig.~6 in \citet{Spite2005}.
For small metallicity of $\feoh < -2$, all the stars with small carbon enhancement of $\cfe < -0.1$ exhibit large nitrogen abundances with the ratio to the carbon abundance in the range of $\nbyc \simeq 0.8 \hyp 1.8$.
No stars are observed with the nitrogen abundance smaller than this despite the detection limit is well below $\abra{N}{H} =-3$ for giants with $\feoh < -2$.
Their large $\nbyc$ ratios are indicative of deep processing by CN cycles, corresponding to the ``mixed'' stars.

On the other hand, the stars without carbon depletion of $\cfe \ga -0.1$ are observed with small nitrogen abundances down to $\nbyc \simeq \abra{N}{Fe} \simeq -0.5$, corresponding to the ``unmixed'' stars group.
Note that ``unmixed'' stars have a rather wide range of nitrogen abundances, some above the solar ratios relative both to iron and carbon.
Among 171 carbon-normal stars with the carbon abundances $ 0.2 \leq \abra{C}{Fe} \leq 0.5$ in giants with $\feoh < -2$ in our sample, there are 14 stars having $\abra{N}{Fe} \ga 0$, mostly with $0 < \nbyc < 0.5$, among 28 stars for which the nitrogen abundances are derived.
This can be one of grounds for the assertion that nitrogen is a primary element, produced via hydrogen and helium during the evolution before the core-collapse at least in some Pop.~III and EMP stars, rather than a secondary element, produced from the pristine carbon and oxygen of the stars.

For $\feoh >-2.0$, on the contrary, there are stars with the small nitrogen enhancement of $\nbyc \la 0.5$ in the range of small carbon enhancement below $\abra{C}{Fe} = -0.1$.
Such low nitrogen enhancement categorised by the ``unmixed'' stars is restricted to the stars of $\abra{C}{Fe} \ga -0.4$.
For $\abra{C}{Fe} \la -0.4$, all the stars are observed with the large nitrogen enhancement of $\nbyc \ga 1.0$.
They show no enhancement of \sit process elements except for HD 110885 without the derived abundances for heavy elements.
We may regard these stars as the ``mixed'' stars considering the depleted lithium abundance and small values of \ciso, as argued by \citet{Gratton2000} and \citet{Spite2006}.
The number count of ``mixed'' stars candidate can be applied to stars with $\feoh > -2$ to double check the contribution of primary processes for nitrogen production, although the sample is small.
The result is almost the same as for metal-poor population where we find a slightly larger proportion of stars produced by primary processes.
Accordingly, we discern the same tendency of separation into the ``mixed'' and ``unmixed'' groups, though the boundary is shifted to smaller carbon enhancement with increasing metallicity;
this fact may have relevance to our finding that the stars in the metallicity below and above $\feoh \simeq -2$ have different origins.

\begin{figure}
  \begin{center}
    \includegraphics[width=84mm]{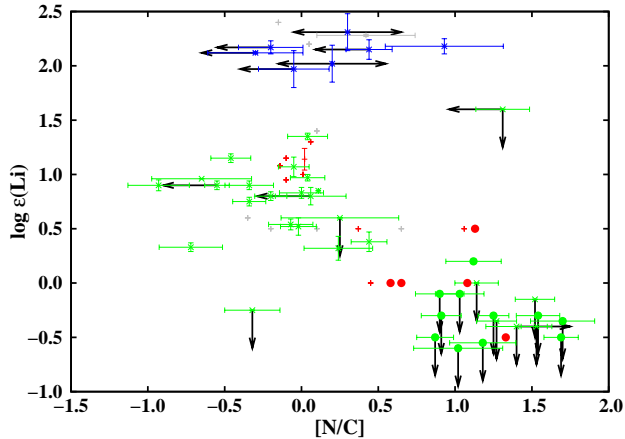}
  \end{center}
  \caption{Correlation between CN processing and Li depletion among the carbon-normal stars with $\feoh \le -2$.
  The definition of the symbols is the same as in Fig.~\ref{fig:nccfe}
  }\label{fig:ncli}
\end{figure}

In conclusion, there are two distinct groups of stars, one with the surface abundances deeply processed by CN cycle reactions and the other not or only slightly processed.
The both groups do exist regardless of the metallicity and consist of comparable number of stars in our sample of carbon-normal stars with the nitrogen abundances derived.  
Fig.~\ref{fig:ncli} shows the distribution on the diagram of $\nbyc$ against $\logli$ with separate symbols used for stars having $\ciso \leq 10$ for the carbon-normal stars of $\feoh \le -2.0$.
This reveals a tight correlation between the processing by CN cycles and the Li depletion for RGB stars \citep[see also][]{Gratton2000,Spite2006};
the ``unmixed'' stars have the normal Li depletion characteristic to the RGB stars without exception, while all the ``mixed'' stars show significant Li depletion.
This implies the operation of extra mixing during the evolution of red giant branch, similarly to the lithium depletion in the RGB discussed above:
it occurs for some of stars but not all, irrespective of metallicity.

% Figure log-e(Li) vs. [N/C]

If, on the other hand, the operation of the CN cycle is ascribed to the binary companion on the erstwhile AGB, it is expected that two populations should be found among dwarfs.
There are two stars showing $\abra{C}{N} < 0$ among the EMP MS group.
One is HE1337+0012 (or G64-12), well studied by many authors, has $\cfe = 0.49$ and $\nfe = 1.42$ \citep{Aoki2006b}, which clearly shows the signature of efficient CN cycles, while lithium is not depleted \citep[$\logli = 2.147$,][]{Charbonnel2005}.
Another is CS22963-004, having a smaller enhancement of C and N than those of HE1337+0012 \citep[$\cfe = 0.40$, $\nfe = 0.80$, obtained by][]{Lai2008}.
Both of these stars show no enhancement of neutron capture elements like Sr and Ba.
Judging from the carbon enhancement, these stars may be assigned not to ``mixed'' stars but to the carbon-normal stars with primary nitrogen enhancement, as observed for RGB stars.
Accordingly, there are no counterparts of ``mixed'' stars among the dwarfs and this may give a support to the hypothesis of extra mixing during the early phase of RGB evolution.
Although we should take into account the detection limit of carbon $\abra{C}{H} \simeq -3$ for dwarfs \citep[][, see also Fig.~\ref{fig:cbyh}]{Aoki2007b}, the carbon abundance of ``mixed'' stars are larger, or as large as this detection limit, at least for some of stars with $\feoh < -2$.

Here we also note the different behaviours between the stars of the metallicity below and above $\feoh \simeq -2$;
For the lower metallicity, the carbon-normal stars are all deeply processed by the CN cycles for stars with the carbon enhancement of $\cfe < -0.1$.
For the higher metallicity, on the other hand, the carbon-normal stars are processed by the CN cycles for stars with $\cfe < -0.4$, while it is not the case for $\cfe > -0.4$.
The difference in the threshold carbon enhancement can be regarded as the evidence of the transition of the carbon yield relative to iron on average.
This transition occurs in rather narrow range of the metallicity of $-2.2 < \feoh < -1.8$, and hence, may well be interpreted as the effects of changeover of the initial mass function, as discussed above, rather than as the result of metallicity effect on supernovae.

\subsection{Nitrogen enhancement vs. carbon enhancement}\label{sec:nemp}

% Figure [N/C] vs. [C+N/Fe]
\begin{figure}
  \begin{center}
    \includegraphics[width=84mm]{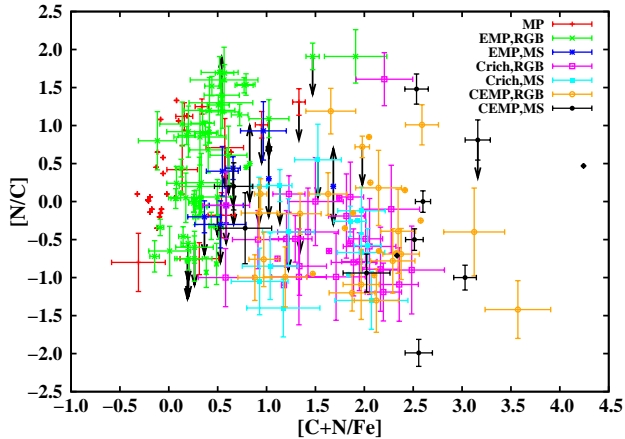}
  \end{center}
  \caption{Abundance relation between \nbyc\ and \abra{C+N}{Fe} for 172 stars taken from the SAGA database.
		   The definition of the symbols is the same as in Fig.~\ref{fig:cbyh} and shown in the top right corner.
		   Large symbols denote stars with low values of \abra{C}{N} in the EMP MS group or NEMP star candidates.
  }\label{fig:cncn}
\end{figure}

In this paper, we define NEMP stars by the following criteria; $\nbyc \ga 1$, $\abra{C+N}{Fe} \ga 1$.
These criteria are based on the requirement that the observed ratio of \nbyc\ is larger than expected from the equilibrium values of CN cycles under the temperature typical of the usual hydrogen shell burning and that the progenitor star has undergone the great carbon enhancement probably by the third dredge-up.
Note that the criterion $\abra{C+N}{Fe} \ga 1.0$ is set to discriminate between NEMP stars and ``mixed'' stars, discussed above.
Fig.~\ref{fig:cncn} shows the enhancement of nitrogen for all the sample stars with measured nitrogen and carbon abundances.
In this figure, we have three more N-rich stars (HE1337+0012, HE0400-2030, and CS22960-053) in addition to four stars (CS22949-037, CS29528-041, HE1031-0020, and CS30322-023) pointed out by \citet{Johnson2007} in their search for nitrogen-enhanced metal-poor (NEMP) stars.
Among these stars, they do not classify CS30322-023 and CS22949-037 as NEMP stars because CS30322-023 is believed to be an AGB star and because CS22949-037 has extremely high \ofe\ ($=1.97$) and no enrichment of \sit process elements.
However, we remove the condition of \sit process elements considering CEMP-no$s$ stars.

% Figure CEMP-nos/CEMP-s and NEMP/CEMP-s vs. M(HBB)
\begin{figure}
  \begin{center}
    \includegraphics[width=84mm]{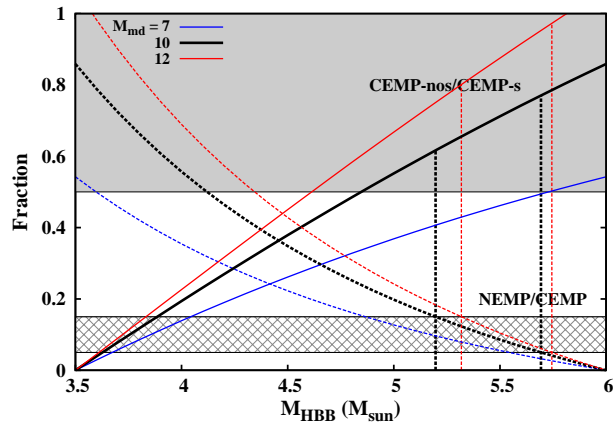}
  \end{center}
  \caption{The frequency, \cempnos $/$ \cemps\ (solid curves) and NEMP $/$ CEMP (dashed curves) as a function of the boundary of initial mass above which the NEMP stars are formed.
		   The model and assumptions are the same as in \citet{Komiya2007}, which assumes the log-normal initial mass function whose dispersion is set at $0.33$.
		   The median mass, $M_{\rm md}$ is set at 7, 10, or $12 \msun$ as shown in the top-left corner.
		   The frequency is determined by the assumption that the secondary of the binary system accretes matter from the primary whose carbon abundance is $\abra{C}{H} = 0$, affected by the He-FDDM event for $M \geq 3.5 \msun$.
		   Nitrogen enhancement is assumed to be converted from carbon after the third dredge-up so that ${\rm C}+{\rm N}$ keeps constant.
		   The shaded and hatched area denotes the required fraction for \cempnos $/$ \cemps\ and NEMP $/$ CEMP constrained by observations, respectively.
		   The solution of $M_{\rm HBB}$ is given by the overwrapping mass range obtained by the required mass range to satisfy both of the observed fractions.
		   Vertical lines, corresponding to the model results, provide the solutions for $M_{\rm md} = 10$ and $12 \msun$.
		   For $M_{\rm md} = 7 \msun$, no overwrapping mass range is found to be consistent with the observed fractions.
  }\label{fig:imf}
\end{figure}

Among newly classified NEMP stars, HE0400-2030 is first reported in \citet{Aoki2007b} that $\cfe = 1.14$ and $\nfe = 2.75$.
It also shows large \sit process element enhancement ($\bafe = 1.64$), although this star is more metal-rich ($\feoh = -1.73$) than other NEMP stars.
CS22960-053 is excluded from the sample of NEMP stars by \citet{Johnson2007} since they derived a nitrogen abundance smaller by two orders of magnitude then \citet{Aoki2007b}.
Here we adopted the abundances for \citet{Aoki2007b} with high resolution.
According to the result of \citet{Aoki2007b}, CS22960-053 ($\feoh = -3.14$) has $\cfe = 2.05$, $\nfe = 3.05$, and $\bafe = 0.86$, and therefore, is a candidate NEMP star.
CS29528-041 ($\feoh = -3.30$) is a known NEMP dwarf reported by \citet{Sivarani2006}, which shows $\cfe = 1.59$ and $\nfe = 3$ with a mildly enhanced barium abundance ($\bafe = 0.97$).
HE1337+0012 (G64-12), stated above, is a EMP dwarf ($\abra{C+N}{Fe} = 0.97$ while $\cfe=0.49$) without enhancement of \sit process elements ($\bafe = -0.25$).
Almost all NEMP stars show $\nfe > 2$, while much more stars are observed as non-NEMP stars because of their huge overabundance of carbon, as shown in the lower right part of the figure.

Accordingly, we have 5 and 2 confirmed NEMP stars with and without \sit process element enhancement, respectively, by setting the criterion as $\bafe = 0.5$ for the boundary of \sit process element enhancement.
We obtain 7 N-rich stars out of 71 carbon-enhanced stars by excluding the data with low-resolution spectra with $R < 10000$.
This still does not change the conclusion that the frequency of NEMP stars is much smaller than that of CEMP stars as pointed out by \citet{Johnson2007}.
At present, the ratio of NEMP stars to CEMP stars is the order of 0.1.
\citet{Pols2008} calculate the frequency of CEMP and NEMP stars using the binary population synthesis \citep{Izzard2004} based on stellar models including the effect of AGB evolution and thermohaline mixing, but excluding the models of He-FDDM.
They obtained a smaller value of the NEMP/CEMP ratio by enhanced efficiency of the effect of the third dredge-up with the initial mass function peaked at low-mass.
However, as seen in the result of \citet{Pols2008}, the frequency is sensitive to the condition of the occurrence of the third dredge-up and hot bottom burning, both of which are not well established in the stellar evolution models.

Based on the observed ratio of \cemps, \cempnos\ that is not taken into account in \citet{Pols2008}, and NEMP stars, we explore the condition to become a NEMP star through the hot bottom burning in AGB stars using the same assumptions as in \citet{Komiya2007}.
We adopt the initial mass function with a median mass of $10 \msun$ and with the dispersion of $0.33$ in log-normal function as suggested by the paper.
Fig.~\ref{fig:imf} shows the frequency of \cempnos\ and NEMP stars with respect to \cemps\ stars as a function of the boundary of initial mass to experience the hot bottom burning to become NEMP stars at $\feoh \la -2.5$.
In this figure, we draw the observed range of their frequencies, discussed above, $\hbox{\cempnos}/\hbox{\cemps} = 0.5-1.0$ and $\hbox{NEMP}/\hbox{\cemps} = 0.05-0.15$.
We also plot the results for the upper and lower mass boundaries for median mass that are still consistent with \citet{Komiya2007}.
The figure suggests that the hot bottom burning should occur in the mass range of $5 \la M / \msun \la 6$ to be compatible with the observed ratio that is deduced from the requirement that abundances in \cemps, \cempnos, and NEMP stars come from AGB companion stars in binaries, although the compatible mass range is rather sensitive to the peak mass of the IMF.

% Figure [Na/Fe] vs. [O/Fe]
\begin{figure}
  \begin{center}
    \includegraphics[width=84mm]{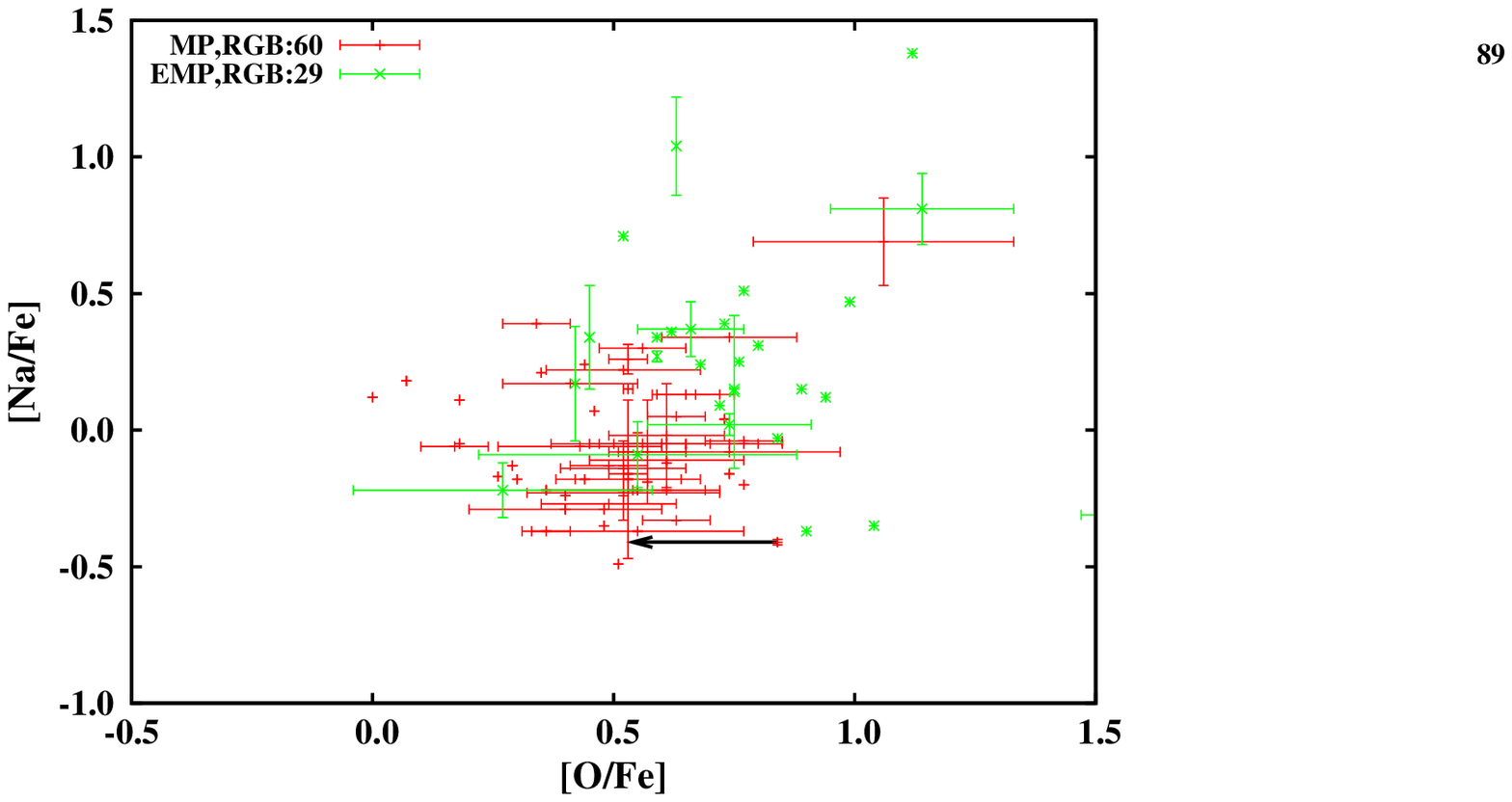}
  \end{center}
  \caption{Abundance relation between oxygen and sodium for 89 giants taken from the SAGA database.
		   The definition of the symbols is the same as in Fig.~\ref{fig:cbyh} except for MP stars for which we only include giants defined as ``RGB''.
		   The number of stars for each group is shown next to the label.
  }\label{fig:ona}
\end{figure}

% Figure [Al/Fe] vs. [Mg/Fe]
\begin{figure}
  \begin{center}
    \includegraphics[width=84mm]{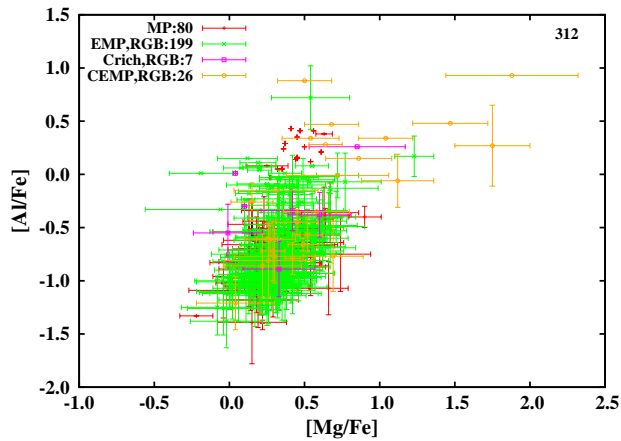}
  \end{center}
  \caption{Abundance relation between magnesium and aluminium for 312 stars taken from the SAGA database.
		   The definition of the symbols is the number of stars are labelled as in Fig.~\ref{fig:ona}.
  }\label{fig:mgal}
\end{figure}

\subsection{Difference between halo and globular cluster stars}
It is well known that most of the Galactic globular clusters exhibit star-to-star abundance variations within one cluster as well as between clusters.
There are the so-called abundance anomalies that are found amongst some stars in clusters irrespective of or dependent on the evolutionary status of the stars.
The source of abundance anomalies is considered as a result of the hydrogen burning for CNO elements and other light elements, although the nucleosynthetic sites are not yet pinpointed.
One of the most well-known examples is the O-Na anti-correlation that is now believed to show in all globular clusters.
Among others, there is the anti-correlation of Mg and Al abundances along the red giant branch and CN variations in both red giants and dwarfs \citep[see, e.g., as reviewed by][]{Kraft1994,Gratton2004}.

Such anti-correlations are not found in field halo giants, as shown in Figs.~\ref{fig:ona} and \ref{fig:mgal}.
This has been discussed and concluded for field stars with $\feoh \ga -2$ \citep[e.g.][]{Pilachowski1996,Gratton2000}.
This previous finding also holds for stars with $\feoh < -2$. 
Furthermore, no field stars have much less than $\ofe = 0$ in EMP stars either, as pointed out by \citet{Gratton2000} for stars with $\feoh > -2$, while it can be as low as $\ofe \simeq -1$, for example, in M13 first reported by \citet{Kraft1992} and in NGC2808 \citep{Carretta2004}.
The figures suggest that the abundances in halo stars show a positive correlation, though with fairly large scatters, rather than an anti-correlation for any pairs of concerned light elements.
In Fig.~\ref{fig:ona}, we exclude C-rich and CEMP groups from the sample.
If we include the data of carbon-enhanced stars into the figure, we will see strong positive correlations with large scatters.
For the relation between Mg and Al in Fig.~\ref{fig:mgal}, the correlation looks similar, but clearer compared with the relation between O and Na.
This trend is irrespective of the carbon enrichment, evolutionary status, and metallicity.
In this figure, we can see the separate population among the MP group that shows slightly larger values of $0 \la \abra{Al}{Fe} \la 0.5$ than the other population with $\abra{Al}{Fe} < 0$.

\section{\sit process in EMP AGB stars}\label{sec:spr}

% Figure [s/Fe] vs. [C/Fe]
\begin{figure*}
  \begin{center}
    \includegraphics[width=\textwidth]{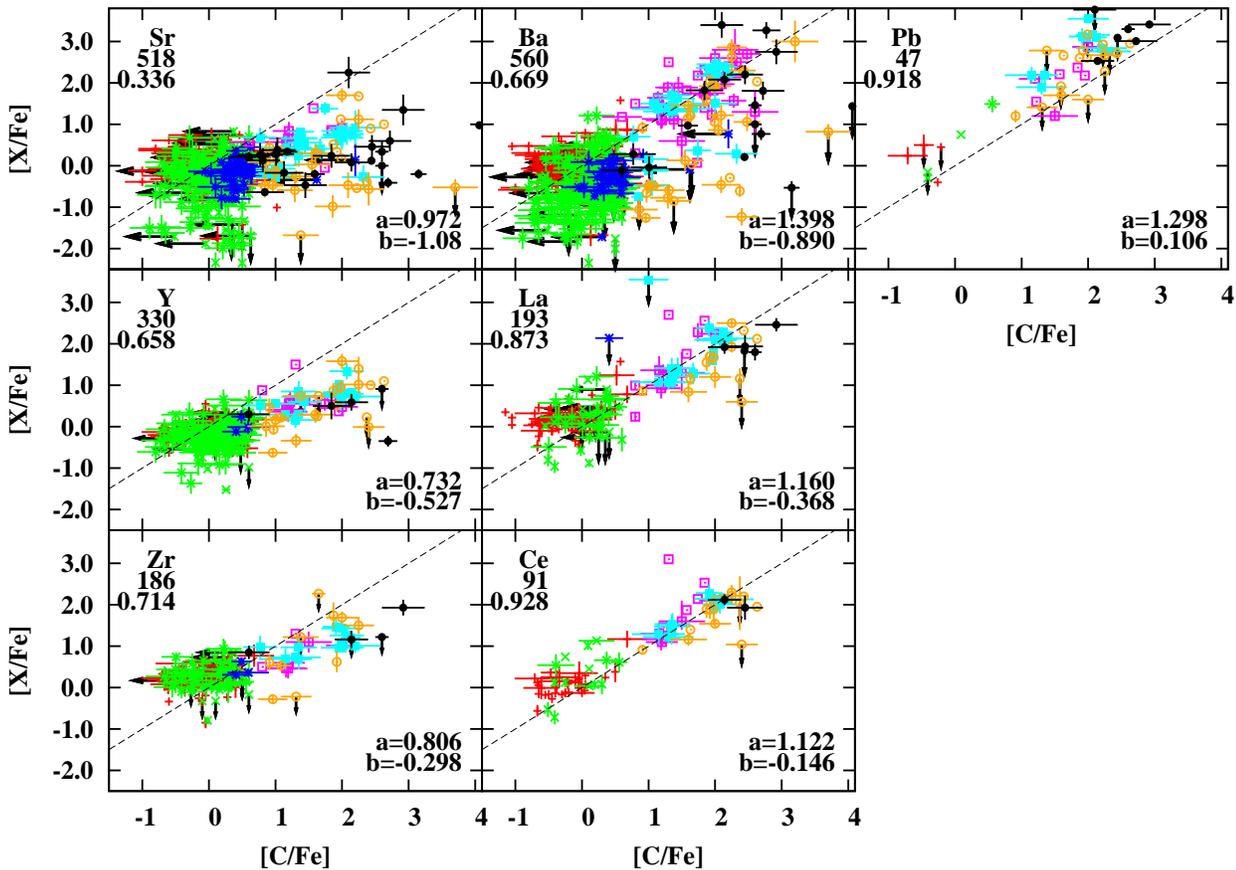}
  \end{center}
  \caption{Correlation between carbon and \sit process element abundances.
		   The line with the slope of unity is displayed by dashed line as a guide.
           Element species, the number of data, and the Pearson coefficient are shown in the top left corner of each panel.
		   The coefficients, $a$ and $b$ of least-squares fit, $\xfe = a \cfe + b$, are shown in the bottom right corner.
		   In calculating the Pearson coefficient and least-squares fit, we exclude the data with upper limits that is shown by arrows in the figure.
		   The definition of the symbols is the same as in Fig.~\ref{fig:cbyh}.
  }\label{fig:sprocess}
\end{figure*}

The production site of the \sit process nucleosynthesis in EMP stars is still in controversy, although AGB stars are thought to be responsible.
In particular, \citet{Suda2004} proposed the \sit process in the helium-flash convective zones triggered by the hydrogen ingestion.
The enhancement of \sit process elements are clearly found on the surface of CEMP dwarfs and red giants, which is thought of as evidence that they are affected by binary mass transfer.
On the other hand, binarity of \cemps\ stars is not fully established with the current observations of radial velocity monitoring.
However, the positive correlation of the carbon abundance and \sit process elements still support the idea that \sit process elements are synthesized in intermediate-mass AGB stars and transferred to binary companions which we observe today.

The abundances of \sit process elements are largely and globally correlated with the carbon abundance as shown in Fig.~\ref{fig:sprocess}.
The degree of correlation is significant in both \sit process elements inferred from the large values of the Pearson coefficient (most elements have $> 0.6$) even for elements containing more than several hundred data points.
The simple explanation for this trend is that most of the carbon enhanced EMP stars plotted in the figure belong to the \cemps\ group, although some stars belong to the \cempnos\ group.
In general, \cemps\ stars have large enhancements of both carbon and neutron capture elements in contrast to \cempnos\ group stars, and contribute to the positive correlation of abundances.
We can see very few neutron capture element data for \cempnos\ stars except for the barium abundance probably due to the difficulty in the line detection of other neutron capture elements.
Among other \sit process elements, strontium does not show a strong linear correlation with the carbon enhancement.
This is partly because the Sr abundance in EMP RGB group stars is largely scattered as shown in Fig.~\ref{fig:nfe}, although it is also the case for barium.
The weak correlation between C and Sr abundances may be ascribed to the existence of weak \sit\ or weak \rit processes.

There also exists the positive correlations of the abundances between carbon and \rit process elements like in Fig.~\ref{fig:sprocess}.
The Pearson coefficients are large for any elements, although the number of data is much smaller than 100, except for europium.
The correlations are simply understood by the fact that \rit process elements are also affected by neutron exposure in the \sit process, which can be seen in Fig.~\ref{fig:euba}.
The correlation between [{\it r}/Fe] and $\cfe$ may be affected by the correlation between [{\it s}/Fe] and $\cfe$ (``{\it r}'' and ``{\it s}'' means the \rit\ and \sit process element abundances) because it is natural to think of the nucleosynthesis in AGB stars that undergo \sit process and dredge-up events during the thermal pulses, which is discussed below.

Interestingly, we can see the increasing slopes fitted by the RMA ($\xfe = a \cfe + b$) with increasing the mass number of \sit process elements as shown by the values of $a$ and $b$ in Fig.~\ref{fig:sprocess}.
In particular, these slopes seem to correspond with the first, second, and third peaks of \sit process elements.
For \rit process elements, we cannot see any trend with the mass number of element species, although the number of samples is small.
For the elements corresponding to the neutron magic number of 50, the implication of the weak \sit [\rit] process may partly account for the decreased slope for non carbon-rich objects in Fig.~\ref{fig:sprocess}.
As far as this trend is concerned, however, weak neutron-capture process is responsible only for stars without enhancement of carbon abundance, i.e., the contribution of this process can be seen in the left side of the panel in Fig.~\ref{fig:sprocess}.
This argument still requires an explanation for the relatively small enhancement of Sr, Y, and Zr abundance for carbon-rich stars having $\cfe > 1$.
In addition, the differences of the slopes of elements between the neutron magic numbers 82 and 126 are still to be explained.

Given the strong correlation between [{\it s}/Fe] and \cfe\, and larger slopes in heavier elements, it is tempting to speculate that the production sites of carbon and \sit process elements are identical or closely related.
If these elements have the same production site, it is natural to insist that both elements are produced in helium flash convective zones in AGB stars by the He-FDDM events for stars with $\feoh \la -2.5$ \citep{Suda2004,Nishimura2009,Suda2010}.
In this scenario, both carbon and \sit process elements are processed in He-flash convection where protons in the H-burning shell are mixed and burned to produce \nucm{13}{C}, and hence neutrons via $\nucm{13}{C}(\alpha, n)$ \nucm{16}{O} reactions.
The larger the rate of mixing during the He-shell flashes, the larger the efficiency of the \sit process in the He-flash convective zones \citep{Nishimura2009}.
Therefore, the positive correlation between the carbon abundance and the \sit process abundances can be naturally explained.
However, it is difficult to explain the larger slope for larger mass number \sit process elements with this scenario because the efficient mixing does not necessarily result in deep dredge-up after the helium shell flash.
The final abundance ratio of [{\it s}/C] in the surface depends on the \sit process element production in the He-flash convective zones and the mass of the convective zone involved in the nucleosynthesis.
This scenario also has a problem in explaining C-rich groups in which the He-FDDM events are not expected to occur, while the correlations are common with CEMP groups.
In any case, the convective nucleosynthesis is worth investigating to determine whether the observed trend for the enhancement of carbon and \sit process elements can be explained.

Another plausible hypothesis to explain this trend is that the number of third dredge-up (TDU) events has a strong correlation with the number of the formations of a \cpocket.
For the \cpocket\ scenario, it is also difficult to explain the different slopes.
In the production site of \sit process elements by \cpocket, the former radiative region to form the \cpocket\ is diluted by the subsequent dredge-up, and hence the abundant \sit process elements no longer remain in the region.
In the actual model of low-mass AGB stars, \citet{Busso2001} find ``{\it the highly nonlinear trend of the \sit process efficiency}'' for the production of neutron-capture elements as a function of pulse number (or the number of \cpocket).
At least in their results, a larger number of \cpocket\ formation does not necessarily produce more heavier elements.

Finally, \sit process by \nucm{22}{Ne}($\alpha, n$)\nucm{25}{Mg} in helium flash convective zones seems unlikely to reproduce the trend.
In massive AGB stars having a large core mass, the temperature at the bottom of the helium-flash convective zone can be high enough to operate the above reaction to produce neutrons.
On the other hand, the mass of the helium convective zone decreases with increasing core mass, typically $\approx \pow{2.4}{-2}$, $\pow{3.7}{-3}$, and $\pow{9.8}{-4} \msun$ for the helium core mass of $0.6$, $0.8$, and $1.0 \msun$, respectively \citep[according the models of $\feoh = -3$ in][]{Suda2010}.
As a consequence, the carbon enhancement by the dredge-ups is expected to decrease with increasing core mass, and hence, increasing efficiency of neutron production under the constant dredge-up efficiency.
At the same time, the smaller mass budget of convective shell increases the neutron density and may enhance the \sit process efficiency to produce heavier elements.
This qualitative discussion leads to the opposite trend for the enhancement of carbon and \sit process elements, although the model calculations are required to confirm the trend.

% Figure [C,Ba/Fe] vs. Teff
\begin{figure}
  \begin{center}
    \includegraphics[width=84mm]{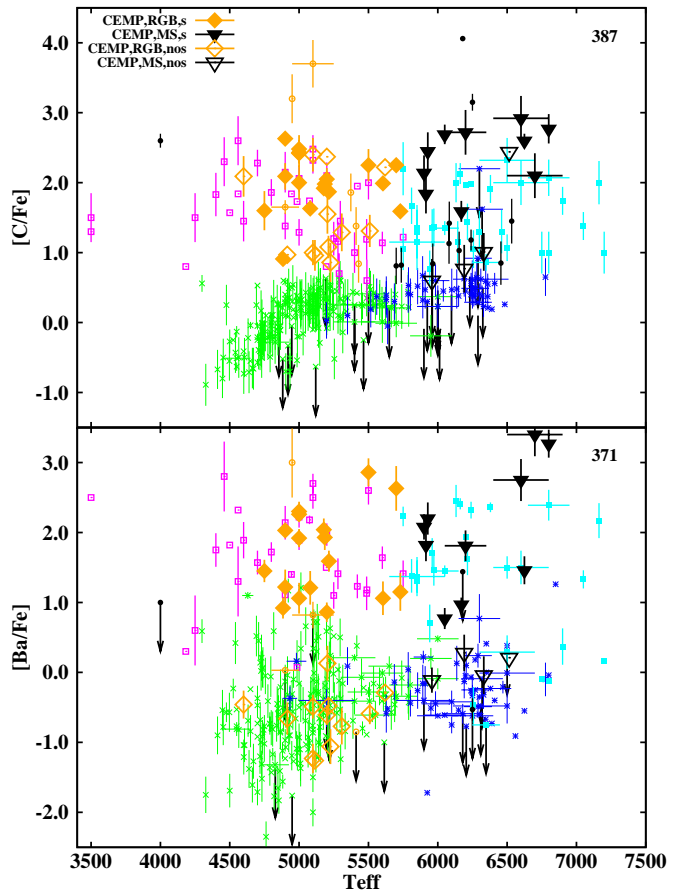}
  \end{center}
  \caption{Carbon (top) and barium (bottom) abundances as a function of effective temperature.
		   The ``MP'' group is removed from this figure for visibility.
		   For ``CEMP'' groups, we define sub-classes of ``\cemps'' and ``\cempnos'' according to the discussion in \S~\ref{sec:cemp} (See also Fig.~\ref{fig:euba}.).
		   These groups are shown by large symbols.
		   If the criterion for \sit process is not applicable, we use ``CEMP''.
		   For other classes including ``CEMP'', the same symbols are used as those in Fig.~\ref{fig:cbyh}.
		   In total, 387 and 371 stars are available for carbon and barium abundance, respectively.
  }\label{fig:dil}
\end{figure}

\section{evidence of binary mass transfer}\label{sec:bin}

The mass transfer in EMP binaries has a great impact on the origins of EMP stars in the viewpoint of the nucleosynthesis in AGB stars.
Our previous works suggest that the nucleosynthesis in helium flash convective zones during thermal pulses plays a role in understanding the chemically peculiar stars among observed EMP stars \citep{Suda2004,Komiya2007,Nishimura2009}.
Many previous works also suggest that some EMP dwarf stars show the influence of binary mass transfer in the viewpoint of the abundances such as carbon and \sit process elements \citep[see, e.g.][]{Lucatello2005}.
If binary mass transfer occurs through roche lobe overflow or wind accretion, it is expected that the effect of pollution in the surface of binary companions survives today.
In typical stars with $\feoh < -3$, the surface convective zones deepen in mass rapidly in the range of effective temperatures of $6500 \geq \teff \geq 5500$ \citep[see, e.g., ][]{Fujimoto1995}.

However, as seen in the previous sections, there are no significant differences in abundance anomalies in EMP stars between dwarfs and giants enough to account for the different mass of convective zone by two orders of magnitudes, although there exists a different sample volume and different limitations on the determinations of abundances.
Fig.~\ref{fig:dil} provides a test for the dilution effect by the surface convection.
We divide the sample into sub classes, \cemps\ and \cempnos\ for stars with different evolutionary status according to Tab.~\ref{tab:class}.
The label ``CEMP'' is used for stars from which the criterion for \sit process cannot be judged due to the lack of abundance data for barium.
We cannot see any transition around $\teff \sim 6000$ K as expected from the dilution in the envelope.
\citet{Masseron2010} also point out the same indifference between dwarfs and giants using stellar luminosity instead of \teff.
No significant difference is also seen for the CEMP and the C-rich groups, i.e., no difference above and below $\feoh = -2.5$.
The almost constant abundances with respect to \teff\ hold for any elements at any metallicity range.
As far as we checked on the database, the direct evidence of dilution can be found only for lithium (see Fig.~\ref{fig:liteff}).
One should note that the decreasing trend of carbon and barium abundances for the EMP RGB group at $\teff = 5000 \hyp 5500$ K is in part due to the observational lower limit for determination of element abundances.
\citet{Aoki2007b} estimate the detection limit of the absolute carbon abundance as a function of effective temperature for the data using the current equipment of high resolution spectroscopy.

The implications of extra mixing in CEMP stars are discussed in the viewpoint of extra mixing.
The apparent lack of the difference in surface abundances between giants and dwarfs may support the existence of diffusion into the stellar interior of the main sequence stars, for example by the thermohaline mixing \citep{Stancliffe2007}.
On the other hand, the very efficient diffusion of elements in the surface of CEMP MS stars is unlikely from the distribution of [C/H] for these stars as discussed by \citet{Aoki2007b} and \citet{Denissenkov2008}.
If the extra mixing is not responsible for stars in the CEMP MS group, it is likely that the accreting matter from the binary companion was larger than or comparable to the depth of the convective envelope of red giants, i.e., $M_{\rm acc} \ga 0.3 \msun$ as insisted by \citet{Aoki2007b} and \citet{Aoki2008}.

% Figures [alpha/H] vs. Teff
\begin{figure*}
\centering
    \includegraphics[width=\textwidth]{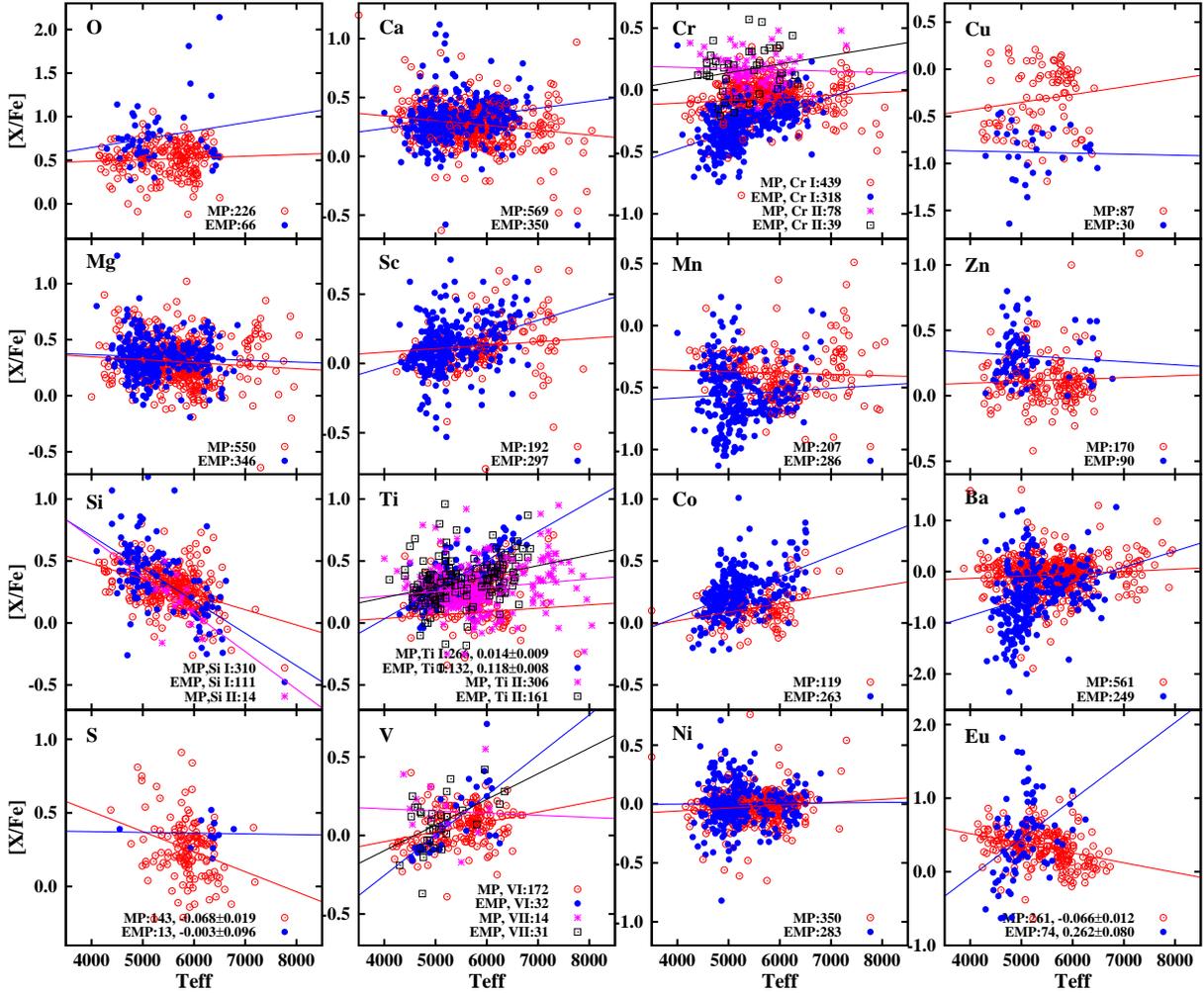}
  \caption{Abundance trends as a function of effective temperature for \aelms\, iron-group elements, barium, and europium as labelled on top right corner in each panel.
		   For O, Mg, Ba, and Eu, the carbon rich groups are excluded.
		   For other elements, whole data are divided into two groups, namely the MP group and the EMP groups by the critical metallicity of $\feoh = -2.5$.
		   For Ti, V, and Cr, the data are further divided into two groups by neutral or ionized species.
		   Least squares fits are shown by lines for each group with the same colour.
		   The number of stars in each group is labelled next to the group name.
		   For some of the groups, slopes and its standard error are provided.
		   HE1424-0241 ($\abra{Si}{Fe} = -1.01$) and CS22949-037 ($\abra{S}{Fe} = 1.78$) are removed from the sample in order to reduce the scale of \abra{Si}{Fe} and \abra{S}{Fe}, respectively.
		   For Ti abundance, some stars with low-resolution spectra are removed, all of which are [Ti/Fe] $< 0.5$ and 5000 K $< \teff <$ 6000 K.
  }\label{fig:talpha}
\end{figure*}

\section{Unexplained correlations}\label{sec:unexp}

Apart from our concentration on the role of low- and intermediate-mass stars, we report the findings of peculiar and unexpected correlations in the sample of the SAGA database. 
Some of them were already reported, but the reasons for the correlations and trends are still to be explained.

In Fig.~\ref{fig:talpha}, we plot the relation between abundances with respect to iron and effective temperature.
For Si, Ti, and Cr, unexpected trends of abundances with \teff\ are reported by the previous works \citep{Preston2006,Lai2008}, while \citet{Bonifacio2009} reported 0.2 - 0.4 dex discrepancy between giants and dwarfs for most of \aelms\ and iron-group elements.
\citet{Preston2006} report the decreasing trend of Si abundance with increasing \teff\ in their study of field horizontal branch stars.
\citet{Lai2008} confirmed this trend and found a similar dependence on \teff\ for Ti and Cr.
\citet{Bonifacio2009} also report the discrepancy of Si abundance between giants and turn-off stars.
In our sample, the decreasing trend with effective temperature is still visible.
Note that the decreasing trend is significant for metal-poor stars, typically $\feoh < -2$, which can be clearly seen in Fig.~\ref{fig:afe}.
For the abundance derived by Si II, the slope is significant, but the number of sample is too small to be confirmed.
The trends for Ti and Cr are not so pronounced in our sample compared with the results presented in \citet{Lai2008}.
For Ti and Cr abundances, all our data having $\teff > 6800$ K come from \citet{Preston2000}.
The data of these high effective temperatures are independent of $\teff$ and do not contribute to the trend.
The reason for these trends is investigated by the above papers in view of spectral analyses, but they do not deduce a firm conclusion.
Because of the difficulty in explaining the trend with stellar evolution processes, they still suspect that these may be due to the problems of spectral analyses.
The large slope for europium in Fig.~\ref{fig:talpha} is not important because the number of sample for dwarfs is too small to measure the trend and because the dispersion of europium abundance is large enough to make the discrepancy invisible.

\citet{Lai2008} discuss the trend by considering the difference of ionization status.
They find the slope only for Cr~I, while a flat trend is found for Cr~II.
It holds for our sample, although the slope is smaller.
They report a slope of 0.09 dex per 500 K for Cr abundance in total, while we find the slopes of 0.0082 for Cr~II in 121 stars and 0.049 dex per 500 K in total.
As shown in Fig.~\ref{fig:talpha}, the abundances derived by Cr~II are systematically larger than that for Cr~I.
The difference amounts to $\sim 0.3$ dex at 6000 K and $\sim 0.4$ dex at 5000 K.
\citet{Sobeck2007} also discuss the different abundances obtained by Cr~I and Cr~II lines.
They insist that the discrepancy may be due to NLTE effect.
In addition to the discrepancy of chromium abundance, the slope of the abundance with respect to \teff\ has a metallicity dependence as also pointed out by \citet{Sobeck2007}.
The slope for Cr~I in the EMP group is 0.071 dex per 500 K and larger than that in the whole metallicity range.

\citet{Lai2008} also find the slope for Ti irrespective of its ionization state.
In our sample, however, we find the difference between \abra{Ti I}{Fe} and \abra{Ti II}{Fe} and between the MP and EMP groups.
For our sample of \abra{Ti I}{Fe} and \abra{Ti II}{Fe} data, we obtain a slope of 0.047 for Ti~I and 0.019 dex per 500 K for Ti~II.
The trend has a weak metallicity dependence as shown by the slightly steeper lines for the EMP group in Fig.~\ref{fig:talpha}.
The slope is 0.11 for Ti~I and 0.043 for Ti~II in our EMP sample.

\citet{Bonifacio2009} reported the abundance discrepancy between giants and dwarfs for C, Mg, Si, Sc, Ti, Cr, Mn, Co, and Zn.
The abundances of these elements are shown in Fig.~\ref{fig:talpha} except for carbon.
The carbon abundance is affected by its detection limit as discussed in the previous section and is not appropriate to discuss here.
For Mg, Mn and Zn, the discrepancies are negligible or not significant in our sample.
For Sc abundance, the discrepancy is small but visible, which is consistent with \citet{Bonifacio2009}.
The discrepancy for Co abundance is clearly visible for the EMP and CEMP groups in Fig.~\ref{fig:iron} and \ref{fig:talpha}.
The Co abundance shows increasing trend with increasing \teff.
The discrepancy can be as large as $\simeq 0.3$ dex between $\teff = 4500$ and 6500 K.

We find unexpected abundance trends for sulphur and vanadium.
For the sulphur abundance, the decreasing trend with increasing \teff\ can be seen, although most of the data concerned are for more metal-rich stars than for the abundance data of other elements.
Sulphur abundances of the sample stars are mainly taken from \citet{Caffau2005,Nissen2004,Nissen2007,TakadaHidai2002}.
The trend comes from the data of Caffau's and Takada-Hidai's for giants.
This trend is apparently found by a small number of sample giants and should be confirmed with a larger sample.

The slope for the vanadium abundance is clear in the figure especially for the EMP and CEMP groups irrespective of the ionization state of the element, although the number of sample is small.
For the MP and C-rich groups, the V~I abundance shows small increasing trend with increasing \teff, which is less than $0.2$ dex between $\teff = 4500$ and 6500 K.

It is to be noted that the dependence of element abundances on the effective temperature does not have a significant effect on the discussions in this study, although minor effect can be seen in the abundance trend.
For example, the increasing trend of Si abundance with decreasing \feoh\ may disappear by removing this dependence, because the ratio of giants and dwarfs is quite different below and above the metallicity of $-2.5$.
In our sample with detected Si abundances, 72 percent of MP group stars ($\feoh > -2.5$) are dwarfs, while only 28 percent of EMP groups are dwarf.
Other possible effects of the discrepancy are discussed separately in \S~\ref{sec:div} and \S~\ref{sec:enrich}.

\section{Summary and Conclusions}\label{sec:sum}

We explore the general characteristics of extremely metal-poor (EMP) stars in the Galactic halo using the SAGA database.
We divide the objects in the database into subclasses depending on carbon, nitrogen, iron, barium, and europium abundance to discriminate the origins of EMP stars, which is summarized in Tab.~\ref{tab:class}.
The frequency of carbon-enhanced stars is 20~\% or larger at $\feoh \la -2.0$ almost irrespective of the definition of carbon-rich by changing from $\cfe \geq 0.5$ to $\cfe \geq 1.0$.
The frequency is nearly constant at the metallicity range of $-4 \la \feoh \la -2$, while it dramatically increases for $\feoh < -4$ and decreases for $\feoh > -2$.
The boundary of carbon-enhanced stars is set at $\cfe \geq 0.7$ in this work, which is reasonable in discriminating the source of carbon in observed stars.
We point out the possible transition of the initial mass function (IMF) at the metallicity of $\feoh = -2$ based on the change of the peak in carbon abundance and the dramatic change of the frequency of carbon-enhanced stars.
The transition of the IMF is also based on the argument that the one in the early Galactic evolution has top heavy function peaked at $\sim 10 \msun$ \citep{Komiya2007}, and should have changed to low-mass peaked function during the evolution and the structure formation.

As for neutron-capture elements, we divide the stars with known barium and europium abundances, if available, into two subclasses by the origin of their nucleosynthetic site.
We define the ``\rit-dominant'' and ``\sit-dominant'' groups depending on the critical value of $\euba = -0.2$ or $\bafe = 0.5$, which is representative of the dominant source for neutron capture elements either by the \rit\ or \sit process.
With respect to the metallicity of stars, we set the boundary of EMP stars at $\feoh = -2.5$ as in the previous paper of \citet{Suda2008}, which is based on a theoretical model of stellar evolution.
According to our definition of carbon-rich EMP stars (CEMP stars), we confirmed different evolutionary origins of carbon-rich stars above and below $\feoh \sim -2.5$.
For more metal-rich stars, the carbon-rich group is identical to classical CH stars, while the origin of CEMP stars should be related to the so-called helium-flash driven deep mixing in EMP stars and binary mass transfer from the former AGB stars.
As for \cempnos\ stars, defined as CEMP stars with $\bafe < 0.5$ as shown in Tab.~\ref{tab:class}, it is suggested that efficiency of \cpocket\ has dependence on initial mass and/or metallicity.

We investigated the possibility of varying abundance scatter as a function of metallicity.
We find that the abundance scatters are almost constant and within the observational errors for \aelms\ and iron-group elements in the metallicity range of $-3.5 \la \feoh \la -2$, while most of the \sit\ and \rit process elements show an increasing scatter with decreasing metallicity.
In particular, the linear regression analysis gives the slopes of nearly unity for the relation between $\alpha$-element abundances and iron abundance.
These observational feature provides the strong constraints on the supernova yields, i.e., the chemical yields of $\alpha$-elements and titanium have weak dependence on mass and metallicity of progenitors.
The relations between the abundances of \aelms\ reveal that most of the sample in the database are distributed in the tight linear correlations except for a few peculiar stars.
Among these peculiar stars, all the $\alpha$-enhanced and depleted stars show relative enhancement of magnesium whose origins are still open question.

The global chemical enrichment in the Galactic halo is discussed using the reduced major axis regression for element abundances relative to metallicity.
For iron group elements, we find non-unity slopes, which suggests that the supernova yields are dependent on metallicity or dependent on mass associated with the variable IMF.
We derived the chemical yield ratios of supernovae as a function of metallicity constrained by observations by assuming the non-variable IMF, and compared it with supernovae models available from the literature.
The chemical enrichment with small scatter for both \aelms\ and iron group elements implies a well-mixed interstellar medium after the pollution by supernovae at any stage of Galactic chemical evolution.
On the other hand, for the observed scatters of the abundances of \rit process elements, it is suggested that the dominant site for \rit process is restricted to the specific mass range of progenitors.

We explore the possibility of extra mixing in EMP stars.
We discuss the lithium depletion in the surface convection during the first ascent on the red giant branch where the mass of the convective envelope becomes maximum at $\teff \approx 5500$ K.
The dilution in the convective envelope is not enough to deplete lithium to explain giants showing lithium depletion by more than 1 dex below the prediction by theoretical models.
The observed scatters and frequency of lithium-depleted stars cannot be explained solely by the effect of binary mass transfer, which implies extra mixing operating at the bottom of the red giant branch phase.
The extra mixing is also suggested for the ``mixed stars''.
Although the definition of these stars can vary with metallicity, there have to be two distinct groups for giants showing normal or depleted carbon abundance.
We confirmed that the anti-correlations of O-Na and Mg-Al found amongst globular cluster stars are not observed for stars with $\feoh < -2$.

We report three nitrogen-enhanced metal-poor stars from our sample.
The number of nitrogen-enhanced stars is seven if we set the definition as $\abra{C}{N} \la -1$, $\abra{C+N}{H} \ga -2$, and $\cfe \ga 0.5$, which is reasonable to insist that the CN cycles efficiently worked in the former AGB companion.
Two of them show no enhancement of \sit process elements, while others show a moderate to large enhancement.
We report that the frequency of nitrogen-rich stars to carbon-rich stars is $\sim 1/10$.
We derive the mass range for the occurrence of hot bottom burning as $5 \la M / \msun \la 6$ with this observational constraint and with the following assumptions;
the observed CEMP and NEMP stars were affected by binary mass transfer whose companions experienced AGB evolution;
the abundance pattern of CEMP and NEMP stars comes from the third dredge-up, helium-flash driven deep mixing \citep{Suda2004}, hot bottom burning, or a combination of them;
the IMF of EMP halo stars has a peak at $10 \hyp 12 \msun$ as suggested by the statistics of CEMP stars by \citet{Komiya2007}.

We find the increasing power law of the correlations between the enhancement of \sit process elements and carbon abundance with increasing the mass number of \sit process elements.
This trend is difficult to explain by the currently known possible site for the \sit process such as \nucm{13}{C} pocket, convective nucleosynthesis driven by hydrogen ingestion into the helium-flash convective zones for EMP stars, and \nucm{22}{Ne} ($\alpha, n$) \nucm{25}{Mg} in the helium-flash convective zones.
None of the candidates can explain the correlation and we need more detailed models for \sit process nucleosynthesis.

As far as we explored the database, there are no significant abundance gaps between dwarfs and giants enough to account for the dilution effect by the convective envelope at the RGB for any choice of elements except for lithium, even though most of these stars are expected to be affected by binary mass transfer.
This may suggest that the effect of dilution by the convective envelope is erased by the effect of accretion from a binary companion.
Therefore, the accreted mass is typically larger than $\sim 0.3 \msun$ that corresponds to the mass of the convective envelope of EMP giants.

Finally, we report the unexpected relations between derived element abundances and the effective temperature for observed metal-poor stars.
The temperature dependence of the abundances are found for Si, Sc, Ti, Cr and Co, as already reported, with a much larger sample size than previously investigated.
On the other hand, we do not find the temperature dependence for Mn and Zn, while it is insisted by \citet{Bonifacio2009}.
We find that the degree of the dependence is smaller than previously reported.
It is confirmed that the trends are found only in EMP stars for titanium and chromium.
We also confirm that there is a discrepancy between abundances derived for Cr~I and Cr~II as previously reported.
We find a similar temperature dependence of the sulphur and vanadium abundance, i.e., the decreasing and increasing trend of abundance with increasing temperature, respectively, which is also theoretically unexpected.

\section*{Acknowledgments}

We would like to acknowledge Michael E. Bennett for reading the manuscript and for giving useful comments.
This work has been partially supported by Grant-in-Aid for Scientific Research (18104003, 19740098), from Japan Society of the Promotion of Science.
T. S. has been supported by a Marie Curie Incoming International Fellowship of the European Community FP7 Program under contract number PIIF-GA-2008-221145.

\appendix

\section{Classification of CEMP and NEMP stars}

\begin{table*}
  \caption{Classification of CEMP stars in the SAGA database, which are based on Table~\ref{tab:class}.
  Stars without measurement for Ba or Eu abundance are classifed as other CEMP stars.
  Numbers in the parentheses denote the number of stars in the category.}\label{tab:cemp}
   \centering
    \begin{minipage}{140mm}
%   \begin{center}
    \begin{tabular}{l*{11}{c}}
      \hline
      Object & \teff & $\log g$ & \feoh & \cfe &  \nfe & \ofe & \bafe & \eufe & {\it s} $/$ {\it r}$^{a}$ & binary$^{b}$ \\
      \hline
% Object        Teff     log g   [Fe/H]   [C/Fe]   [N/Fe]   [O/Fe]   [Ba/Fe]  <(Ba)   [Eu/Fe]  <(Eu)
\cemps\ (14) & & & &     &      &   &      &  & & \\

CS22183-015 & 5178 & 2.69 & $-2.75$ & 1.92 & 1.77 & - & 2.04 & 1.7 & \is & - \\
CS22942-019 & 5000 & 2.4 & $-2.64$ & 2 &    0.8 &  - & 1.92 & 0.79 & \is & 2800 \\
CS22948-027 & 5000 & 1.9 & $-2.47$ & 2.43 & 1.75 & - & 2.26 & 1.88 & \is & 426.5 \\
CS29497-034 & 4900 & 1.5 & $-2.9$ &  2.63 & 2.38 & - & 2.03 & 1.8 & \is & 4130 \\
CS30301-015 & 4750 & 0.8 & $-2.64$ & 1.6 &  1.7 &  - & 1.45 & 0.2 & \is & - \\
CS31062-012 & 5901 & 4.5 & $-2.53$ & 2.14 & 1.2 &  1.09 & 2.08 & 1.62 & \is & - \\
HE0024-2523 & 6625 & 4.3 & $-2.72$ & 2.6 &  2.1 &  0.4 &  1.46 & 1.1 & \is & 3.14 \\
HE0131-3953 & 5928 & 3.83 & $-2.68$ & 2.45 & - & - & 2.2 &  1.62 & \is & - \\
HE0336+0113 & 5700 & 3.5 & $-2.68$ & 2.25 & 1.6 &  - & 2.63 & 1.18 & \is & - \\
HE1031-0020 & 5080 & 2.2 & $-2.86$ & 1.63 & 2.48 & - & 1.21 & $<$0.87 & \is & - \\
HE1410-0004 & 5605 & 3.5 & $-3.02$ & 1.99 & - & 1.18 & 1.06 & $<$2.40 & \is & - \\ % added by hand
HE1509-0806 & 5185 & 2.5 & $-2.91$ & 1.98 & 2.23 & - & 1.93 & $<$0.93 & \is & - \\
HE2158-0348 & 5215 & 2.5 & $-2.7$ &  1.87 & 1.52 & - & 1.59 & 0.8 & \is & - \\
LP625-44 &    5500 & 2.5 & $-2.7$ &  2.25 & 0.95 & 1.85 & 2.86 & 1.76 & \is & 4382.91 \\

\cempnos (13) & & & &     &      &   &      & &  & \\

BD+44 493 &   5510 & 3.7 &  $-3.68$ & 1.31 & 0.32 & 1.59 & $-0.59$ & $<$0.13 & \ir & - \\
BS16077-007 & 5958 & 3.68 & $-2.72$ & 0.6 &  $<$0.8 &  $<$1.39 & $-0.11$ & $<$0.81 & \ir & - \\
BS16929-005 & 5212 & 2.78 & $-3.17$ & 1.08 & 0.32 & - & $-0.48$ & $<$0.03 & \ir & - \\
CS22877-001 & 5100 & 2.2 &  $-2.72$ & 1 &    0 &    - & $-0.49$ & $<$0.6 & \ir & - \\
CS22892-052 & 4884 & 1.81&  $-2.92$ & 0.91&  1.01 & 0.72 & 0.92 & 1.51 & \ir & - \\  % added by hand
CS22949-037 & 4915 & 1.7 &  $-3.79$ & 0.97 & 2.16 & 1.96 & $-0.66$ & 0.04 & \ir & - \\
CS22957-027 & 5100 & 1.9 &  $-3.12$ & 2.4 &  1.62 & - & $-1.23$ & $<$0.97 & \ir & 3125 \\
CS29502-092 & 5114 & 2.51 & $-3.18$ & 0.96 & 0.81 & 0.75 & $-1.26$ & $<-0.31$ & \ir & - \\
HE0007-1832 & 6515 & 3.8 &  $-2.7$ &  2.44 & 1.73 & - &  0.21 & $<$1.84 & \ir & - \\
%HE0132-2429 & 5294 & 2.75 & $-3.55$ & 0.62 & 1.07 & $<$1.67 & $-0.85$ & $<$1.18 & \ir & - \\  % added by hand
HE0557-4840 & 4900 & 2.2 &  $-4.75$ & 1.65 & $<$2.37 & $<$3.09 & $<$0.03 & $<$2.04 & \ir & - \\  % added by hand
HE1012-1540 & 5620 & 3.4 &  $-3.43$ & 2.22 & 1.25 & 2.25 & $-0.29$ & $<$1.62 & \ir & yes \\
HE1150-0428 & 5200 & 2.55 & $-3.3$ &  2.37 & 2.52 & - & $-0.61$ & $<$1.45 & \ir & - \\
HE1300+0157 & 5411 & 3.38 & $-3.73$ & 1.38 & $<$1.22 & 1.76 & $<-0.85$ & $<$1.56 & \ir & - \\  % added by hand

Other CEMP (34) & & & &     &      &   &      & &  & \\

CS22887-048 & 6455 & 3.99 & $-2.79$ & 0.85 & 0.5 &  - & - & - &  -  & - \\
CS22958-042 & 6250 & 3.5 & $-2.85$ & 3.15 & 2.15 & 1.35 & $<-0.53$ & - & \ir & - \\
CS22960-010 & 5737 & 4.85 & $-2.65$ & 0.82 & - & - & - & - & - & - \\
CS22960-053 & 5200 & 2.1 & $-3.14$ & 2.05 & 3.06 & - & 0.86 & - & \is & - \\
CS29498-043 & 4400 & 0.6 & $-3.75$ & 1.9 &  2.3 &  2.9 &  $-0.45$ & - & \ir & - \\
CS29527-015 & 6240 & 4 & $-3.55$ & 1.18 & - & - & - & - & - & - \\
CS29528-028 & 6800 & 4 & $-2.86$ & 2.77 & $<$3.58 & - & 3.27 & - & \is & - \\
CS29528-041 & 6170 & 4 & $-3.06$ & 1.59 & 3.07 & $<$1.4 &  0.97 & - & \is & - \\
CS31080-095 & 6050 & 4.5 & $-2.85$ & 2.69 & 0.7 &  2.35 & 0.77 & - & \is & - \\
HE0012-1441 & 5730 & 3.5 & $-2.52$ & 1.59 & 0.64 & - & 1.15 & - & \is & yes \\
HE0107-5240 & 5100 & 2.2 & $-5.39$ & 3.7 &  2.28 & 2.3 &  $<$0.82 & $<$2.78 & - & - \\
HE0450-4705 & 5429 & 3.34 & $-3.1$ &  0.84 & - & - & - & - & - & - \\
HE1005-1439 & 5000 & 1.9 & $-3.17$ & 2.48 & 1.79 & - & 1.06 & - & \is & - \\
HE1045-1434 & 4950 & 1.8 & $-2.5$ &  3.2 &  2.8 &  1.8 &  3 & - & \is & - \\
HE1124-2335 & 5226 & 2.68 & $-2.93$ & 0.86 & - & - & $-1.06$ & - & \ir & - \\
HE1135-0344 & 6154 & 4.03 & $-2.63$ & 1.03 & - & - & - & - & - & - \\
HE1148-0037 & 5964 & 4.16 & $-3.46$ & 0.84 & - & - & - & - & - & - \\
HE1217-0540 & 5700 & 4.2 & $-2.94$ & 0.81 & - & - & - & - & - & - \\
HE1221-1948 & 6083 & 3.81 & $-3.36$ & 1.42 & - & - & - & - & - & - \\
HE1245-1616 & 6191 & 4.04 & $-2.97$ & 0.77 & - & - & 0.28 & - & \ir & - \\
HE1249-3121 & 5373 & 3.4 & $-3.23$ & 1.86 & - & - & - & - & - & - \\
HE1300-0641 & 5308 & 2.96 & $-3.14$ & 1.29 & - & - & $-0.77$ & - & \ir & - \\
HE1300-2201 & 6332 & 4.64 & $-2.6$ &  1.01 & - & - & $-0.04$ & - & \ir & - \\
HE1305-0331 & 6081 & 4.22 & $-3.26$ & 1.13 & - & - & - & - & - & - \\
HE1327-2326 & 6180 & 3.7 & $-5.7$ &  4 &    4.56 & 3.7 &  $<$1.46 & $<$4.44 & - & - \\
%HE1337+0012 & 6270 & 4.4  & $-3.37$ & 0.49 & 1.42 & 0.73 & $-0.25$ & - & \ir & - \\  % added by hand
HE1351-1049 & 5204 & 2.85 & $-3.45$ & 1.55 & - & - & 0.13 & - & \ir & - \\
HE1413-1954 & 6533 & 4.59 & $-3.19$ & 1.45 & - & - & - & - & - & - \\
HE1430-1123 & 5915 & 3.75 & $-2.7$ &  1.84 & - & - & 1.82 & - & \is & - \\
HE1528-0409 & 5000 & 1.8 & $-2.61$ & 2.42 & 2.03 & - & 2.3 &  - & \is & - \\
HE2330-0555 & 4900 & 1.7 & $-2.78$ & 2.09 & 1 &    - & 1.22 & - & \is & - \\
G77-61      & 4000 & 5.1 & $-4.03$ & 2.6 &  2.6 &  1.8 &  $<$1 &   $<$3 & - & 245 \\
SDSS0126+06 & 6600 & 4.1 & $-3.11$ & 2.92 & - & - & 2.75 & - & \is & - \\
SDSS0924+40 & 6200 & 4 & $-2.51$ & 2.72 & - & - & 1.81 & - & \is & - \\
SDSS1707+58 & 6700 & 4.2 & $-2.52$ & 2.1 &  - & - & 3.4 &  - & \is & - \\
      \hline
   \end{tabular}
%  \end{center}
   \medskip
$^{a}$ dominant source for neutron-capture elements following Table~\ref{tab:class}. If Eu abundance is not available, \sit\ and \rit dominance is judged by $\bafe \geq$ and $< 0.5$, respectively. \\
$^{b}$ binarity or binary period in days if available
   \end{minipage}
\end{table*}

\begin{table*}
  \caption{The same as Table~\ref{tab:cemp}, but for the C-rich groups.
  }\label{tab:crich}
   \centering
    \begin{minipage}{140mm}
    \begin{tabular}{l*{10}{c}}
      \hline
      Object & \teff & $\log g$ & \feoh & \cfe &  \nfe & \ofe & \bafe & \eufe & \is $/$ \ir & binary \\
      \hline
%Object     Teff     log g     [Fe/H]     [C/Fe]     [N/Fe]     [O/Fe]     [Ba/Fe]     [Eu/Fe]
C-rich-{\it s} (18) & & & &     &      &   &      &  & & \\

BD-01 2582 & 5196 & 3 & $-2.21$ & 0.8 &  -  & 0.54 & 1.5 & 0.67 & \is & - \\
CS22880-074 & 5850 & 3.8 & $-1.93$ & 1.3 & $-0.1$ &  -  & 1.31 & 0.5 & \is & - \\
CS22881-036 & 6200 & 4 & $-2.05$ & 1.96 & 1 &  -  & 1.93 & 1 & \is & - \\
CS22898-027 & 5750 & 3.58 & $-2.26$ & 2.2 & 0.9 &  -  & 2.23 & 1.88 & \is & - \\
CS22947-187 & 5489 & 3.44 & $-2.25$ & 0.6 & 0.55 &  -  & 1.18 & 0.73 & \is & - \\
CS22964-161A & 6050 & 3.7 & $-2.39$ & 1.35 &  -  &  -  & 1.45 & 0.69 & \is & 252.481 \\
CS22964-161B & 5850 & 4.1 & $-2.41$ & 1.15 &  -  &  -  & 1.37 & 0.69 & \is & 252.481 \\
CS29497-030 & 7163 & 4.2 & $-2.2$ & 2 & 1.88 & 1.67 & 2.17 & 1.44 & \is & 346 \\
CS29526-110 & 6500 & 3.2 & $-2.38$ & 2.2 & 1.4 &  -  & 2.11 & 1.73 & \is & yes \\
CS31062-050 & 5500 & 2.7 & $-2.3$ & 2 & 1.2 &  -  & 2.6 & 1.91 & \is & yes \\
HD196944 & 5250 & 1.8 & $-2.4$ & 1.2 & 1.3 & 0.85 & 1.1 & 0.17 & \is & yes \\
HE0143-0441 & 6240 & 3.7 & $-2.31$ & 1.98 & 1.73 &  -  & 2.32 & 1.46 & \is & - \\
HE0202-2204 & 5280 & 1.65 & $-1.98$ & 1.16 &  -  &  -  & 1.41 & 0.49 & \is & - \\
HE0338-3945 & 6162 & 4.09 & $-2.42$ & 2.13 & 1.55 & 1.4 & 2.41 & 1.94 & \is & - \\
HE1105+0027 & 6132 & 3.45 & $-2.42$ & 2 &  -  &  -  & 2.45 & 1.81 & \is & - \\
HE1135+0139 & 5487 & 1.8 & $-2.31$ & 1.19 &  -  &  -  & 1.13 & 0.33 & \is & - \\
HE1305+0007 & 4560 & 1 & $-2.03$ & 1.84 & 1.9 & 0.8 & 2.32 & 1.97 & \is & - \\
HE2148-1247 & 6380 & 3.9 & $-2.32$ & 1.91 & 1.65 &  -  & 2.36 & 1.98 & \is & yes \\
%count(*)
%18

C-rich-no{\it s} (2) & & & &     &      &   &      & & & \\

CS22941-012 & 7200 & 4.2 & $-2.03$ & 1 &  -  & 0.4 & 0.16 & $<$1.31 & \ir & - \\
HD135148 & 4183 & 1.24 & $-2.17$ & 0.8 &  -  & 0.4 & 0.3 & 0.71 & \ir & 1416 \\
%count(*)
%2

Other C-rich (45) & & & &     &      &   &      & & & \\

CS22166-016 & 5388 & 3.26 & $-2.4$ & 1.02 &  -  & - & $-0.37$ &  -  & \ir & - \\
CS22876-042 & 6750 & 4.2 & $-2.06$ & 1 &  -  & 0.54 & $-0.1$ &  -  & \ir & - \\
CS22879-029 & 6300 & 3.9 & $-1.93$ & 1.3 & $<$0.9 &  -  &  -  &  -  & - & - \\
CS22884-097 & 6460 & 4 & $-1.94$ & 1.3 & $<$1.85 &  -  &  -  &  -  & - & - \\
CS22891-171 & 5297 & 3.07 & $-2.45$ & 1.45 & 0.6 &  -  &  -  &  -  & - & - \\
CS22945-024 & 5289 & 3.04 & $-2.26$ & 0.7 & $-0.3$ &  -  &  -  &  -  & - & - \\
CS22949-008 & 6144 & 3.82 & $-1.92$ & 1.15 & 0.3 &  -  &  -  &  -  & - & - \\
CS22956-028 & 6900 & 3.9 & $-2.08$ & 1.74 &  -  & 0.5 & 0.37 &  -  & \ir & 1290 \\
CS22964-214 & 6800 & 4.5 & $-2.3$ & 1 &  -  & 0.7 & $-0.12$ &  -  & \ir & - \\
CS29495-042 & 5400 & 3.32 & $-2.3$ & 1 & 0.5 &  -  &  -  &  -  & - & - \\
CS29503-010 & 6500 & 4.5 & $-1.06$ & 1.07 & $<$1.28 &  -  & 1.5 &  -  & \is & - \\
CS29509-027 & 7050 & 4.2 & $-2.02$ & 1.38 &  -  & 0.55 & 1.33 &  -  & \is & 196 \\
CS29512-073 & 5751 & 3.62 & $-2.1$ & 1.05 & 0 &  -  &  -  &  -  & - & - \\
CS30338-089 & 5000 & 2.1 & $-2.45$ & 2.06 & 1.27 &  -  & 2.22 &  -  & \is & - \\
HD13826     & 3500 & 0.5 & $-2.4$  & 1.5  & 1.5  & 0.2 &  -   &  -  & - & - \\
HD187216    & 3500 & 0.4 & $-2.48$ & 1.3 & 0.2 &  -  & 2.5 &  -  & \is & - \\
HD5223      & 4500 & 1 & $-2.06$ & 1.57 &  -  &  -  & 1.82 &  -  & \is & - \\
HE0206-1916 & 5200 & 2.7 & $-2.09$ & 2.1 & 1.61 &  -  & 1.97 &  -  & \is & - \\
HE0212-0557 & 5075 & 2.15 & $-2.27$ & 1.74 & 1.09 &  -  & 2.18 &  -  & \is & - \\
HE0231-4016 & 5972 & 3.59 & $-2.08$ & 1.36 &  -  &  -  & 1.47 &  -  & \is & - \\
HE0322-1504 & 4460 & 0.8 & -2 & 2.3 & 2.2 & 1 & 2.8 &  -  & \is & - \\
HE0400-2030 & 5600 & 3.5 & $-1.73$ & 1.14 & 2.75 &  -  & 1.64 &  -  & \is & - \\
HE0430-4404 & 6214 & 4.27 & $-2.08$ & 1.44 &  -  &  -  & 1.62 &  -  & \is & - \\
HE0441-0652 & 4900 & 1.4 & $-2.47$ & 1.38 & 0.89 &  -  & 1.11 &  -  & \is & - \\
HE0507-1430 & 4560 & 0.8 & $-2.4$ & 2.6 & 1.7 & 1.1 & 1.3 &  -  & \is & - \\
HE0507-1653 & 5000 & 2.4 & $-1.38$ & 1.29 & 0.8 &  -  & 1.89 &  -  & \is & - \\
HE0534-4548 & 4250 & 1.5 & $-1.8$ & 1.5 & 1.1 & 0.1 & 0.6 &  -  & \is & - \\
HE1157-0518 & 4900 & 2 & $-2.34$ & 2.15 & 1.56 &  -  & 2.14 &  -  & \is & - \\
HE1319-1935 & 4600 & 1.1 & $-1.74$ & 1.45 & 0.46 &  -  & 1.89 &  -  & \is & - \\
HE1330-0354 & 6257 & 4.13 & $-2.29$ & 1.05 &  -  &  -  & $-0.47$ &  -  & \ir & - \\
HE1343-0640 & 5942 & 3.97 & $-1.9$ & 0.77 &  -  &  -  & 0.7 &  -  & \is & - \\
HE1410+0213 & 4985 & 2 & $-2.16$ & 1.73 & 1.78 &  -  & 0.07 &  -  & \ir & - \\
HE1429-0551 & 4700 & 1.5 & $-2.47$ & 2.28 & 1.39 &  -  & 1.57 &  -  & \is & - \\
HE1434-1442 & 5420 & 3.15 & $-2.39$ & 1.95 & 1.4 &  -  & 1.23 &  -  & \is & - \\
HE1443+0113 & 4945 & 1.95 & $-2.07$ & 1.84 &  -  &  -  & 1.4 &  -  & \is & - \\
HE1447+0102 & 5100 & 1.7 & $-2.47$ & 2.48 & 1.39 &  -  & 2.7 &  -  & \is & - \\
HE1523-1155 & 4800 & 1.6 & $-2.15$ & 1.86 & 1.67 &  -  & 1.72 &  -  & \is & - \\
HE2150-0825 & 5960 & 3.67 & $-1.98$ & 1.35 &  -  &  -  & 1.7 &  -  & \is & - \\
HE2221-0453 & 4400 & 0.4 & $-2.22$ & 1.83 & 0.84 &  -  & 1.75 &  -  & \is & - \\
HE2227-4044 & 5811 & 3.85 & $-2.32$ & 1.67 &  -  &  -  & 1.38 &  -  & \is & - \\
HE2228-0706 & 5100 & 2.6 & $-2.41$ & 2.32 & 1.13 &  -  & 2.5 &  -  & \is & - \\
HE2232-0603 & 5750 & 3.5 & $-1.85$ & 1.22 & 0.47 &  -  & 1.41 &  -  & \is & - \\
HE2240-0412 & 5852 & 4.33 & $-2.2$ & 1.35 &  -  &  -  & 1.37 &  -  & \is & - \\
SDSS0036-10 & 6500 & 4.5 & $-2.41$ & 2.32 &  -  &  -  & 0.29 &  -  & \ir & - \\
SDSS2047+00 & 6600 & 4.5 & $-2.05$ & 2 &  -  &  -  & 1.5 &  -  & \is & - \\
%count(*)
%43
      \hline
   \end{tabular}
   \end{minipage}
\end{table*}

\begin{table*}
  \caption{Seven NEMP stars defined in this study following Table~\ref{tab:class}.
  The meanings of the columns are the same as in Tab.~\ref{tab:cemp}.
  }\label{tab:nemp}
   \centering
    \begin{minipage}{140mm}
    \begin{tabular}{l*{11}{c}}
      \hline
       Object & \teff & $\log g$ & \feoh & \cfe &  \nfe & \ofe & \bafe & \eufe & {\it s}$/${\it r} & binary \\
      \hline
HE1337+0012 & 6270 & 4.4  & $-3.37$ & 0.49 & 1.42 & 0.73 & $-0.25$ & - & \ir & - \\
HE0400+2030 & 5600 & 3.5  & $-1.73$ & 1.14 & 2.75 & -    & 1.64 & - & \is & - \\
CS22960-053 & 5200 & 2.1 & $-3.14$ & 2.05 & 3.06 & - & 0.86 & - & \is & - \\
CS22949-037 & 4915 & 1.7 &  $-3.79$ & 0.97 & 2.16 & 1.96 & $-0.66$ & 0.04 & \ir & - \\
CS29528-041 & 6170 & 4.0 & $-3.06$ & 1.59 & 3.07 & $<$1.4 &  0.97 & - & \is & - \\
HE1031-0020 & 5080 & 2.2 & $-2.86$ & 1.63 & 2.48 & - & 1.21 & $<$0.87 & \is & - \\
CS30322-023 & 4300 & 1.0 & $-3.25$ & 0.56 & 2.47 & 0.63 &  0.59 & $-0.51$ & \is & - \\
      \hline
   \end{tabular}
    \end{minipage}
   \medskip
\end{table*}

\newpage

\section{Characteristics of Lithium Depletion and Binarity}

The peculiar feature of Li abundance, discussed in \S~\ref{sec:mixing}, is also shared by some carbon-normal dwarfs.
We can see three lithium depleted EMP dwarfs without carbon enhancement in Fig.~\ref{fig:li}.
Among them, G 158-100 and G 82-23 have low effective temperatures ($\teff \approx 4900$ K) and should be at the lower main sequence.
These stars are thought to have small total masses, and hence, have deep convective envelopes to deplete lithium.
The star HD 340279 is a typical turn-off star and does not show any peculiarity except for lithium depletion, i.e., neither carbon enhancement \citep[$\cfe = 0.21$][]{Akerman2004} nor peculiar abundance patterns for any other elements (Be, C, O, S, and Fe) whose abundances are available from the literature.
As for normal metal-poor stars, \citet{Shi2007} reported two stars having \loge $\sim 1.5$ among 19 metal-poor halo dwarfs, and \citet{Gratton2000} reported HD 196892 whose Li abundance is \loge $= 1.00$.

There are 6 possible or confirmed binaries among the Li-depleted stars.
Among the dwarfs in binary systems, HE1353-2735 is a double lined spectroscopic binary that shows the ordinary lithium abundance of $\logli = 2.06$ \citep{Depagne2000}.
This can be naturally explained by its evolutionary status that both of the binary components in HE1353-2735 are unevolved dwarfs, although the secondary component has a relatively low effective temperature (5200 K).
On the other hand, CS29497-030, CS22948-027, G 77-61, and HE0024-2523 shows the lithium depletion with binarity confirmed by the radial velocity monitoring.
Among them, CS29497-030 and CS22948-027 have a period of 342 days \citep{Sneden2003} and 426.5 days \citep{Barbuy2005}, respectively.
They are the typical cases of possible AGB mass transfer because they share the common feature of large C and Pb enhancement.
However, CS22948-027 is a evolved giant and its lithium depletion can be simply explained by the deepening of the convective envelope \citep[$\logli < 1.0 $,][]{Aoki2002a}.
Another CEMP-MS star, HE0024-2523, that has a very small binary period of 3.14 days \citep{Lucatello2003} may belong to the same group.
This star is also expected to reflect the envelope abundances of an AGB companion.
G 77-61 is another CEMP-MS star with Li depletion, while the binary period is determined to be 245 days \citep{Plez2005}.
This star shows the low upper limit for Ba and no signature of active \sit process.
It is not clear that Li in G 77-61 was affected by a binary companion because this star is expected to be wholly convective due to its small total mass of $\sim 0.3 \msun$.
Of course, we still cannot exclude the possibility of mass transfer for this star because its large carbon and nitrogen abundances are to be explained.

Possible scenarios to explain the Li-depleted dwarfs are discussed in \citet{Ryan2001}.
They argue that Li-depleted stars are low mass secondaries of binary companions whose initial masses are smaller than the turn-off masses ($\sim 0.8 \msun$) and their surface lithium abundance is depleted by a deep convective envelope or by binary mass transfer.
To find evidence of binary mass transfer, \citet{Ryan2002} explore the possibility of angular momentum transfer for Li-depleted dwarfs by measuring the rotational broadening of these objects.
They find a line broadening for three of the four lithium-depleted dwarfs, arguing that these stars are affected by angular momentum and mass transfer.
The same discussion will apply for the EMP counterparts.
It is to be noted that this scenario is to relate the Li-depleted stars with blue straggler stars (BSSs), but not with CEMP stars.
It is to be stressed, however, that the Li-depleted stars may also be related to carbon-rich stars because of the large frequency of carbon-rich stars among BSSs.
If we define the BSSs by $\teff \geq 6500$ K, the sum of Crich and CEMP stars is 20, while the number of MP and EMP stars is 71.
If we set the boundary at $\teff = 6800$ K, the number of carbon-enhanced and carbon-normal stars is 7 and 44, respectively.
Note that, at $\teff \geq 6500$ K, only 3 stars have derived carbon abundances for the stars in the EMP MS group and none of the stars in the MP group have derived carbon abundances.  

\bibliographystyle{mn}
\bibliography{reference,reference_mps_obs,reference_gc}

\end{document}